\documentclass{aa}  
\usepackage{graphicx}
\usepackage{supertabular}
\usepackage{longtable}
\usepackage{txfonts}
\usepackage{natbib}
\bibpunct{(}{)}{;}{a}{}{,}
\begin{document}

\title{Stellar abundances and ages for metal-rich Milky Way globular clusters}

   {\subtitle{Stellar parameters and elemental abundances for 9 HB
       stars in NGC\,6352{\thanks{Based on observations collected at
           the European Southern Observatory, Chile, ESO
           No. 69.B-0467}}\thanks{Table\,\ref{eqw.tab} is only available in
the on-line version of the paper or in electronic form
at the CDS via anonymous ftp to cdsarc.u-strasbg.fr (130.79.128.5)
or via http://cdsweb.u-strasbg.fr/cgi-bin/qcat?J/A+A/}}

   \author{S. Feltzing
          \inst{1}
          \and
          F. Primas \inst{2}
          \and
	  R.A. Johnson \inst{3}
          }

   \offprints{S. Feltzing}

   \institute{Lund Observatory, Box 43, SE-221 00 Lund, Sweden\\
              \email{sofia@astro.lu.se}
	 \and 
	     European Southern Observatory, Karl-Schwarzschild Str. 2, 85748 Garching b. M\"unchen, Germany \\
	     \email{fprimas@eso.org}
         \and
             Astrophysics, Oxford University, Denys Wilkinson Building, Keble Road, Oxford,
	     OX1 3RH, UK\\
             \email{raj@astro.ox.ac.uk}
             }

   \date{Received 2008-05-06; Accepted 2008-09-16}

  \abstract
  {
Metal-rich globular clusters provide important tracers of the
formation of our Galaxy.  Moreover, and not less important, they are
very important calibrators for the derivation of properties of
extra-galactic metal-rich stellar populations. Nonetheless, only a few
of the metal-rich globular clusters in the Milky Way have been studied
using high-resolution stellar spectra to derive elemental
abundances. Additionally, Rosenberg et al. identified a small group of
metal-rich globular clusters that appeared to be about 2 billion years
younger than the bulk of the Milky Way globular clusters. However, it
is unclear if like is compared with like in this dataset as we do not
know the enhancement of $\alpha$-elements in the clusters and the
amount of $\alpha$-elements is well known to influence the derivation
of ages for globular clusters.  }
   {To derive elemental abundances for the metal-rich globular cluster NGC\,6352
and to present our methods to be used in up-coming
studies of other metal-rich globular
clusters. }
   {We present a study of elemental abundances for $\alpha$- and iron-peak
elements for nine HB stars in the metal-rich globular cluster
NGC\,6352. The elemental abundances are based on high-resolution, high
signal-to-noise spectra obtained with the UVES spectrograph on
VLT. The elemental abundances have been derived using standard LTE
calculations and stellar parameters have been derived from the
spectra themselves by requiring ionizational as well as excitational
equilibrium.}
   {We find that NGC\,6352 has {[Fe/H]$= -0.55$}, is enhanced in the 
$\alpha$-elements to about +0.2 dex for Ca, Si, and Ti relative to Fe.
For the iron-peak elements we find  solar values. Based on 
the spectroscopically derived stellar parameters 
we find that an $E(B-V)=0.24$ and $(m-M)\simeq14.05$ better fits the data
than the nominal values. 
An investigation
of $\log gf$-values for suitable Fe\,{\sc i} lines lead us to the 
conclusion that the commonly used correction to the May et al.\,(1974)
data should not be employed.}

   \keywords{(Galaxy:) globular clusters: individual:NGC\,6352, 
Stars: horizontal-branch,
Stars: abundances 
               }

   \maketitle
%

\section{Introduction}

The globular clusters in a galaxy trace (part of) the formation
history of their host galaxy, in particular merger events have been
shown to trigger intense periods of formation of stellar clusters
\citep[e.g.][]{forbes2006}. The perhaps most spectacular evidence of
such an event is provided by the Antennae galaxies
\citep{whitmore1995,whitmore1999}.  Results for the recent merger
system NGC\,1052/1316 appear to show that indeed some of the clusters
that form in a merger event between gas-rich galaxies may result in
what we today identify as globular clusters
\citep{forbes2006,goudfrooij2001,pierce2005}.

Even though globular clusters are thought to probe important episodes
in the formation of galaxies there is increasing evidence that they
may not be a fair representation of the underlying stellar
populations.  For example, \citet{vandalfsen2004} point out the
increasing evidence that the metallicity distribution functions for
globular clusters in other galaxies less and less resemble the
metallicity distribution functions of the field stars in their host
galaxies.

Nevertheless, globular clusters provide one of the most powerful tools
for studying the past history of galaxies outside the Local Group and
in order to fully utilize this it becomes important to find local
templates that can be used to infer the properties of the
extra-galactic clusters. Such templates can be provided by the Milky
Way globular clusters and clusters in the LMC and SMC. There is a
large literature on this, especially for the metal-poor clusters
\citep[i.e. for clusters with iron abundances less than --1 dex, see
  e.g.][and references therein]{gratton2004}. However, for the
metal-rich clusters with with iron abundances larger than $-$1 dex
(which are extremely important for studies of e.g. bulges and other
metal-rich components of galaxies) the situation is less developed.

The Milky Way has around 150 globular clusters. These show a bimodal
distribution in colour as well as in metallicity
\citep[e.g.][]{zinn1985}. Such bimodalities are quite commonly
observed also in other galaxies.  The source of the bimodality
could be a period of heightened star formation, perhaps triggered by a
major merger or a close encounter with another (large) galaxy. For
example, \citet{casuso2006} advocates a picture where the metal-rich
globular clusters in the Milky Way formed during times of enhanced
star formation (perhaps triggered by a close passing by by the LMC
and/or SMC) and that some, but not all, of these new young clusters
were ``expelled'' to altitudes more akin to the thick than the thin
disk or that the clusters actually formed at these higher
altitudes. That second possibility is somewhat related to the model by
\citet{kroupa2002} which was developed to explain the scale height of
the Milky Way thick disk. In contrast, \citet{vandalfsen2004}
advocates a fairly simple chemical evolution model of the
``accreting-box'' sort to explain the bimodal metallicity distribution
of the globular clusters in the Milky Way. This model is able to
reproduce the observed metallicity distribution function but offers no
explicit explanation of {\it why} the different epochs of heightened
star formation happened.

To put constraints on these types of models it thus becomes
interesting to study the age-structure for the globular clusters in
the Milky Way. \citet{rosenberg1999} found that a small group
of metal-rich clusters, NGC\,6352, 47 Tuc, NGC\,6366, and NGC\,6388
(all with [Fe/H] $> -0.9$), show apparent young ages, around 2 Gyr
younger than the bulk of the cluster system.  As discussed in detail
in \citet{rosenberg1999} the ages of this group are model dependent,
but, the internal consistency is remarkable and intriguing. However,
it is not clear if like is compared with like in this group of
clusters.  The reasons are (at least) two, first this group includes a
mixture of disk and halo clusters, secondly knowledge of the
$\alpha$-enhancement is not available for all of the clusters. In fact
these concerns are connected. We know, from the local field dwarfs,
that the chemical evolution in the halo and the disk are different,
i.e.  the majority of the stars in the halo have a large $\alpha$-enhancement, while
in the disk we see a decline of the $\alpha$-enhancement starting
somewhere around the metallicities of these clusters \citep[see
  e.g.][]{bensby2005}.  Thus it could well be that the halo and disk
clusters have distinct profiles as concerns their elemental
abundances. In that case the derivation of the ages of the clusters in
relation to each other might be erroneous as $\alpha$-enhancement
clearly affect age determinations \citep[see
  e.g.][]{salasnich2000,kim2002}.

We have therefore constructed a program to provide a homogeneous set
of elemental abundances for a representative set of metal-rich
globular clusters, including both halo and bulge clusters. The two
globular clusters NGC\,6352 and NGC\,6366 provide an unusually
well-suited pair to target for a detailed abundance analysis.
NGC\,6352 is a member of the {\sl disk} cluster population while
NGC\,6366, although it is metal-rich, unambiguously, due to its
kinematics, belong to the {\sl halo} population. 

Further, both clusters are ideal for spectroscopic studies since they
are sparsely populated. This means that it is easy to position the
slit on individual stars even in the very central parts of the
cluster.  47 Tuc on the other hand is around 100 times more crowded
and spectroscopy of single stars becomes increasingly difficult. The
fourth cluster, NGC\,6388, is also very centrally concentrated and
therefore less amenable to spectroscopic studies. For both NGC\,6352
and NGC\,6366 the background contamination is minimal so that the
selected horizontal branch (HB) stars should all be members.

Good colour-magnitude diagrams exist for both clusters; for NGC\,6352
based on HST/WFPC2 observations and for NGC\,6366 a good ground-based
CMD exists \citep{alonso1997}. Combined with our new elemental
abundances we would thus be in a position to do a relative age dating
of these two clusters.

We have obtained spectra for nine HB stars in NGC\,6352 and eight in
NGC\,6366. In addition we also have data for six HB and red giant
branch stars (RGB) in NGC\,6528 from our own observations which will
be combined with observations of additional stars present in the VLT
archive. Additional archival material exist for other metal-rich
globular clusters. Also for NGC\,6528 decent CMDs exist
\citep[e.g.][]{feltzing2002}.

Here we report on the first determinations of elemental
abundances for one of the globular clusters, NGC\,6352, in the
program. We also spend extra time explaining the methods that we will
use also for the other cluster, especially as concerns the choice of
atomic data for the abundance analysis.

The paper is organized as follows: in Sect.\,\ref{sect:selection1} we
describe the selection of target stars for the spectroscopic
observations in NGC\,6352.  Section\,\ref{sect:spec} deals with the
observations, data reduction and analysis of the stellar
spectra. Section\,\ref{sect:abun} describes in detail our abundance
analysis, including a discussion of the atomic data used. In
Sect.\,\ref{sect:res} the elemental abundance results are
presented. The results are discussed in Sect.\,\ref{sect:disc} in the
context of other metal-rich globular clusters and the Milky Way
stellar populations in general.  Section\,\ref{sect:sum} provides a
summary of our findings.

\section{Selection of stellar sample for our spectroscopic programme for NGC\,6532}
\label{sect:selection1}

\begin{table*}
\caption[]{Data for our sample.  The first column gives our designation for
  the stars (compare Fig.\,\ref{fig.image}), second and third give
  alternative designations of the stars from \citet{alcaino1971}
  (marked by A$\times$$\times$$\times$) and \citet{hartwick1972}
  (marked by H$\times$$\times$$\times$). Column four and five give the
  stellar coordinates \citep[taken from the 2MASS
    survey,][]{skrutskie2006}. Columns six and seven give the
  HST/WFPC2 in-flight magnitudes and colours.  The last column lists
  the $K$ magnitude for the stars from the 2MASS survey
  \citep{skrutskie2006}.}
\label{phot.tab}
\begin{tabular}{llllllllllll}
\hline\hline 
\noalign{\smallskip}
Star  & A$\times$$\times$$\times$ & H$\times$$\times$$\times$ & $\alpha$ &  $\delta$ & $V_{\rm 555}$ & 
 $V_{\rm 555}-I_{\rm 814}$ &  $K$\\
\noalign{\smallskip}
\hline
\noalign{\smallskip}
NGC\,6352-01 &  --   & --   &261.378121 &-48.425865& 15.32 & 1.28 & 12.437 \\
NGC\,6352-02 &  --   & H220 &261.400858 &-48.420418& 15.34 & 1.33 & 12.181\\
NGC\,6352-03 &  A61  & H56  &261.392403 &-48.428165& 15.24 & 1.18 & 12.320\\
NGC\,6352-04 &  A58  & H234 &261.409088 &-48.432890& 15.30 & 1.20 & 12.325\\
NGC\,6352-05 &  A56  & H237 &261.405980 &-48.436588& 15.22 & 1.17 & 12.342\\
NGC\,6352-06 &  A155 & H250 &261.384160 &-48.440037& 15.28 & 1.16 & 12.445\\
NGC\,6352-07 &  A152 & H252 &261.377315 &-48.441895& 15.26 & 1.20 & 12.420\\
NGC\,6352-08 &  A151 & --   &261.376366 &-48.443417& 15.25 & 1.16 & 12.406\\
NGC\,6352-09 &  A150 & H253 &261.373965 &-48.443913& 15.30 & 1.21 & 12.354\\
\noalign{\smallskip}
\hline
\end{tabular}
\end{table*}

\begin{figure*}
\centering
\resizebox{12cm}{!}{\includegraphics[bb = 40 60 557 550]{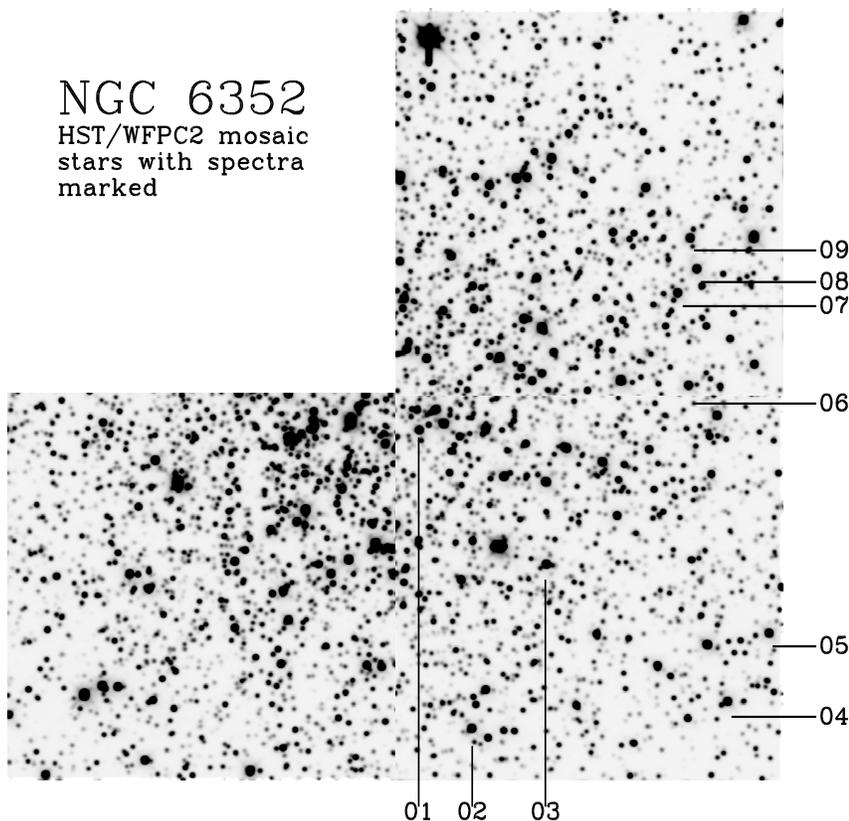}}
\hfill
\parbox[b]{55mm}{
\caption{HST/WFPC2 mosaic image of NGC\,6352 (the PC image is excluded).
The stars with UVES spectra are marked with the corresponding numbers from
Table\,\ref{phot.tab}. This image is created from the following three
HST/WFPC2 datasets: u28q0404t, u28q0405t, and u28q040bt.
}\label{fig.image}}
\end{figure*}

Stars for the spectroscopic observations were selected based on their
position in the CMD.  Only a few stars in NGC\,6352 have previously
been studied with spectroscopy and hence there was no prior knowledge
of cluster membership. Therefore we decided to select only stars on
the HB in order to maximize the possibility for them to be
members. Selecting HB stars rather then RGB and AGB stars has the
further advantage that the stars will have fairly high effective
temperatures ($T_{\rm eff}$) which significantly will facilitate the
analysis of the stellar spectra. At lower temperatures the amount of
molecular lines start to become rather problematic \citep[see e.g. the
  discussion in][]{barbuy2000,carretta2001,cohen1999}.

The HB in NGC\,6352 is situated at $V_{\rm 555} \sim 15.2$. Data for
the target stars for the spectroscopic programme are listed in Table
\ref{phot.tab}.  In Fig.\,\ref{fig.image} we show a mosaic image based
on HST/WFPC2 images with the stars observed in the spectroscopic
programme labeled by their corresponding numbers from Table
\ref{phot.tab}. The table also includes a cross-identification with
designations used in other major studies of NGC\,6352
\citep{alcaino1971,hartwick1972}.

\section{Spectroscopy}
\label{sect:spec}

\subsection{Observations and data reduction}

Observations were carried out in service mode as part of observing
programme 69.B-0467 with the UVES spectrograph on Kueyen. We used the
red CCD with a standard setting centered at 580.0 nm. With this
setting we cover the stellar spectra from 480.0 to 680.0 nm with a gap
between 576.0 nm and 583.5 nm. Each star was observed for 4800 s in a
single exposure.

The spectra were pipeline calibrated as part of the service mode
operation.  As our spectra are of moderate S/N (in the red up to 80,
but in the blue more like 60) we have visually inspected the reduced
and extracted one-dimensional spectra for known foibles and found them
to not suffer from any of these problems.

\subsection{Radial velocity measurements and cluster membership}

Radial velocities were measured from the stellar spectra using the
{\sc rv} suite of programs inside {\sc iraf}\footnote{IRAF is
  distributed by National Optical Astronomy Observatories, operated by
  the Association of Universities for Research in Astronomy, Inc.,
  under contract with the National Science Foundation, USA.}. From the
observed radial velocities helio centric velocities and velocities
relative to the local standard of rest (LSR) were calculated and are
listed in Table\,\ref{rv.tab}. We find the cluster to have a mean
velocity relative to the LSR of --120.7 km s$^{-1}$ with $\sigma=3.7$
km s$^{-1}$.  All of our program stars have velocities that deviate
less then 2$\sigma$ from the mean velocity. Hence they are all
members.

The most recent value for $V_{\rm LSR}$ in the catalogue of globular
clusters \citep[][catalogue
\footnote{We have used the latest revision (2003) available at
  http://www.physics.mcmaster.ca/Globular.html}]{harris1996} is $-116.7$\,km
s$^{-1}$.  This is in reasonably good agreement with our new result
based on data for nine stars. The \citet{harris1996} value is based
on a weighted average from three studies
\citep{rutledge1997,zw84,hesser1986}. \citet{rutledge1997} found
$V_{\rm Helio} = -122.8$\,km s$^{-1}$ for a sample of 23 stars. Using
the following equation

$$V_{\rm LSR} = V_{\rm Helio} + 11.0 \cos b \cos l + 14.0 \cos b \sin l + 7.5 \sin b $$

\noindent
\citep[][]{1989ApJ...339..106R}, with $l=341.4$ and $b=-7.2$ for NGC\,6352,
this corresponds to a $V_{\rm LSR} =-117.9$\,km s$^{-1}$. We note that
\citet{rutledge1997} estimate their external errors for the
measurement of the radial velocities for stars in NGC\,6352 to be on
the order of 10\,km s$^{-1}$. More recently, \citet{carrera2007}
find $V_{\rm Helio} = -114$\,km\,s$^{-1}$ based on 23 stars, which is
equivalent to $V_{\rm LSR} = -109$\,km\,s$^{-1}$.  No estimate of
external errors are given in their study. Their value is more similar
to that measured by \citet{hartwick1972}, $V_{\rm
  Helio}=-112.2$\,km\,s$^{-1}$, than to ours. There are no stars in
common between our study and \citet{carrera2007}\footnote{We thank
  the authours for making the coordinates of their sample available to
  us so that we could check for common stars. None were found.}.

Hence, it does appear that our estimate of $V_{\rm LSR}$ for NGC\,6352
is somewhat high when compared to other estimates available in the
literature. However, as we do not have a good estimate of zero-point
errors for the various studies and as no doubt different types of
stars have been used in the various studies, e.g.  we use only HB
stars whilst some of the earliest studies clearly will have relied on
very cool giants where e.g. motions in the stellar atmospheres might
play a role \citep{carney2003}, and since we have no information on binarity for any of
these stars the current value should be regarded as being in good
agreement with previous estimates.

\begin{table}
\caption[]{Measured and derived velocities. The second column
gives the radial velocity of the star as measured from the
stellar spectrum. The third column the derived
 helio centric velocity and the fourth the 
velocity relative to the local standard of rest (LSR). 
The last line gives the mean helio centric
velocity for all the stars and the corresponding standard deviation 
as well as the mean LSR velocity with its corresponding 
standard deviation.}
\label{rv.tab}
\begin{tabular}{llllllllllll}
\hline\hline 
\noalign{\smallskip}
Star  & $V_{\rm 0bs}$ & $  V_{\rm Helio}$ & $  V_{\rm LSR}$\\  
      & km s$^{\rm -1}$ &km s$^{\rm -1}$ &km s$^{\rm -1}$ \\
\noalign{\smallskip}
\hline
\noalign{\smallskip}
NGC\,6352-01 &  --154.30 & --146.18 & --127.36  \\ 
NGC\,6352-02 &  --147.13 & --142.05 & --123.24 \\ 
NGC\,6352-03 &  --141.65 & --136.72 & --117.90 \\ 
NGC\,6352-04 &  --140.71 & --135.87 & --117.06 \\ 
NGC\,6352-05 &  --144.62 & --139.90 & --121.09 \\ 
NGC\,6352-06 &  --137.72 & --133.75 & --114.94 \\ 
NGC\,6352-07 &  --143.60 & --139.52 & --120.71 \\ 
NGC\,6352-08 &  --144.81 & --140.63 & --121.82 \\ 
NGC\,6352-09 &  --148.53 & --141.31 & --122.50 \\
\noalign{\smallskip} 
\hline
\noalign{\smallskip}
NGC\,6352    &           & --139.5 $\sigma=3.7$ & --120.7 $\sigma=3.7$\\
\noalign{\smallskip}
\hline 
\end{tabular}
\end{table}

\subsection{Measurement of equivalent widths}

Equivalent widths were measured using the {\sc splot} task in {\sc
  iraf}. For each line the local continuum was estimated with the help
of synthetic spectra generated using appropriate stellar parameters
and a line-list, typical for a K giant, from VALD, see \citet{vald1},
\citet{vald2}, and \citet{vald3}. The equivalent widths used in the
abundance analysis are listed in Table\,\ref{eqw.tab}\footnote{
  Table\,\ref{eqw.tab} only appear in the online
  material. Table\,\ref{eqw.tab} is also available in electronic form
  at the CDS via anonymous ftp to cdsarc.u-strasbg.fr (130.79.128.5)
  or via http://cdsweb.u-strasbg.fr/cgi-bin/qcat?J/A+A/.}.

\onllongtab{3}{
\begin{longtable}{llrrrrrrrrrrr}
\caption[]{Measured equivalent widths used in the abundance
  analysis. Column one lists the element and the ionizational state
  for the line, col. 2 the wavelength (in {\AA}), col. 3 the
  excitation energy of the lower level involved in the transition,
  col. 4 the oscillator strength (see Sect.\,\ref{sect:abun} for
  references). Cols. 5 to 13 then lists the measured equivalenth
  widths in m{\AA} for each star, as indicated in the header. The stars are
  identified in Fig.\,\ref{fig.image} and Table\,\ref{phot.tab}). }\label{eqw.tab}\\
\hline
\hline 
\noalign{\smallskip}
El & $\lambda$ & $\chi_{\rm l}$ & $\log gf$ &   NGC\,6352-01 & -02 & -03 & -04 & -05 & -06 & -07 & -08 & -09\\
\noalign{\smallskip}
\hline
\endfirsthead
\caption{Continued.}\\
\hline\hline
\noalign{\smallskip}
El & $\lambda$ & $\chi_{\rm l}$ & $\log gf$ &   NGC\,6352-01 & -02 & -03 & -04 & -05 & -06 & -07 & -08 & -09\\
\noalign{\smallskip}
\hline
\endhead
\noalign{\smallskip}
\hline
\endfoot
\noalign{\smallskip}
\hline
\endlastfoot
Na\,{\sc i} & 5682.65 & 2.10 & -0.71 &  89.6 &  98.7 & 119.2 &  93.1 & 115.6 &  89.9 &  88.6 & 102.5 &  92.4\\
Na\,{\sc i} & 5688.22 & 2.10 & -0.40 & 113.5 & 112.6 & 128.4 & 111.6 & 128.4 & 110.3 & 110.2 & 119.3 & 116.4\\
Na\,{\sc i} & 6154.22 & 2.10 & -1.57 &  36.3 &  32.2 &  43.2 &  30.9 &  49.4 &  31.3 &  28.8 &  41.2 &  30.3\\
Na\,{\sc i} & 6160.75 & 2.10 & -1.27 &  49.8 &  53.5 &  70.0 &  43.7 &  70.2 &  44.1 &  47.6 &  60.2 &  56.1\\
Mg\,{\sc i} & 5711.09 & 4.33 & -1.87 & 112.7 & 115.3 & 116.9 & 118.1 & 116.8 & 118.3 & 116.5 & 118.5 & 121.7\\
Al\,{\sc i} & 6696.03 & 3.14 & -1.63 &  52.0 &  45.4 &  46.8 &       &  47.0 &  45.6 &  41.0 &  35.5 &  48.3\\
Al\,{\sc i} & 6698.67 & 3.14 & -1.92 &  21.3 &  23.6 &  31.9 &  23.8 &  17.5 &  24.8 &  23.5 &       &      \\
Si\,{\sc i} & 5128.03 & 5.08 & -2.60 &  16.0 &  18.3 &  18.9 &  22.2 &  15.5 &       &       &       &  22.0\\
Si\,{\sc i} & 5517.53 & 5.08 & -2.38 &  16.2 &  15.1 &  18.0 &  18.8 &  22.0 &  16.4 &  17.6 &  17.2 &  17.9\\
Si\,{\sc i} & 5621.60 & 5.08 & -2.50 &   8.0 &  10.9 &   6.3 &   8.2 &   7.7 &   5.6 &  12.7 &   6.1 &      \\
Si\,{\sc i} & 5645.61 & 4.92 & -2.04 &  40.4 &  44.1 &  41.0 &  40.8 &  37.0 &  40.6 &  41.6 &  45.1 &  44.5\\
Si\,{\sc i} & 5665.55 & 4.93 & -1.94 &  50.7 &  49.2 &       &  46.7 &  48.9 &  43.8 &  51.9 &  44.8 &  45.3\\
Si\,{\sc i} & 5684.48 & 4.95 & -1.55 &  66.5 &  65.7 &  67.4 &  57.2 &  61.1 &  64.9 &  65.3 &  65.8 &  67.0\\
Si\,{\sc i} & 5701.12 & 4.93 & -1.95 &  45.0 &  42.9 &  38.0 &  42.6 &  35.9 &  49.9 &  44.1 &  45.0 &  44.0\\
Si\,{\sc i} & 5948.54 & 5.08 & -1.13 &  94.1 &  87.1 &  91.3 &  92.6 &  91.7 &  83.6 &  90.5 &  88.6 &  94.0\\
Si\,{\sc i} & 6125.03 & 5.61 & -1.52 &  30.6 &  33.7 &  27.4 &  33.1 &  28.8 &  30.5 &  31.3 &  28.5 &  33.2\\
Si\,{\sc i} & 6142.49 & 5.62 & -1.50 &  34.9 &  37.7 &  31.1 &  29.8 &  30.4 &  32.9 &  33.2 &  32.4 &  38.2\\
Si\,{\sc i} & 6145.02 & 5.61 & -1.46 &  33.5 &  36.4 &  32.8 &  31.8 &  36.1 &  36.9 &  35.4 &  29.9 &  34.5\\
Si\,{\sc i} & 6155.14 & 5.62 & -0.72 &  76.6 &  73.5 &  70.6 &  70.7 &  67.4 &  69.2 &  75.0 &  73.2 &  72.0\\
Si\,{\sc i} & 6555.46 & 5.98 & -1.00 &  48.4 &  44.5 &       &  30.6 &       &       &  31.8 &  14.2 &  38.3\\
Ca\,{\sc i} & 5260.39 & 2.52 & -1.71 &  44.6 &  46.7 &  39.6 &  42.1 &       &  34.6 &  36.9 &  36.4 &  39.6\\
Ca\,{\sc i} & 5512.98 & 2.93 & -0.44 &  89.5 &  88.7 &  86.2 &  89.2 &  87.5 &  87.5 &  86.7 &  85.9 &  91.9\\
Ca\,{\sc i} & 5581.97 & 2.52 & -0.55 & 101.2 & 102.5 &  97.6 & 103.4 & 102.2 &  97.9 & 101.8 &  96.7 & 103.9\\
Ca\,{\sc i} & 5601.28 & 2.53 & -0.52 & 124.1 & 116.0 & 108.7 & 109.3 & 105.3 & 112.9 & 105.0 & 110.0 & 102.1\\
Ca\,{\sc i} & 6161.30 & 2.52 & -1.26 &  76.4 &  76.3 &  65.1 &  71.6 &  62.5 &  74.4 &  69.3 &  68.2 &  75.8\\
Ca\,{\sc i} & 6166.44 & 2.52 & -1.14 &  81.1 &  83.0 &  77.6 &  76.9 &  75.4 &  77.0 &  76.6 &  73.9 &  85.5\\
Ca\,{\sc i} & 6169.04 & 2.52 & -0.79 & 101.4 & 100.6 &  92.6 &  98.0 & 100.0 &  95.6 &  96.9 &  96.5 & 109.1\\
Ca\,{\sc i} & 6169.56 & 2.52 & -0.47 & 118.9 & 110.9 & 113.3 & 115.0 & 112.5 & 111.1 & 109.3 & 110.1 & 120.1\\
Ca\,{\sc i} & 6439.08 & 2.52 &  0.39 & 164.5 & 168.8 & 160.0 & 172.4 & 168.6 & 168.4 & 163.4 & 170.0 & 164.6\\
Ca\,{\sc i} & 6455.61 & 2.52 & -1.29 &  65.1 &  70.1 &  64.9 &  57.2 &  62.6 &  64.0 &       &  61.7 &  60.4\\
Ca\,{\sc i} & 6471.67 & 2.52 & -0.68 & 102.6 & 105.1 & 102.4 &  99.0 &  94.7 &  96.9 & 101.1 &  94.7 & 107.1\\
Ca\,{\sc i} & 6493.79 & 2.52 & -0.10 & 139.9 &       &       &       &       &       &       &       &      \\
Ca\,{\sc i} & 6499.65 & 2.52 & -0.81 &  96.9 &  96.9 &  99.0 & 100.4 &  94.1 &  99.5 &  95.5 &  96.2 & 102.9\\
Ti\,{\sc i} & 4885.08 & 1.88 &  0.41 &       & 100.7 &  92.4 &  99.1 &  92.5 & 101.7 &  91.9 &  91.6 & 106.8\\
Ti\,{\sc i} & 4913.62 & 1.87 &  0.22 &  76.0 &  79.1 &  76.1 &  79.5 &  75.9 &  76.5 &  74.3 &  72.2 &  79.8\\
Ti\,{\sc i} & 4915.23 & 1.88 & -0.96 &  15.5 &  21.5 &  11.2 &  20.6 &  16.9 &  17.0 &  15.9 &  11.5 &      \\
Ti\,{\sc i} & 4981.73 & 0.84 &  0.56 & 155.4 & 152.6 & 141.6 & 148.2 & 139.4 & 150.2 & 135.3 & 146.3 & 154.0\\
Ti\,{\sc i} & 4997.09 & 0.00 & -2.06 &  83.8 &  84.0 &  74.6 &  79.8 &  73.3 &  73.8 &  71.1 &  72.6 &  75.2\\
Ti\,{\sc i} & 5000.99 & 1.99 &  0.02 &  62.5 &  55.5 &  55.5 &  71.1 &  51.5 &  62.2 &  54.2 &  55.9 &  61.8\\
Ti\,{\sc i} & 5016.16 & 0.85 & -0.52 & 103.5 & 106.1 & 100.0 & 106.4 &  95.2 & 102.2 & 101.7 &  98.8 & 101.6\\
Ti\,{\sc i} & 5020.02 & 0.83 & -0.35 & 118.0 & 118.3 & 116.0 & 117.9 & 117.8 & 116.1 & 120.0 & 113.5 & 116.5\\
Ti\,{\sc i} & 5022.87 & 0.83 & -0.38 & 114.7 &       & 107.8 & 113.2 & 113.9 & 104.7 & 108.2 & 104.7 & 110.3\\
Ti\,{\sc i} & 5024.85 & 0.82 & -0.55 & 110.6 & 107.3 & 101.8 & 104.3 & 104.1 & 100.7 & 102.7 &  96.2 & 109.8\\
Ti\,{\sc i} & 5043.59 & 0.83 & -1.67 &  45.4 &  48.9 &  46.7 &  58.4 &  45.4 &  43.3 &  45.3 &  41.6 &  49.2\\
Ti\,{\sc i} & 5062.10 & 2.16 & -0.40 &       &  29.1 &  25.2 &  23.1 &  23.9 &  23.2 &  25.8 &  21.7 &  25.1\\
Ti\,{\sc i} & 5071.46 & 1.46 & -1.00 &  43.3 &  50.4 &  44.2 &       &  46.6 &  47.5 &  43.4 &       &  47.0\\
Ti\,{\sc i} & 5087.06 & 1.43 & -0.78 &       &  57.7 &  54.8 &  63.9 &  47.8 &  54.4 &       &  48.3 &  59.7\\
Ti\,{\sc i} & 5113.45 & 1.44 & -0.73 &  54.3 &  55.4 &  50.5 &  56.2 &  49.4 &  49.2 &  46.7 &  46.0 &  55.0\\
Ti\,{\sc i} & 5145.46 & 1.46 & -0.51 &  61.0 &  63.6 &  60.7 &  66.6 &  64.0 &  60.7 &  60.8 &  57.8 &  67.0\\
Ti\,{\sc i} & 5147.47 & 0.00 & -1.95 &  85.0 &  88.1 &  78.6 &  84.1 &  79.0 &  81.4 &  78.4 &  73.1 &  87.7\\
Ti\,{\sc i} & 5201.08 & 2.09 & -0.69 &  22.6 &  18.7 &  23.7 &  25.4 &  27.3 &  21.0 &  22.7 &  19.0 &  26.2\\
Ti\,{\sc i} & 5210.38 & 0.04 & -0.82 & 138.7 & 148.5 & 140.5 & 145.2 & 131.5 & 137.5 & 136.8 & 134.0 & 142.5\\
Ti\,{\sc i} & 5219.70 & 0.02 & -2.24 &  71.7 &  71.5 &  70.2 &  71.5 &  68.3 &  65.3 &  66.1 &  61.7 &  71.5\\
Ti\,{\sc i} & 5223.62 & 2.09 & -0.50 &  32.3 &  28.6 &  34.4 &  29.6 &  33.9 &  27.8 &  30.1 &  26.8 &  30.0\\
Ti\,{\sc i} & 5426.26 & 0.02 & -2.95 &  28.6 &  27.9 &  29.9 &  34.2 &  24.9 &  31.6 &  27.4 &  23.7 &  28.4\\
Ti\,{\sc i} & 5471.20 & 1.44 & -1.40 &  21.8 &  24.8 &  26.0 &  21.8 &  15.3 &  24.6 &  23.5 &  20.7 &  21.1\\
Ti\,{\sc i} & 5474.23 & 1.46 & -1.23 &  23.0 &  32.2 &  26.7 &  36.8 &  31.4 &  26.5 &  38.1 &  30.2 &  34.4\\
Ti\,{\sc i} & 5490.15 & 1.46 & -0.88 &  53.1 &  54.9 &  47.8 &  55.0 &  48.8 &  46.6 &  49.0 &  44.2 &  50.2\\
Ti\,{\sc i} & 5662.15 & 2.31 & -0.05 &  40.9 &  45.3 &  36.5 &  45.9 &  43.1 &  42.3 &  42.3 &  34.0 &  47.6\\
Ti\,{\sc i} & 5679.91 & 2.47 & -0.41 &  12.7 &  11.5 &   9.9 &  11.2 &       &   7.0 &  11.6 &   4.4 &  13.3\\
Ti\,{\sc i} & 5689.46 & 2.29 & -0.36 &  20.4 &  22.2 &  15.7 &  25.4 &  20.2 &  20.3 &  20.4 &  16.9 &  22.6\\
Ti\,{\sc i} & 5702.65 & 2.29 & -0.59 &  13.0 &  20.2 &  23.7 &  22.8 &  15.1 &       &  13.3 &  19.6 &  17.6\\
Ti\,{\sc i} & 5716.44 & 2.29 & -0.72 &  11.9 &  15.6 &   9.1 &  12.3 &  14.5 &  14.3 &  11.4 &  10.9 &  13.3\\
Ti\,{\sc i} & 5866.46 & 1.07 & -0.78 &  87.6 &  90.5 &  84.5 &  83.7 &  83.0 &  82.4 &  85.0 &       &  89.3\\
Ti\,{\sc i} & 5880.26 & 1.05 & -1.70 &       &       &  25.2 &       &       &       &       &       &      \\
Ti\,{\sc i} & 5918.53 & 1.06 & -1.46 &       &  36.5 &  35.9 &  38.0 &  35.2 &  30.5 &  33.4 &       &  39.1\\
Ti\,{\sc i} & 5922.11 & 1.04 & -1.41 &  58.7 &  54.1 &  50.3 &  55.9 &       &  46.4 &  46.5 &  41.3 &  58.7\\
Ti\,{\sc i} & 5937.80 & 1.06 & -1.84 &  22.1 &  23.6 &  16.6 &  29.9 &  20.1 &  19.0 &  21.0 &  15.4 &  22.9\\
Ti\,{\sc i} & 5953.17 & 1.89 & -0.27 &  73.4 &  66.1 &  63.4 &  71.0 &  60.8 &  62.5 &  66.1 &  59.4 &  66.7\\
Ti\,{\sc i} & 6126.22 & 1.07 & -1.37 &  53.7 &  57.3 &  47.1 &  58.9 &  44.4 &  54.1 &  47.3 &  43.5 &  59.6\\
Ti\,{\sc i} & 6258.11 & 1.44 & -0.30 &  87.4 &  85.6 &  80.2 &       &       &  76.3 &  80.2 &  75.3 &  93.3\\
Ti\,{\sc i} & 6261.11 & 1.43 & -0.42 &  74.1 &  77.4 &  79.3 &  85.0 &  71.3 &  85.0 &  81.3 &  72.9 &  81.1\\
Ti\,{\sc i} & 6743.13 & 0.89 & -1.63 &  44.8 &  36.5 &  42.3 &       &  40.9 &  44.8 &  44.9 &  34.8 &  53.9\\
Ti\,{\sc ii}& 4849.18 & 1.13 & -3.00 &  83.7 &  83.1 &  78.1 &  77.5 &  83.0 &  80.7 &  81.7 &  80.3 &  83.1\\
Ti\,{\sc ii}& 4865.61 & 1.11 & -2.79 &  87.3 &  88.6 &  83.2 &  75.8 &  81.6 &  86.3 &  74.6 &  85.8 &  84.4\\
Ti\,{\sc ii}& 4874.01 & 3.09 & -0.80 &  63.0 &  60.9 &  56.0 &  57.8 &  60.2 &  61.7 &  69.5 &  58.2 &  58.1\\
Ti\,{\sc ii}& 4911.19 & 3.12 & -0.61 &  73.4 &  75.8 &  86.1 &  76.5 &  67.3 &  76.5 &  77.5 &  72.8 &  80.9\\
Ti\,{\sc ii}& 5005.15 & 1.56 & -2.72 &  58.5 &  50.6 &  52.6 &  52.0 &  51.6 &  53.2 &  52.2 &  53.2 &  53.6\\
Ti\,{\sc ii}& 5013.33 & 3.09 & -1.91 &       &       &       &  80.3 &  70.4 &  80.4 &  80.3 &  77.5 &  81.0\\
Ti\,{\sc ii}& 5013.67 & 1.58 & -2.19 &  88.8 &  84.1 &  83.6 &  97.4 &  88.0 &  89.8 &  90.4 &  91.5 &  98.4\\
Ti\,{\sc ii}& 5154.07 & 1.56 & -1.75 & 108.6 & 100.3 &       & 110.2 &  92.3 & 113.4 &  99.7 & 108.1 & 104.3\\
Ti\,{\sc ii}& 5185.91 & 1.89 & -1.49 &  98.1 &  98.7 & 101.4 & 102.2 & 104.5 &  96.7 &  99.9 & 103.8 & 103.8\\
Ti\,{\sc ii}& 5336.77 & 1.58 & -1.59 & 110.1 & 106.3 & 111.2 & 110.1 & 109.9 & 106.7 & 108.5 & 109.6 & 108.2\\
Ti\,{\sc ii}& 5381.01 & 1.56 & -1.92 & 101.6 &  98.5 & 105.5 &  98.2 &  99.5 & 100.6 & 103.8 &  97.7 &  94.3\\
Ti\,{\sc ii}& 5418.75 & 1.58 & -2.00 &  88.2 &  84.1 &  86.8 &  86.8 &  82.0 &  88.0 &  90.1 &  88.8 &  86.0\\
Ti\,{\sc ii}& 6491.56 & 2.06 & -1.89 &  76.3 &  69.4 &  63.8 &  69.2 &  66.9 &  69.8 &  72.0 &  68.3 &  64.9\\
Ti\,{\sc ii}& 6559.58 & 2.04 & -2.13 &  58.5 &  54.7 &  55.7 &  52.3 &  53.8 &  52.5 &  53.5 &  51.1 &  51.4\\
Ti\,{\sc ii}& 6606.94 & 2.06 & -2.76 &  23.1 &  29.8 &  26.4 &  25.7 &       &  22.5 &  25.1 &       &  32.9\\
Ti\,{\sc ii}& 6680.13 & 3.09 & -1.78 &  19.8 &  19.4 &  18.8 &  19.2 &  12.0 &  14.3 &  18.9 &       &      \\
Cr\,{\sc i} & 5238.96 & 2.71 & -1.43 &  16.1 &       &       &  20.4 &  16.3 &  14.9 &       &  12.5 &      \\
Cr\,{\sc i} & 5296.69 & 0.98 & -1.41 & 117.0 & 110.7 & 110.3 & 119.0 & 112.1 & 112.4 & 114.9 & 108.9 & 113.9\\
Cr\,{\sc i} & 5300.74 & 0.98 & -2.12 &  79.1 &  83.3 &  79.2 &  85.2 &  75.3 &  74.9 &  72.3 &  71.1 &  84.3\\
Cr\,{\sc i} & 5304.18 & 3.46 & -0.78 &  15.5 &   9.3 &       &   7.7 &       &   8.8 &   9.0 &  11.9 &  11.9\\
Cr\,{\sc i} & 5318.77 & 3.44 & -0.77 &  13.6 &  16.4 &  11.8 &  13.8 &  10.4 &  13.4 &  17.1 &       &  19.4\\
Cr\,{\sc i} & 5348.31 & 1.00 & -1.29 & 123.7 & 116.7 & 121.0 & 119.9 & 113.9 & 117.6 & 114.5 & 115.9 & 119.6\\
Cr\,{\sc i} & 6330.09 & 0.94 & -2.90 &  37.6 &  42.0 &  36.4 &  38.5 &  34.9 &  37.6 &  28.5 &  33.8 &  41.9\\
Cr\,{\sc i} & 6630.03 & 1.03 & -3.60 &  17.0 &  13.4 &  10.6 &  11.3 &  11.7 &       &  15.0 &   6.7 &  12.9\\
Cr\,{\sc ii}& 4848.25 & 3.86 & -1.14 &  64.2 &       &  79.7 &       &  71.1 &  57.0 &  67.3 &  68.1 &  69.7\\
Cr\,{\sc ii}& 5237.32 & 4.07 & -1.18 &  57.8 &  57.4 &  58.2 &  58.3 &  55.0 &       &  58.5 &  55.2 &  53.5\\
Cr\,{\sc ii}& 5305.85 & 3.82 & -2.06 &  30.4 &  26.3 &  27.2 &  24.0 &  25.8 &  29.2 &       &  31.1 &  30.7\\
Cr\,{\sc ii}& 5310.68 & 4.07 & -2.24 &  13.5 &       &  12.8 &       &  14.9 &       &  18.9 &  11.0 &  12.6\\
Cr\,{\sc ii}& 5313.56 & 4.07 & -1.55 &  43.0 &  33.7 &  34.5 &       &  39.5 &  33.7 &  39.2 &       &  38.6\\
Cr\,{\sc ii}& 5502.06 & 4.16 & -1.97 &  20.7 &  20.7 &  24.7 &  20.4 &  22.2 &  23.7 &  22.5 &       &  13.1\\
Fe\,{\sc i} &4802.88 & 3.64 & -1.51 &  49.8 &  64.0 &  57.2 &  53.8 &  58.5 &       &       &  55.5 &  58.3\\
Fe\,{\sc i} &4834.50 & 2.42 & -3.41 &  54.2 &  55.9 &  56.4 &  57.3 &  48.0 &  54.8 &  51.0 &  54.8 &  60.1\\
Fe\,{\sc i} &4839.54 & 3.27 & -1.82 &  77.0 &  70.4 &  77.7 &  79.7 &  74.9 &  74.7 &       &  79.2 &  83.8\\
Fe\,{\sc i} &4848.88 & 2.27 & -3.14 &  61.5 &  63.6 &  55.2 &  69.0 &  58.8 &  54.4 &  54.5 &  54.4 &  56.6\\
Fe\,{\sc i} &4849.66 & 3.57 & -2.68 &       &       &       &       &       &       &       &       &  13.6\\
Fe\,{\sc i} &4874.35 & 3.07 & -3.03 &  27.8 &  38.9 &  34.2 &  26.4 &  28.9 &  24.4 &  36.2 &  23.5 &  27.9\\
Fe\,{\sc i} &4882.14 & 3.42 & -1.64 &  75.2 &  88.5 &  81.6 &  89.8 &  81.9 &  83.7 &  89.0 &  78.5 &  93.4\\
Fe\,{\sc i} &4892.85 & 4.21 & -1.29 &  55.3 &  57.8 &  52.4 &  57.2 &  54.8 &  51.0 &  56.7 &  46.9 &  56.6\\
Fe\,{\sc i} &4896.43 & 3.88 & -2.05 &  39.7 &  40.0 &  34.2 &  43.5 &  33.8 &  35.8 &  37.7 &  34.3 &  41.0\\
Fe\,{\sc i} &4907.73 & 3.43 & -1.84 &  67.1 &  67.0 &  65.1 &  64.6 &  67.7 &  63.9 &  64.5 &  65.9 &  71.6\\
Fe\,{\sc i} &4911.77 & 3.92 & -1.79 &  51.5 &  56.1 &       &  62.1 &  49.3 &  48.1 &  51.9 &  39.5 &  56.1\\
Fe\,{\sc i} &4917.23 & 4.19 & -1.18 &  66.8 &  68.8 &  64.8 &  65.8 &  57.9 &  60.7 &  66.1 &  66.1 &  66.6\\
Fe\,{\sc i} &4927.41 & 3.57 & -2.07 &  61.0 &  59.4 &  59.3 &  61.3 &  53.1 &  57.6 &  55.5 &  46.0 &  50.7\\
Fe\,{\sc i} &4946.38 & 3.37 & -1.17 & 103.6 & 105.6 & 105.6 & 103.6 & 102.2 & 106.9 & 105.2 & 100.6 & 105.9\\
Fe\,{\sc i} &4950.10 & 3.41 & -1.67 &  82.4 &  87.5 &  80.1 &       &  87.0 &  81.7 &  74.2 &  83.6 &  85.1\\
Fe\,{\sc i} &4961.92 & 3.63 & -2.19 &  32.8 &  31.3 &  33.4 &  32.3 &  35.4 &  34.1 &  28.9 &  30.1 &  34.7\\
Fe\,{\sc i} &4962.57 & 4.18 & -1.18 &  57.7 &  51.9 &  54.1 &  55.3 &  52.3 &  52.9 &  52.9 &  49.6 &  54.4\\
Fe\,{\sc i} &4969.91 & 4.21 & -0.71 &  78.7 &  78.1 &  76.9 &  79.5 &  72.8 &  80.2 &  80.2 &  73.6 &  83.1\\
Fe\,{\sc i} &4979.58 & 3.64 & -2.58 &  24.2 &  28.7 &  21.2 &  26.3 &  19.6 &  22.3 &  19.9 &  20.8 &  24.2\\
Fe\,{\sc i} &4985.25 & 3.92 & -0.56 &  95.2 &  94.4 &  83.3 &  92.9 &  91.5 &  95.9 &  98.7 &  93.5 &  97.0\\
Fe\,{\sc i} &4985.54 & 2.86 & -1.33 & 116.6 & 119.2 & 111.7 & 115.0 & 105.6 & 111.9 & 111.2 & 112.4 & 116.7\\
Fe\,{\sc i} &4986.22 & 4.21 & -1.39 &  49.1 &  49.2 &  52.1 &  55.6 &  43.3 &  48.3 &  47.5 &  40.6 &  50.2\\
Fe\,{\sc i} &4999.11 & 4.18 & -1.74 &  37.6 &  31.6 &  48.2 &  40.3 &  25.0 &  35.3 &  28.8 &       &  34.9\\
Fe\,{\sc i} &5001.86 & 3.88 &  0.01 & 118.6 & 114.0 & 119.1 & 115.0 & 113.6 & 116.3 & 115.9 & 118.0 & 113.5\\
Fe\,{\sc i} &5004.04 & 4.20 & -1.40 &  47.8 &  45.2 &  48.6 &  57.8 &  44.0 &  44.4 &  48.7 &  45.5 &  51.7\\
Fe\,{\sc i} &5012.69 & 4.28 & -1.79 &  36.6 &  39.0 &  35.2 &  37.9 &  32.2 &  36.2 &  32.3 &  32.0 &  39.7\\
Fe\,{\sc i} &5014.94 & 3.94 & -0.30 & 114.5 & 110.9 & 112.2 & 118.2 & 116.2 & 111.3 & 111.5 & 111.8 & 118.1\\
Fe\,{\sc i} &5016.47 & 4.25 & -1.69 &  18.7 &       &  31.4 &  32.4 &  36.3 &  34.2 &  32.7 &  26.0 &  31.5\\
Fe\,{\sc i} &5028.12 & 3.57 & -1.12 &  95.8 &  94.6 &  91.2 &  95.8 &  90.2 &  91.1 &  91.5 &  84.7 &  92.6\\
Fe\,{\sc i} &5031.91 & 4.37 & -1.67 &  28.9 &  26.2 &  27.9 &  35.4 &  21.1 &  19.3 &  21.1 &  20.3 &  27.1\\
Fe\,{\sc i} &5044.21 & 2.85 & -2.06 &  90.2 &  91.0 &  95.1 &  90.7 &  92.2 &  90.1 &  90.7 &  88.4 &  93.2\\
Fe\,{\sc i} &5054.64 & 3.64 & -1.92 &  40.5 &  42.5 &  35.5 &  45.5 &  37.0 &  38.4 &  43.3 &  32.8 &  40.6\\
Fe\,{\sc i} &5056.84 & 4.26 & -1.96 &  30.8 &  35.1 &  25.1 &  29.2 &  21.1 &  24.5 &  29.6 &  21.4 &  35.7\\
Fe\,{\sc i} &5058.49 & 3.64 & -2.83 &  13.9 &  16.5 &  13.3 &  13.5 &   8.1 &  13.6 &   8.7 &   6.3 &  12.0\\
Fe\,{\sc i} &5067.16 & 4.22 & -0.97 &  72.0 &  75.5 &  70.3 &  70.0 &  67.0 &  69.5 &  68.1 &  68.6 &  71.9\\
Fe\,{\sc i} &5083.33 & 0.95 & -2.96 & 146.9 & 152.7 & 141.0 & 147.2 & 146.8 & 144.9 & 142.1 & 138.1 &      \\
Fe\,{\sc i} &5088.15 & 4.15 & -1.68 &  36.0 &  35.0 &  25.1 &  35.5 &  32.5 &  32.9 &  32.5 &  29.0 &  37.6\\
Fe\,{\sc i} &5090.77 & 4.26 & -0.40 &  81.9 &  91.8 &  83.5 &  90.7 &  85.8 &  90.5 &  92.4 &  86.5 &  88.9\\
Fe\,{\sc i} &5104.44 & 4.28 & -1.59 &  37.1 &  39.2 &  35.7 &  36.3 &  37.9 &  36.4 &  34.2 &  35.1 &  42.3\\
Fe\,{\sc i} &5109.66 & 4.30 & -0.98 &  67.7 &  68.8 &  71.1 &       &  73.7 &  66.9 &  64.2 &  71.7 &  63.0\\
Fe\,{\sc i} &5115.77 & 3.57 & -2.74 &  28.3 &  30.7 &  14.6 &  26.9 &  18.4 &  22.3 &  26.2 &  16.3 &  25.3\\
Fe\,{\sc i} &5127.37 & 0.91 & -3.31 & 130.2 & 129.4 & 129.7 & 123.7 & 132.1 & 133.3 & 130.7 & 129.6 & 136.9\\
Fe\,{\sc i} &5127.67 & 0.05 & -6.12 &  55.7 &  56.9 &  50.3 &  62.0 &  55.2 &  51.6 &  48.7 &  49.5 &  55.9\\
Fe\,{\sc i} &5141.75 & 2.42 & -2.24 & 103.5 & 104.2 & 101.6 & 103.9 & 105.5 & 101.9 & 102.2 &  99.1 & 106.9\\
Fe\,{\sc i} &5143.72 & 2.19 & -3.79 &  44.4 &  38.3 &  37.1 &  44.7 &  33.2 &  35.2 &  35.3 &  32.7 &  43.0\\
Fe\,{\sc i} &5145.09 & 2.19 & -2.88 &  79.3 &  75.5 &  69.9 &  69.2 &  69.9 &  71.7 &  71.7 &  70.0 &  76.4\\
Fe\,{\sc i} &5159.05 & 4.28 & -0.82 &  64.7 &  63.5 &  68.3 &  67.5 &  59.1 &  61.8 &  65.5 &  64.6 &  65.8\\
Fe\,{\sc i} &5187.91 & 4.14 & -1.37 &  63.8 &  57.4 &  58.0 &  55.5 &  55.9 &  55.5 &  57.0 &  55.2 &  54.5\\
Fe\,{\sc i} &5197.93 & 4.30 & -1.64 &  34.9 &  35.1 &  32.5 &  41.8 &  40.4 &  28.8 &  34.8 &  30.4 &  32.1\\
Fe\,{\sc i} &5198.71 & 2.22 & -2.14 & 123.9 & 125.5 & 119.1 & 118.4 & 120.6 & 113.6 & 120.1 & 112.5 & 121.0\\
Fe\,{\sc i} &5209.88 & 3.23 & -3.26 &  37.4 &  23.7 &  27.0 &  17.8 &  15.8 &  14.0 &       &       &      \\
Fe\,{\sc i} &5217.38 & 3.21 & -1.16 & 117.0 & 117.2 & 109.8 & 111.6 & 111.0 & 113.6 & 110.3 & 108.8 & 111.4\\
Fe\,{\sc i} &5223.18 & 3.63 & -1.78 &  30.2 &  39.0 &  29.7 &       &  32.9 &  28.3 &  37.7 &  29.8 &  31.8\\
Fe\,{\sc i} &5242.49 & 3.63 & -0.97 &       &  92.3 &  91.7 &  92.3 &  93.4 &  90.1 &  93.6 &  90.2 &  85.6\\
Fe\,{\sc i} &5249.10 & 4.47 & -1.48 &  35.3 &  38.4 &  24.2 &  39.8 &  30.8 &  36.6 &  34.3 &  30.7 &  36.8\\
Fe\,{\sc i} &5253.03 & 2.27 & -3.84 &  30.5 &  26.7 &  27.9 &  32.2 &  27.9 &  27.8 &  32.7 &  23.9 &  30.1\\
Fe\,{\sc i} &5262.88 & 3.25 & -2.66 &  25.5 &       &  25.9 &  29.4 &  20.0 &  31.3 &  28.4 &  21.1 &  21.9\\
Fe\,{\sc i} &5263.86 & 3.57 & -2.14 &  71.4 &  66.7 &  66.1 &  67.2 &  49.5 &       &  52.7 &  54.7 &  66.2\\
Fe\,{\sc i} &5267.26 & 4.37 & -1.77 &  26.3 &  23.8 &  23.7 &  30.3 &  19.8 &  20.4 &  31.4 &  20.0 &  19.2\\
Fe\,{\sc i} &5285.12 & 4.43 & -1.64 &       &  33.4 &  26.8 &  23.5 &  21.9 &  23.6 &  21.1 &  20.1 &  25.0\\
Fe\,{\sc i} &5288.53 & 3.68 & -1.51 &  65.2 &  62.3 &  69.3 &  65.6 &  62.9 &  64.7 &  66.3 &  62.6 &  65.8\\
Fe\,{\sc i} &5293.95 & 4.14 & -1.87 &  32.3 &  28.4 &  26.1 &  26.1 &  28.7 &  26.1 &  27.5 &  25.1 &  28.0\\
Fe\,{\sc i} &5294.54 & 3.64 & -2.76 &  15.6 &  16.5 &  10.5 &  16.0 &  18.1 &  13.9 &  12.1 &  12.4 &  11.6\\
Fe\,{\sc i} &5295.31 & 4.41 & -1.59 &  28.3 &  28.2 &  23.4 &  25.8 &  23.4 &  26.5 &  19.9 &  21.3 &  23.9\\
Fe\,{\sc i} &5307.36 & 1.61 & -2.99 & 121.1 & 119.3 & 111.6 & 123.7 & 119.7 & 117.8 & 117.6 & 112.9 & 122.6\\
Fe\,{\sc i} &5321.10 & 4.43 & -0.95 &  39.2 &  41.3 &  38.8 &  39.7 &  37.8 &  39.5 &  39.0 &  38.9 &  41.7\\
Fe\,{\sc i} &5322.04 & 2.28 & -2.80 &  84.0 &  83.5 &  81.0 &  80.7 &  81.1 &  77.9 &  78.7 &  76.4 &  82.8\\
Fe\,{\sc i} &5364.87 & 4.44 &  0.23 & 107.2 & 110.8 & 108.2 & 106.3 & 107.5 & 109.1 & 111.7 & 108.2 & 109.9\\
Fe\,{\sc i} &5365.39 & 3.57 & -1.02 &  86.0 &  92.4 &  87.6 &  82.4 &  91.2 &  83.3 &  87.9 &  84.5 &  85.3\\
Fe\,{\sc i} &5367.46 & 4.41 &  0.44 & 113.1 & 121.8 & 115.6 & 114.3 & 119.4 & 109.0 & 116.5 & 112.9 & 121.2\\
Fe\,{\sc i} &5373.69 & 4.47 & -0.76 &  62.2 &  59.0 &  60.5 &  59.8 &  59.3 &  56.1 &  63.0 &  57.3 &  58.9\\
Fe\,{\sc i} &5379.57 & 3.69 & -1.51 &  65.3 &  69.3 &  69.8 &  64.4 &  65.1 &  64.2 &  64.9 &  57.6 &  66.1\\
Fe\,{\sc i} &5383.36 & 4.31 &  0.64 & 127.8 & 134.7 & 130.6 & 128.6 & 130.4 & 131.6 & 133.0 & 124.0 & 130.9\\
Fe\,{\sc i} &5386.33 & 4.15 & -1.67 &  34.5 &  35.8 &  25.0 &  28.4 &  24.0 &  29.0 &  29.5 &  23.8 &  29.1\\
Fe\,{\sc i} &5389.47 & 4.41 & -0.41 &  86.2 &  86.7 &  78.7 &  78.1 &  83.8 &  82.0 &  81.8 &  77.6 &  81.5\\
Fe\,{\sc i} &5395.21 & 4.44 & -2.07 &  15.6 &  16.8 &  18.6 &  26.4 &  16.0 &  12.2 &       &  15.2 &  23.6\\
Fe\,{\sc i} &5398.27 & 4.44 & -0.63 &  73.7 &  54.1 &  69.1 &  68.8 &  66.3 &  67.0 &  73.2 &       &  75.7\\
Fe\,{\sc i} &5410.90 & 4.47 &  0.40 & 115.2 & 108.0 & 112.8 & 103.3 & 107.3 & 105.3 & 109.2 & 105.0 & 106.6\\
Fe\,{\sc i} &5412.78 & 4.43 & -1.89 &  22.5 &  17.2 &       &  18.7 &       &  11.6 &       &  16.3 &  19.6\\
Fe\,{\sc i} &5436.59 & 2.27 & -2.96 &  67.3 &  69.8 &       &       &  64.2 &  66.8 &  67.7 &  65.1 &  72.3\\
Fe\,{\sc i} &5441.34 & 4.31 & -1.63 &  27.8 &  33.3 &  27.1 &  29.6 &  28.8 &  29.5 &  28.2 &  25.9 &  30.7\\
Fe\,{\sc i} &5445.04 & 4.38 & -0.02 & 106.2 & 103.2 & 104.9 & 102.0 & 103.9 & 107.2 & 108.4 & 101.7 & 106.9\\
Fe\,{\sc i} &5452.08 & 3.64 & -2.86 &  25.1 &  28.1 &       &       &       &  27.6 &  12.1 &  21.5 &  20.7\\
Fe\,{\sc i} &5460.87 & 3.07 & -3.58 &   9.2 &       &  15.6 &  12.4 &   8.4 &  19.4 &   9.2 &   8.3 &  10.2\\
Fe\,{\sc i} &5461.55 & 4.44 & -1.80 &  29.7 &       &  23.7 &  26.0 &  25.1 &  22.4 &  22.1 &  23.3 &  28.2\\
Fe\,{\sc i} &5463.27 & 4.43 &  0.11 & 103.8 & 100.0 & 104.4 &       &  97.6 &  95.6 &  96.9 &  96.5 &  98.8\\
Fe\,{\sc i} &5464.28 & 4.14 & -1.40 &  38.8 &  38.6 &  32.3 &  39.2 &  42.7 &  35.8 &  33.5 &  35.5 &  43.8\\
Fe\,{\sc i} &5466.99 & 3.57 & -2.23 &  35.3 &  33.1 &  30.2 &  37.6 &  38.7 &  33.4 &  37.1 &  35.5 &  36.6\\
Fe\,{\sc i} &5470.09 & 4.44 & -1.81 &  21.4 &  20.1 &  17.0 &  19.3 &       &  24.6 &  21.2 &  21.3 &  23.7\\
Fe\,{\sc i} &5501.46 & 0.95 & -3.05 & 158.5 &       & 152.8 & 162.2 & 150.3 &       & 153.6 & 146.0 & 158.4\\
Fe\,{\sc i} &5522.44 & 4.20 & -1.45 &  42.6 &  41.6 &  41.9 &  41.4 &  36.7 &  39.1 &  29.7 &  39.6 &  41.0\\
Fe\,{\sc i} &5539.28 & 3.64 & -2.66 &  22.8 &  15.0 &  19.4 &  24.7 &  18.8 &  20.0 &  23.6 &  16.2 &      \\
Fe\,{\sc i} &5543.93 & 4.21 & -1.14 &  58.9 &  57.7 &  59.3 &  59.2 &  56.5 &  54.7 &  60.5 &  56.4 &  48.8\\
Fe\,{\sc i} &5546.50 & 4.37 & -1.21 &  50.6 &  53.2 &  47.3 &  52.1 &  51.1 &  52.5 &  51.6 &  48.4 &  51.5\\
Fe\,{\sc i} &5554.89 & 4.54 & -0.44 &  86.4 &  84.1 &  80.3 &  83.6 &  84.0 &  82.1 &  84.1 &  84.9 &  88.1\\
Fe\,{\sc i} &5560.20 & 4.43 & -1.09 &  51.0 &  50.3 &  47.3 &  45.1 &  53.3 &  43.5 &  45.6 &  44.9 &  49.5\\
Fe\,{\sc i} &5576.10 & 3.43 & -0.90 & 118.2 & 119.3 & 112.0 & 112.7 &  94.9 & 110.0 & 117.3 & 107.5 & 116.6\\
Fe\,{\sc i} &5587.57 & 4.14 & -1.85 &  34.9 &  37.4 &  33.7 &  33.9 &  31.2 &  30.8 &  33.8 &  32.4 &  38.3\\
Fe\,{\sc i} &5618.63 & 4.20 & -1.28 &  50.5 &  43.8 &  47.5 &  44.5 &  54.5 &       &  51.2 &  49.6 &      \\
Fe\,{\sc i} &5619.59 & 4.38 & -1.60 &  34.0 &  35.7 &  31.1 &  30.0 &  31.9 &  26.9 &  31.0 &  30.1 &  32.5\\
Fe\,{\sc i} &5624.02 & 4.38 & -1.48 &  42.0 &  34.2 &  40.0 &  45.1 &  49.5 &  47.0 &  49.6 &       &  50.3\\
Fe\,{\sc i} &5633.94 & 4.99 & -0.27 &  61.1 &  59.1 &  55.8 &  57.8 &  55.6 &  57.6 &  58.0 &  54.0 &  60.0\\
Fe\,{\sc i} &5635.83 & 4.25 & -1.79 &  25.4 &  31.5 &  29.9 &  30.2 &  28.0 &  30.5 &  22.7 &  26.9 &  33.2\\
Fe\,{\sc i} &5638.26 & 4.22 & -0.87 &  71.3 &  74.5 &  71.7 &  75.1 &  72.6 &  74.1 &  74.6 &  73.2 &  76.2\\
Fe\,{\sc i} &5649.98 & 5.09 & -0.92 &  28.2 &  30.2 &  29.4 &  28.0 &  20.1 &  24.0 &  26.6 &  29.3 &  31.1\\
Fe\,{\sc i} &5650.70 & 5.08 & -0.96 &  26.1 &       &  27.2 &  31.7 &  30.7 &  26.6 &  27.3 &  26.3 &  32.8\\
Fe\,{\sc i} &5651.47 & 4.47 & -1.90 &  14.7 &  20.8 &       &  18.9 &  17.2 &  16.0 &  15.1 &  12.7 &  15.3\\
Fe\,{\sc i} &5652.31 & 4.26 & -1.85 &  27.0 &  26.0 &  20.1 &  25.2 &  23.0 &  24.1 &  21.4 &  26.2 &  23.8\\
Fe\,{\sc i} &5653.86 & 4.38 & -1.64 &  34.3 &  36.3 &  32.6 &  38.3 &  28.4 &  35.7 &  32.0 &  32.5 &  39.3\\
Fe\,{\sc i} &5661.34 & 4.28 & -2.02 &  26.5 &  26.9 &  22.0 &  24.5 &  26.8 &  21.7 &  22.9 &  21.7 &  26.7\\
Fe\,{\sc i} &5662.51 & 4.17 & -0.57 &  92.8 &  93.0 &  90.1 &  89.1 &  89.4 &  95.4 &  87.2 &  93.4 &  99.2\\
Fe\,{\sc i} &5667.51 & 4.17 & -1.58 &  50.6 &  55.0 &  50.8 &  55.6 &  41.3 &  53.7 &  55.6 &  45.0 &  51.8\\
Fe\,{\sc i} &5679.02 & 4.65 & -0.82 &  40.0 &  52.4 &  46.8 &       &  50.7 &  47.9 &  53.4 &  46.4 &  52.6\\
Fe\,{\sc i} &5680.24 & 4.18 & -2.58 &  14.3 &       &       &       &       &       &       &       &      \\
Fe\,{\sc i} &5691.49 & 4.30 & -1.52 &  47.4 &  44.3 &  36.6 &  33.2 &  32.2 &  35.0 &  38.1 &  34.0 &  37.1\\
Fe\,{\sc i} &5701.54 & 2.56 & -2.12 & 105.7 & 102.0 &  97.4 &  99.5 & 104.1 & 103.5 &  98.2 & 100.9 & 109.1\\
Fe\,{\sc i} &5705.46 & 4.30 & -1.50 &  38.3 &  36.2 &  31.6 &  36.7 &  31.7 &  38.2 &  33.5 &  34.1 &  38.0\\
Fe\,{\sc i} &5717.83 & 4.28 & -1.13 &  61.0 &  65.0 &  57.6 &  66.4 &  70.1 &  64.3 &  62.6 &  65.5 &  67.3\\
Fe\,{\sc i} &5731.76 & 4.25 & -1.20 &  54.1 &  55.4 &  54.3 &  53.6 &  60.0 &  55.7 &  56.5 &  57.6 &  57.8\\
Fe\,{\sc i} &5741.84 & 4.25 & -1.85 &  26.6 &  28.4 &  23.6 &  30.3 &  22.9 &  22.6 &  26.4 &  25.4 &      \\
Fe\,{\sc i} &5752.02 & 4.54 & -0.66 &  48.8 &  49.0 &  46.8 &  49.3 &  51.2 &  48.8 &  49.7 &  48.1 &  50.5\\
Fe\,{\sc i} &5753.13 & 4.26 & -0.69 &  79.4 &  77.1 &  76.9 &  74.6 &  77.9 &  79.2 &  80.3 &  77.3 &  82.5\\
Fe\,{\sc i} &5852.21 & 4.54 & -1.23 &  40.6 &  44.0 &  39.1 &  39.8 &  38.9 &  36.5 &  42.2 &       &  36.4\\
Fe\,{\sc i} &5853.14 & 1.48 & -5.28 &  20.5 &  17.9 &  15.6 &  20.4 &  19.4 &  15.6 &   2.6 &  18.2 &  16.6\\
Fe\,{\sc i} &5855.08 & 4.61 & -1.48 &  21.8 &  18.9 &  15.8 &  20.1 &       &  16.1 &  18.4 &  14.7 &  19.4\\
Fe\,{\sc i} &5856.08 & 4.29 & -1.33 &  31.9 &  29.9 &  30.2 &  31.6 &  23.6 &  28.5 &  30.6 &  24.8 &  31.7\\
Fe\,{\sc i} &5859.57 & 4.54 & -0.30 &  71.3 &  70.3 &  70.8 &  68.7 &  67.6 &  66.0 &  68.2 &  66.8 &  71.2\\
Fe\,{\sc i} &5905.67 & 4.65 & -0.73 &  50.5 &  50.5 &  45.3 &  46.3 &  41.5 &  49.1 &  40.5 &  44.8 &  51.6\\
Fe\,{\sc i} &5927.78 & 4.65 & -1.09 &  46.2 &  37.0 &  38.5 &  40.5 &  38.4 &  37.1 &  37.9 &  27.8 &  43.0\\
Fe\,{\sc i} &5929.67 & 4.54 & -1.41 &  38.4 &  36.1 &  31.8 &  32.1 &  33.9 &  32.9 &  32.6 &  26.2 &  39.3\\
Fe\,{\sc i} &5930.17 & 4.65 & -0.23 &  81.3 &  81.7 &  80.8 &  82.9 &  80.4 &  82.3 &  79.0 &  86.3 &  84.4\\
Fe\,{\sc i} &5934.65 & 3.92 & -1.17 &  76.4 &  75.7 &  74.3 &  81.1 &  75.3 &  79.6 &  76.5 &  73.9 &  79.7\\
Fe\,{\sc i} &5956.71 & 0.86 & -4.61 &  88.8 &  88.7 &  82.3 &  87.9 &  80.6 &  83.0 &       &  79.9 &  81.7\\
Fe\,{\sc i} &5984.81 & 4.73 &  0.17 &  79.1 &  75.2 &  79.2 &  74.7 &  78.1 &  72.4 &  77.6 &  76.7 &  81.8\\
Fe\,{\sc i} &6003.01 & 3.88 & -1.12 &  80.8 &  87.0 &  81.7 &  83.5 &  82.3 &  84.6 &  82.3 &  82.4 &  86.7\\
Fe\,{\sc i} &6012.20 & 2.22 & -4.20 &  36.4 &  37.0 &  28.8 &  30.2 &  29.5 &  34.6 &  36.2 &  29.7 &  31.9\\
Fe\,{\sc i} &6024.05 & 4.54 & -0.12 &  95.6 &  93.7 & 101.2 &  97.6 &  91.4 &  93.5 &  99.4 &  94.4 &  97.5\\
Fe\,{\sc i} &6027.05 & 4.07 & -1.09 &  66.6 &  68.3 &  63.5 &  62.7 &  65.4 &  62.0 &  69.4 &  61.1 &  68.2\\
Fe\,{\sc i} &6056.00 & 4.73 & -0.46 &  65.7 &  64.9 &  61.5 &  61.7 &  58.4 &  60.4 &  61.4 &  62.4 &  65.7\\
Fe\,{\sc i} &6065.48 & 2.61 & -1.53 & 136.8 & 135.7 & 135.2 & 131.2 & 141.6 & 128.3 & 125.9 & 124.8 & 147.1\\
Fe\,{\sc i} &6079.00 & 4.65 & -1.12 &  33.2 &  41.9 &  36.1 &  39.0 &  30.0 &  33.7 &  42.3 &  40.7 &  38.9\\
Fe\,{\sc i} &6082.72 & 2.22 & -3.57 &  55.3 &       &  49.5 &  50.9 &  51.5 &  47.3 &  51.6 &  50.8 &  54.0\\
Fe\,{\sc i} &6085.26 & 2.75 & -2.71 &  66.4 &  70.8 &  64.4 &  69.2 &  61.7 &  65.2 &  64.7 &  55.0 &  69.6\\
Fe\,{\sc i} &6093.64 & 4.60 & -1.40 &  21.8 &  27.9 &  24.3 &  24.5 &  23.6 &  22.2 &  27.0 &  21.8 &  28.1\\
Fe\,{\sc i} &6094.37 & 4.65 & -1.84 &  16.6 &  17.3 &  14.5 &  14.9 &  12.2 &  16.9 &  16.3 &  29.0 &  21.5\\
Fe\,{\sc i} &6096.66 & 3.98 & -1.83 &  37.8 &  37.6 &  33.8 &  38.2 &  40.1 &       &  36.2 &  26.3 &  36.6\\
Fe\,{\sc i} &6120.24 & 0.91 & -5.95 &   9.8 &  15.6 &   9.5 &  14.4 &       &  11.1 &   8.6 &   7.9 &  17.9\\
Fe\,{\sc i} &6127.90 & 4.14 & -1.40 &  50.4 &  52.6 &  46.5 &  49.1 &  47.9 &  52.5 &  47.6 &  45.2 &  55.6\\
Fe\,{\sc i} &6137.69 & 2.58 & -1.40 & 149.9 & 154.0 &       & 159.8 & 150.4 & 152.4 & 149.3 & 149.5 & 141.8\\
Fe\,{\sc i} &6151.61 & 2.17 & -3.27 &  70.1 &  70.5 &  70.4 &  68.0 &  63.8 &  70.8 &  68.3 &  64.0 &  71.1\\
Fe\,{\sc i} &6157.72 & 4.07 & -1.16 &  70.7 &  66.8 &  67.0 &  59.3 &  65.6 &  66.3 &  67.5 &  62.5 &  68.7\\
Fe\,{\sc i} &6159.37 & 4.60 & -1.97 &  12.7 &  15.8 &  11.2 &  15.3 &   8.2 &   6.7 &  14.7 &   9.4 &  10.3\\
Fe\,{\sc i} &6165.36 & 4.14 & -1.47 &  44.0 &  45.1 &  44.4 &  45.2 &  38.3 &  43.4 &  46.6 &  41.3 &  46.3\\
Fe\,{\sc i} &6173.34 & 2.22 & -2.88 &  97.4 &  93.7 &  84.6 &  89.9 &  91.7 &  86.6 &  86.9 &  86.5 &  93.3\\
Fe\,{\sc i} &6180.20 & 2.72 & -2.59 &  77.0 &  71.4 &  68.9 &  73.0 &  71.4 &  72.2 &  70.5 &  66.9 &  76.5\\
Fe\,{\sc i} &6187.99 & 3.94 & -1.62 &  51.8 &  50.3 &  44.9 &  53.6 &  46.7 &  43.7 &  47.2 &  46.7 &  51.3\\
Fe\,{\sc i} &6200.31 & 2.60 & -2.44 &  94.2 &  87.4 &  85.3 &  83.0 &  88.2 &  85.6 &  88.0 &  82.2 &  89.6\\
Fe\,{\sc i} &6213.43 & 2.22 & -2.48 &  91.5 & 108.1 & 102.1 & 105.1 & 101.7 &  99.9 & 102.0 & 102.6 & 108.3\\
Fe\,{\sc i} &6219.28 & 2.19 & -2.42 & 118.5 & 116.7 & 101.1 & 112.0 & 110.3 & 115.1 & 111.7 & 110.0 & 119.8\\
Fe\,{\sc i} &6226.73 & 3.88 & -2.12 &  26.7 &  26.8 &  22.2 &  25.2 &  23.6 &  27.7 &  25.4 &  19.2 &  29.2\\
Fe\,{\sc i} &6229.22 & 2.84 & -2.87 &  50.3 &  49.1 &  49.2 &  49.6 &  52.6 &  49.7 &  50.1 &  46.2 &  49.9\\
Fe\,{\sc i} &6232.64 & 3.65 & -0.96 &  90.5 &  87.0 &  86.3 &  85.6 &  87.7 &  84.7 &  89.4 &  82.2 &  89.8\\
Fe\,{\sc i} &6246.31 & 3.60 & -0.88 & 112.9 & 111.2 & 112.0 & 112.7 & 103.5 & 112.3 & 112.5 & 106.6 & 118.8\\
Fe\,{\sc i} &6252.55 & 2.40 & -1.69 & 144.3 & 142.5 & 143.7 & 138.1 & 138.8 & 138.5 & 157.1 & 134.0 & 152.4\\
Fe\,{\sc i} &6265.14 & 2.18 & -2.55 & 110.7 & 113.3 & 114.0 & 113.4 & 114.3 & 108.9 & 111.5 & 115.5 & 114.9\\
Fe\,{\sc i} &6270.22 & 2.85 & -2.61 &  72.3 &  64.8 &  68.0 &  60.4 &  59.6 &  62.6 &  66.3 &  60.4 &  69.0\\
Fe\,{\sc i} &6297.79 & 2.22 & -2.73 &  99.5 & 102.9 &  96.1 &       & 100.3 &       &  97.1 &  97.4 & 105.5\\
Fe\,{\sc i} &6302.49 & 3.68 & -0.91 &  98.7 & 109.6 &  84.2 &  77.3 &       &  86.4 &       &  94.6 &  64.5\\
Fe\,{\sc i} &6311.50 & 2.83 & -3.23 &  40.3 &  47.1 &  38.3 &       &  40.2 &  44.3 &  38.9 &  40.1 &  37.0\\
Fe\,{\sc i} &6322.68 & 2.58 & -2.43 & 100.3 &  72.7 &  89.4 &  95.3 &  89.2 &  97.4 &  98.6 &  96.0 &  98.7\\
Fe\,{\sc i} &6330.84 & 4.73 & -1.74 &  33.6 &  24.3 &  24.4 &  22.8 &  22.8 &  23.8 &  25.3 &  24.0 &  25.8\\
Fe\,{\sc i} &6335.33 & 2.19 & -2.18 & 126.2 & 123.2 & 115.5 & 118.6 & 113.3 & 119.0 & 119.3 & 116.9 & 122.4\\
Fe\,{\sc i} &6336.82 & 3.68 & -1.05 & 101.7 & 105.2 &  99.6 & 101.6 &  97.6 &  97.2 & 102.6 &  99.4 & 101.8\\
Fe\,{\sc i} &6344.14 & 2.43 & -2.92 &  83.9 &  79.6 &  72.6 &  76.7 &  72.9 &  73.6 &  79.7 &  70.5 &  81.0\\
Fe\,{\sc i} &6355.02 & 2.84 & -2.29 &  93.1 &  97.5 &  95.6 &       &       &  88.3 &  87.8 &       &  92.2\\
Fe\,{\sc i} &6380.75 & 4.19 & -1.38 &  57.6 &  55.1 &  52.2 &  51.8 &  51.6 &  56.2 &  57.4 &  75.6 &  60.1\\
Fe\,{\sc i} &6392.53 & 2.27 & -4.03 &  23.7 &  32.4 &  21.3 &  29.4 &  28.0 &  28.4 &  26.3 &  28.4 &  25.0\\
Fe\,{\sc i} &6393.60 & 2.43 & -1.58 & 148.0 & 147.2 & 152.6 & 143.0 & 156.9 & 140.4 & 140.0 & 146.5 & 166.0\\
Fe\,{\sc i} &6411.64 & 3.65 & -0.72 & 121.3 & 120.7 & 115.3 & 112.7 & 116.3 & 117.6 & 120.9 & 112.6 & 118.7\\
Fe\,{\sc i} &6421.35 & 2.27 & -2.03 & 144.1 & 141.5 &       & 137.3 & 135.6 & 141.5 & 141.1 & 139.3 & 145.1\\
Fe\,{\sc i} &6430.84 & 2.17 & -2.01 & 145.4 & 142.2 & 138.0 & 139.9 & 139.7 & 135.0 & 138.9 & 135.2 & 147.1\\
Fe\,{\sc i} &6475.62 & 2.55 & -2.94 &  75.0 &  77.3 &  72.0 &  76.1 &  72.2 &  71.8 &  73.1 &  68.6 &  75.4\\
Fe\,{\sc i} &6481.87 & 2.27 & -2.96 &  86.8 &  80.7 &  88.5 &  80.7 &  82.1 &  86.2 &  78.5 &  81.3 &  87.8\\
Fe\,{\sc i} &6498.94 & 0.96 & -4.70 &  95.4 &  90.4 &  86.1 &  95.1 &  87.4 &  91.7 &  82.7 &  75.2 & 100.4\\
Fe\,{\sc i} &6533.92 & 4.55 & -1.46 &  36.3 &       &  29.2 &  32.4 &  23.0 &  30.9 &  29.8 &  35.8 &      \\
Fe\,{\sc i} &6546.23 & 2.75 & -1.54 & 127.7 & 127.8 & 124.0 & 115.4 & 118.6 & 118.0 & 120.0 & 113.6 & 125.5\\
Fe\,{\sc i} &6569.21 & 4.73 & -0.42 &  72.9 &  73.4 &       &  77.2 &  76.1 &  69.6 &  60.6 &  70.1 &  74.7\\
Fe\,{\sc i} &6574.22 & 0.99 & -5.02 &  60.7 &  62.5 &  57.2 &  58.2 &  53.7 &  53.7 &  54.6 &  49.5 &  60.0\\
Fe\,{\sc i} &6575.01 & 2.58 & -2.71 &  86.0 &  81.7 &  85.5 &  84.7 &  85.0 &  83.2 &  81.9 &  82.0 &  86.6\\
Fe\,{\sc i} &6592.91 & 2.73 & -1.47 & 136.1 &       & 138.5 & 134.3 & 129.6 & 130.8 & 128.0 & 127.1 & 134.5\\
Fe\,{\sc i} &6593.87 & 2.43 & -2.42 & 111.3 & 107.7 & 109.2 & 104.1 & 102.5 & 115.1 & 112.1 & 106.9 & 109.3\\
Fe\,{\sc i} &6597.56 & 4.79 & -1.07 &  36.1 &  36.3 &  38.7 &  33.2 &  33.8 &  36.8 &  37.4 &  32.1 &  40.6\\
Fe\,{\sc i} &6627.56 & 4.55 & -1.58 &  18.2 &  23.2 &  28.1 &  26.9 &  25.2 &  19.7 &  25.2 &  23.8 &  20.7\\
Fe\,{\sc i} &6646.93 & 2.60 & -3.99 &  14.2 &  17.2 &  13.9 &  18.0 &  17.4 &  20.3 &  18.0 &  19.5 &  21.9\\
Fe\,{\sc i} &6677.99 & 2.68 & -1.42 & 144.4 & 148.9 & 145.4 & 129.0 & 143.1 & 142.6 & 139.8 & 139.0 & 148.1\\
Fe\,{\sc i} &6703.56 & 2.75 & -3.06 &  43.7 &  49.2 &  50.0 &  50.3 &  46.1 &  48.0 &  45.7 &  47.3 &  51.1\\
Fe\,{\sc i} &6710.31 & 1.48 & -4.88 &  32.8 &  33.7 &  34.0 &       &  29.4 &  33.4 &  30.3 &  29.4 &  31.8\\
Fe\,{\sc i} &6713.74 & 4.79 & -1.50 &  16.8 &  19.5 &  16.2 &  19.7 &       &  17.4 &  13.2 &  13.5 &  12.2\\
Fe\,{\sc i} &6715.38 & 4.60 & -1.64 &  21.0 &  24.9 &  30.3 &  22.0 &  23.9 &  25.0 &  27.8 &  18.5 &  27.7\\
Fe\,{\sc i} &6725.35 & 4.10 & -2.30 &  16.1 &  15.1 &       &  17.0 &  15.2 &  11.0 &  17.6 &  14.4 &  20.0\\
Fe\,{\sc i} &6726.66 & 4.60 & -1.00 &  37.7 &  39.2 &  33.3 &  30.5 &  38.4 &  39.2 &       &  33.3 &  43.5\\
Fe\,{\sc i} &6739.52 & 1.55 & -4.95 &  22.9 &  26.1 &  16.8 &  24.5 &  17.8 &  25.7 &  24.4 &  16.6 &  29.5\\
Fe\,{\sc i} &6750.15 & 2.42 & -2.62 &  97.6 &  98.7 &  96.5 &  69.4 &  93.5 &  95.2 &  95.6 &  91.3 &  98.3\\
Fe\,{\sc i} &6786.86 & 4.19 & -2.07 &  25.1 &  24.9 &  19.7 &  14.5 &  23.4 &  26.2 &  25.6 &  25.6 &  28.9\\
Fe\,{\sc ii}&4993.35 & 2.81 & -3.52 &  49.7 &  51.3 &  51.3 &  43.4 &  53.8 &  48.0 &  49.1 &  49.1 &  50.1\\
Fe\,{\sc ii}&5234.63 & 3.22 & -2.28 &  95.8 &  95.7 & 101.4 &  96.3 &  97.0 &  97.8 &  99.3 &  97.7 &  91.2\\
Fe\,{\sc ii}&6416.93 & 3.89 & -2.88 &  32.7 &  37.9 &  37.1 &       &  39.1 &  35.9 &  41.0 &  36.0 &  45.1\\
Fe\,{\sc ii}&5991.37 & 3.15 & -3.65 &       &       &  36.1 &       &       &       &       &       &  36.1\\
Fe\,{\sc ii}&6084.11 & 3.19 & -3.88 &  22.5 &  22.7 &       &  24.1 &  27.8 &  24.9 &  26.3 &  24.2 &      \\
Fe\,{\sc ii}&6149.25 & 3.88 & -2.84 &  37.1 &       &  39.8 &  37.4 &  34.7 &  36.0 &  38.3 &  35.8 &  40.3\\
Fe\,{\sc ii}&6247.55 & 3.89 & -2.43 &  59.5 &  54.3 &  62.6 &  52.8 &  58.9 &  59.3 &  56.4 &  60.6 &  57.1\\
Fe\,{\sc ii}&6369.46 & 2.89 & -4.23 &       &  27.0 &  25.5 &  22.6 &  24.3 &  23.2 &  25.7 &  22.3 &  23.3\\
Fe\,{\sc ii}&6432.68 & 2.89 & -3.69 &  52.3 &  47.4 &  40.3 &       &  50.7 &  48.0 &  44.8 &  47.7 &  48.7\\
Fe\,{\sc ii}&6456.38 & 3.90 & -2.19 &  63.4 &  66.5 &  66.4 &  65.8 &  72.4 &  66.5 &  63.1 &  75.2 &  61.1\\
Fe\,{\sc ii}&6516.08 & 2.89 & -3.43 &  61.7 &  66.3 &  65.5 &  62.9 &  63.8 &  67.4 &  66.5 &       &      \\
Fe\,{\sc ii}&4923.92 & 2.89 & -1.50 & 194.7 & 177.7 & 172.3 & 207.2 & 205.7 & 198.1 & 202.5 & 208.3 & 191.4\\
Fe\,{\sc ii}&5132.66 & 2.80 & -4.09 &  29.1 &       &  27.1 &  23.7 &  37.4 &       &  34.6 &       &  26.3\\
Fe\,{\sc ii}&5197.57 & 3.23 & -2.35 &  95.8 &  98.7 &  93.8 &  97.3 & 100.7 &  93.2 &  97.2 &  95.2 &  96.8\\
Fe\,{\sc ii}&5264.81 & 3.23 & -3.13 &  57.9 &       &       &  53.3 &  53.9 &  59.4 &  58.8 &  60.0 &  57.5\\
Fe\,{\sc ii}&5284.10 & 2.89 & -3.20 &  73.8 &  81.0 &  80.7 &  79.3 &  83.2 &  76.2 &  77.5 &  76.4 &  78.3\\
Fe\,{\sc ii}&5325.55 & 3.22 & -3.32 &  52.4 &  52.8 &  49.5 &  47.8 &  50.9 &  49.0 &  55.3 &  54.8 &  53.4\\
Fe\,{\sc ii}&5414.07 & 3.22 & -3.64 &  31.7 &  32.0 &  35.5 &  34.2 &  34.2 &  30.4 &  38.7 &  33.7 &  35.2\\
Fe\,{\sc ii}&5425.25 & 3.19 & -3.39 &  47.2 &  46.5 &  44.0 &  42.5 &  44.6 &  51.9 &  52.5 &  52.1 &  48.6\\
Fe\,{\sc ii}&5534.83 & 3.25 & -2.86 &  60.7 &  62.7 &  73.2 &  67.9 &  73.2 &  63.8 &  60.9 &  69.4 &  68.9\\
Ni\,{\sc i} &4866.26 & 3.53 & -0.21 &  81.8 &  92.2 &  86.0 &  84.8 &  85.0 &  85.1 &  83.9 &  86.2 &  87.3\\
Ni\,{\sc i} &5082.33 & 3.65 & -0.54 &  67.9 &  73.3 &  63.6 &  74.7 &  67.9 &  68.9 &  70.9 &  59.1 &  72.5\\
Ni\,{\sc i} &5094.40 & 3.83 & -1.11 &  26.4 &  27.5 &  27.7 &  30.4 &  28.8 &  19.3 &  29.1 &  27.0 &  33.6\\
Ni\,{\sc i} &5587.85 & 1.93 & -2.44 &  79.8 &  82.7 &  77.8 &  76.4 &  77.5 &  77.7 &  81.7 &  72.2 &  81.8\\
Ni\,{\sc i} &5593.73 & 3.89 & -0.78 &  43.9 &  42.0 &  44.1 &  41.0 &  39.8 &  42.2 &  43.5 &  33.7 &  42.5\\
Ni\,{\sc i} &5748.34 & 1.67 & -3.24 &  54.3 &  49.6 &  51.9 &  50.1 &  52.1 &  48.7 &  91.4 &  46.0 &  53.0\\
Ni\,{\sc i} &6108.10 & 1.67 & -2.43 &  89.4 &  96.3 &  96.0 &  89.1 &  91.5 &  91.2 &  13.4 &  87.7 &  96.1\\
Ni\,{\sc i} &6111.06 & 4.08 & -0.82 &  31.9 &  31.1 &  32.3 &  34.6 &  35.1 &  32.9 &  17.6 &  28.8 &  30.9\\
Ni\,{\sc i} &6130.13 & 4.26 & -0.88 &  16.3 &  18.7 &  12.8 &  14.9 &  12.3 &  20.7 &  62.9 &  47.0 &  19.0\\
Ni\,{\sc i} &6176.80 & 4.08 & -0.26 &  63.2 &  64.2 &  57.8 &  65.1 &  64.0 &  61.5 &  25.3 &  20.6 &  55.6\\
Ni\,{\sc i} &6177.23 & 1.82 & -3.55 &  30.8 &  25.2 &  24.5 &  27.6 &  28.7 &  26.6 &  22.4 &  25.6 &  23.4\\
Ni\,{\sc i} &6204.60 & 4.08 & -1.10 &  25.6 &  23.9 &  23.9 &  19.5 &  23.4 &  23.8 &  22.9 &  18.5 &  26.0\\
Ni\,{\sc i} &6223.98 & 4.10 & -0.91 &  25.0 &  22.1 &  25.5 &  24.3 &  24.7 &  67.6 &  65.4 &  66.3 &  72.0\\
Ni\,{\sc i} &6327.59 & 1.67 & -3.06 &  69.2 &  70.1 &  67.1 &  64.3 &  65.2 &  27.8 &  27.5 &  31.7 &  23.2\\
Ni\,{\sc i} &6378.24 & 4.15 & -0.81 &  31.6 &  36.0 &  19.1 &  33.1 &  30.0 &  24.5 &  27.0 &  23.3 &  30.4\\
Ni\,{\sc i} &6635.11 & 4.41 & -0.72 &  20.6 &  24.2 & 120.0 &  30.4 &  26.2 & 122.9 & 123.7 & 116.5 & 127.1\\
Ni\,{\sc i} &6643.62 & 1.67 & -1.91 & 125.5 & 128.5 & 104.5 & 126.3 & 119.4 & 100.9 & 105.1 & 101.1 & 108.8\\
Ni\,{\sc i} &6767.76 & 1.82 & -2.10 & 109.3 & 101.8 &  47.2 & 100.5 & 101.3 &  51.6 &  60.1 &  55.0 &  67.6\\
Ni\,{\sc i} &4857.40 & 3.74 & -0.83 & 107.1 &  86.1 &  62.8 &  65.8 &  43.9 &  59.9 &  51.8 &  45.1 &  61.1\\
Ni\,{\sc i} &4953.21 & 3.74 & -0.58 &  63.7 &  68.0 &  62.4 &  58.5 &  60.1 &  41.9 &  37.7 &  38.0 &  47.3\\
Ni\,{\sc i} &4998.22 & 3.60 & -0.69 &  59.8 &  61.1 &  46.9 &  44.7 &  59.4 &  43.1 &  40.5 &  34.9 &  43.2\\
Ni\,{\sc i} &6007.32 & 1.67 & -3.41 &  46.5 &  47.8 &  38.5 &  38.7 &  37.7 &  43.5 &  43.8 &  42.4 &  51.6\\
Ni\,{\sc i} &6086.29 & 4.26 & -0.46 &  41.3 &  44.3 &  45.1 &  47.0 &  38.5 &  24.1 &  26.2 &  24.5 &  30.0\\
Ni\,{\sc i} &6175.37 & 4.09 & -0.50 &  48.0 &  46.4 &  23.8 &  28.3 &  45.8 &  67.6 &  62.3 &  53.3 &  66.4\\
Ni\,{\sc i} &6186.72 & 4.10 & -0.88 &  29.8 &  29.0 &  59.7 &  67.9 &  21.2 &  19.5 &  18.9 &  15.4 &  19.9\\
Ni\,{\sc i} &6482.81 & 1.93 & -2.76 &  66.9 &  64.4 &  17.4 &  16.1 &  58.7 &  51.1 &  51.8 &  47.8 &  60.4\\
Ni\,{\sc i} &6598.61 & 4.23 & -0.91 &  20.9 &  18.8 &  49.1 &  54.4 &  19.1 &       &       &       &      \\
Ni\,{\sc i} &6772.32 & 3.65 & -0.94 &  58.1 &  56.0 &       &       &  53.3 &       &       &       &      \\
Zn\,{\sc i} &4810.53 & 4.08 & -0.29 &  94.4 & 100.6 &  86.9 &  94.8 &       &  88.1 &  86.6 &  75.2 &  86.1\\
Zn\,{\sc i} &6362.35 & 5.79 &  0.09 &  21.8 &  19.6 &  27.6 &  14.3 &  18.7 &  15.0 &  22.4 &  22.9 &  25.7\\
\hline
\end{longtable}
}
\section{Abundance analysis}
\label{sect:abun}

We have performed a standard Local Thermodynamic Equilibrium (LTE)
analysis to derive chemical abundances from the measured values of
$W_{\lambda}$ using the MARCS stellar model atmospheres
\citep{gustafsson1975, edvardsson1993, asplund1997}.

When selecting spectral lines suitable for analysis in a giant star
spectrum we made much use of the VALD database \citep{vald3, vald2,
  vald1}}. VALD also provided damping constants as well as term
   designations which were used in the calculation of the line
   broadening.

\subsection{$\log gf$-values -- general comments}

The elemental abundance is, for not too strong lines, basically
proportional to the oscillator strength ($\log gf$) of the line, hence
correct $\log gf$-values are important for the accuracy of the
abundances. Oscillator strengths may be determined in two ways (apart
from theoretical calculations) -- either through measurements in
laboratories or from a stellar, most often solar, spectrum. The latter
types of $\log gf$-values are normally called astrophysical.  The
astrophysical $\log gf$-values are determined by requiring the line
under study to yield the, pre-known, abundance of that element for the
star used. Since the Sun is the star for which we have the best
determined elemental abundances normally a solar spectrum is used. An
advantage of the astrophysical $\log gf$-values is that, if the solar
spectrum is taken with the same equipment as the stellar spectra are,
any irregularities in the recorded spectrum that arise from the
instrumentation and particular model atmospheres used will, to first
order, cancel.

The laboratory data have a specific value in that they allow absolute
determination of the stellar abundances. Obviously also these data
have associated errors and therefore one should expect some line-to-line
scatter in the final stellar abundances. Furthermore,
the absolute scale of a set of laboratory $\log gf$-values can be
erroneous and then the resulting abundances will be erroneous with the
same systematic error as present in the laboratory data (see e.g. our
discussion as concerns the $\log gf$-values for Ca\,{\sc i}).

We have chosen different options for different elements depending on
the data available. Our ambition has been to create a
line-list that is homogeneous for each element and which can be used in
forthcoming studies of giant stars in other globular clusters.

Whenever possible we have chosen homogeneous data sets of laboratory
data.  When these do not exist we have chosen between different
options: to use purely theoretical data (if they exist), to use only
astrophysical data, or use a combination of laboratory and
astrophysical data. In cases when we have chosen the last option we
have always checked the consistency between the two sets and in
general found them to be internally consistent (see below). For each
element we detail which solution we opted for and why.

As our spectra are roughly of the same resolution as the spectra in
\citet{bensby2003} and we do not have our own solar observations we
decided to use astrophysical $\log gf$-values for these lines by
\citet{bensby2003}. Their analysis is based on a solar spectrum
recorded with FEROS which has a resolution comparable to that of our
UVES spectra.

\subsubsection{$\log gf$ -- for individual elements}

\begin{figure}
\centering
\resizebox{\hsize}{!}{\includegraphics{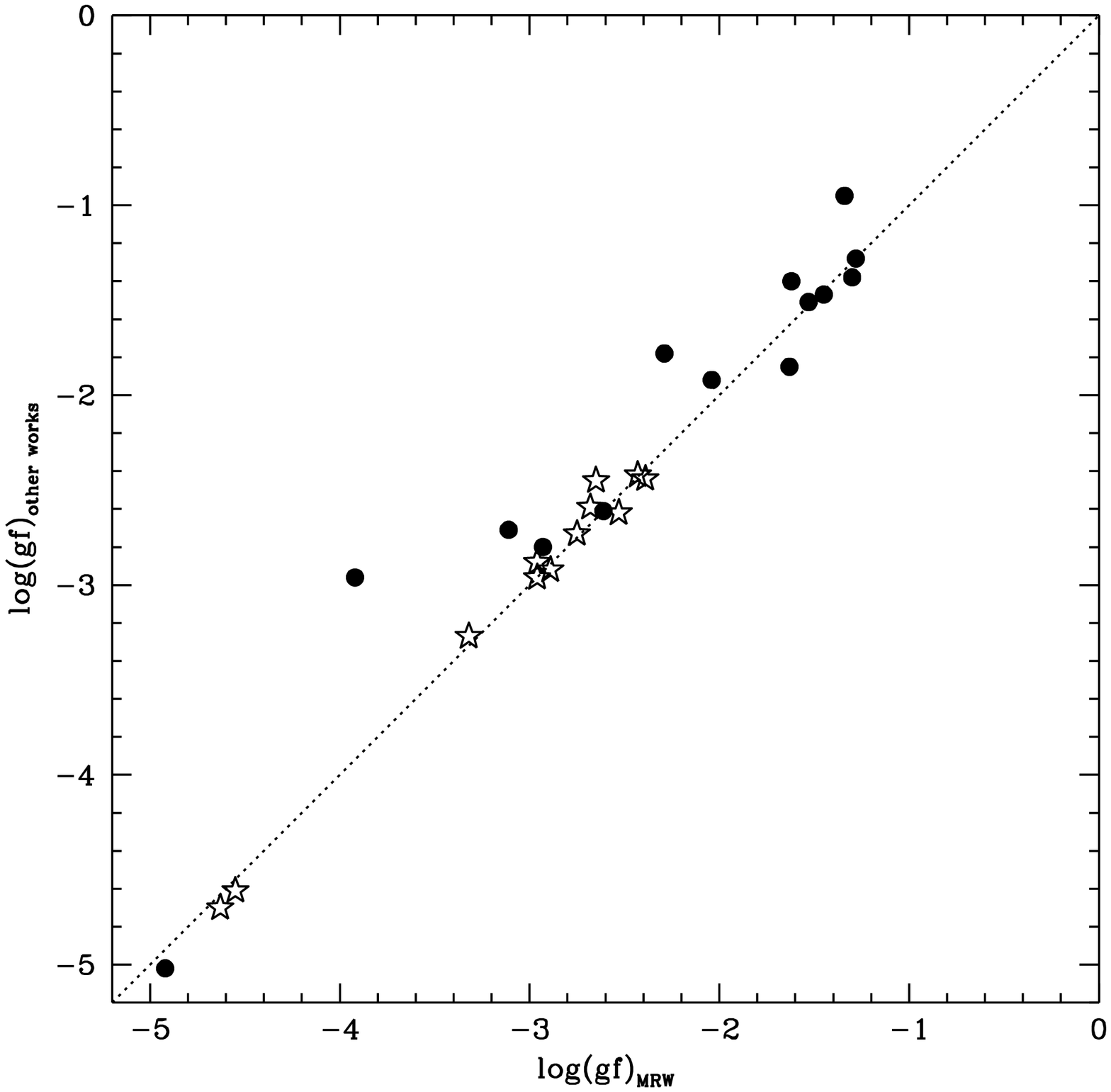}}
\caption{Comparison of the $\log gf$-values for Fe\,{\sc i} lines from
  \citet{may1974} (uncorrected) and values from other
  studies. $\bullet$ mark $\log gf$-values for lines measured by both
  \citeauthor{may1974} and \citet{obrian1991} and the open stars mark
  $\log gf$-values for lines measured by \citet{may1974} and a
  combined data set from \citet{bard1994}, \citet{blackwell1979b},
  \citet{blackwell1982c}, \citet{blackwell1982d}, and \citet{obrian1991}.
The dotted line marks the one-to-one relation.}
\label{fig.comp.may}
\end{figure}

\paragraph{Na\,{\sc i}} To our knowledge there exist no 
laboratory data for the lines in our spectra, however, they can be
readily calculated from theory.  We use the theoretically calculated
$\log gf$-values from \citet{lambert1968} as this provides a
homogeneous data-set for all our Na\,{\sc i} lines.

\paragraph{Si\,{\sc i}} There exist no consistent 
set of $\log gf$-values for those Si\,{\sc i} lines that we are able
to measure in our HB spectra. \citet{garz1973} provides a fairly long
list of laboratory Si\,{\sc i} $\log gf$-values in the visual,
however, most of the wavelength region that we have available is not
covered. Out of the 16 Si\,{\sc i} lines we can measure in our spectra
only 5 have $\log gf$-values from \citet{garz1973}. A further four
lines have astrophysical $\log gf$-values from \citet{bensby2003}.  We
have therefore chosen to measure the remaining lines in a solar
spectrum and derive our own astrophysical $\log gf$-values for the
lines not measured by \citet{garz1973} in order to have two
homogeneous sets of $\log gf$-values and in this way reduce the
line-to-line scatter. We note that the agreement between the
abundances derived from lines with \citet{garz1973} $\log gf$-values
compares very well with those derived using astrophysical $\log
gf$-values. The mean difference between the two sets for all
stars is 0.03 dex. For NGC\,6352-08 the difference is larger, about
0.18 dex. We have no direct explanation for this difference.

\paragraph{Ca\,{\sc i}} For Ca we have decided to use the 
laboratory $\log gf$-values from \citet{smith1981} as this data
set has a high internal consistency. We note, however, that the
absolute scale of this set of $\log gf$-values might be in error as at
least two recent studies, \citet{bensby2003} and \citet{chen2000},
have been unable to reproduce the solar Ca\,{\sc i} abundance
using these $\log gf$-values.  We note that the absolute \citet{smith1981}
 scale is based on the $\log gf$-value for the line at
534.9 nm. Hence if this value should change in the future (as the
solar analyses indicates) then our results should simply be changed by
the difference between the \citet{smith1981} $\log gf$-value for this
line and the new one.

\paragraph{Ti\,{\sc i}} The majority of 
the $\log gf$-values for Ti\,{\sc i} are laboratory data from
\citet{blackwell1982a}, \citet{blackwell1982b}, \citet{blackwell1983},
and \citet{blackwell1986} with corrections according to
\citet{grevesse1989}. For lines not measured by the Oxford group we
apply values from \citet{nitz1998} and \citet{kuehne1978}.

\paragraph{Ti\,{\sc ii}} $\log gf$-values for Ti\,{\sc ii} 
are taken from Tables 1 and 3 in \citet{pickering2001}. Of the 21
values 5 are from Table 3 in Pickering et al. which are purely
theoretical values.

\paragraph{Fe\,{\sc i}} Our main source for identifying suitable Fe\,{\sc i} lines
has been the compilation by \citet{nave1994}. We note that this
compilation, although comprehensive, does not provide a critical
assessment of the quality of the data. Therefore, we have, whenever
possible consulted, and referenced, the original source for the $\log
gf$-values.

One of the most important sources for experimental $\log gf$-values
for medium strong Fe\,{\sc i} lines is the work by
\citet{may1974}. Commonly, following \citet{fuhr1988}, a correction
factor is applied to the \citet{may1974} $\log gf$-values. However,
\citet{bensby2003} found that when the correction factor was applied
to the \citeauthor{may1974} data their $\log gf$-values did result in
an overabundance for the sun of 0.12 dex. Other $\log gf$-values did
not produce such a large overabundance. In Fig.\,\ref{fig.comp.may} we
show a non-exhaustive comparison of \citeauthor{may1974} $\log
gf$-values and data from several other sources (in particular
\citet{obrian1991} and several works by Blackwell and collaborators,
see figure text). We find that the uncorrected \citet{may1974} values
agree very well indeed with data from other studies. This support the
conclusion by \citet{bensby2003} that the correction factors should
not be applied to the \citet{may1974} $\log gf$-values. We thus use
the original values from \citet{may1974}.

\paragraph{Fe\,{\sc ii}} In order to get a homogeneous data-set
we have chosen to use the theoretically calculated $\log gf$-values
from \citet{raassen1998}. They have been shown to agree very well
with data from the {\sc ferrum} project, see  \citet{karlsson2001}
and \citet{nilsson2000}.

\paragraph{Ni\,{\sc i}} We have 28 Ni lines available for abundance
analysis in our spectra. For 7 of these laboratory $\log gf$-values
are available from \citet{wickliffe1997}. For the remainder
(i.e. the majority) no homogeneous data set is available. We thus
decided to follow \citet{bensby2003} and use astrophysical $\log
gf$-values for these lines.

\begin{figure}
\centering
\resizebox{\hsize}{!}{\includegraphics[angle=0]{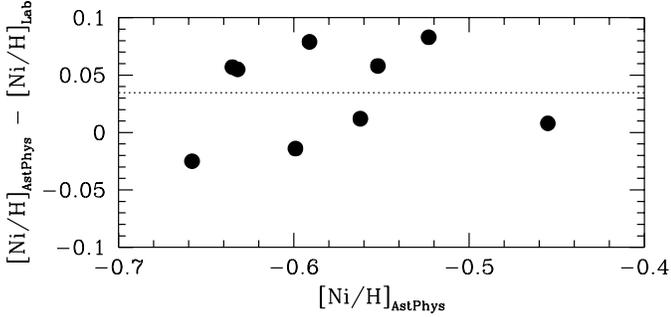}}
\caption{Comparison of resulting nickel abundances for each star when either 
only lines with astrophysical $\log gf$-values are used ([Ni/H]$_{\rm AstPhys}$) 
or when laboratory $\log gf$-values are used ([Ni/H]$_{\rm Lab}$). The 
dotted line indicates the mean difference. }
\label{fig.compnigf}
\end{figure}

In
Fig.\,\ref{fig.compnigf} we compare the resulting [Ni/H] values when
only astrophysical or only laboratory $\log gf$-values are used. The
difference between the two line sets is small (in the mean $< 0.05$ dex)
and will thus not influence our final conclusions in any
significant way. However, they show the desirability in obtaining
larger  sets of laboratory $\log gf$-values for the analysis of
stellar spectra.

\paragraph{Al\,{\sc i}, Mg\,{\sc i}, Cr\,{\sc i}, Cr\,{\sc ii}, and Zn\,{\sc i}}
No laboratory measurements exist for the lines we use for these
elements and we thus use astrophysical $\log gf$-values based on FEROS
spectra taken from \citet{bensby2003}.

\subsubsection{Line broadening parameters}

Collisional broadening is taken into account in the calculation of the
stellar abundances. The abundance program from Uppsala includes
cross-sections from \citet{anstee}, \citet{barklem1}, \citet{barklem3}
1998), \citet{barklem2}, and \citet{barklem4} for over 5000 lines. In particular
the abundances for all but one Ca\,{\sc i} line, all Cr\,{\sc i} lines,
most of the Ni\,{\sc i}, Ti\,{\sc i}, and Fe\,{\sc i} lines are calculated in
this fashion. At the time of our first calculations we did not have 
data for the Fe\,{\sc ii} lines. We thus had a chance to test the influence on
the final Fe abundances as derived from the Fe\,{\sc ii} lines due to
the inclusion of the more detailed treatment for the collisions and found it
to be negligible. 

For the remainder of the lines we apply the classical Uns\"old
approximation for the collisional broadening and use a correction term
($\gamma_{\rm 6}$). For those few Fe\,{\sc i} lines with no
cross-sections we follow \citet{bensby2003} and take the $\gamma_{\rm
  6}$ from \citet{simmons1982} if $\chi_{\rm l}<2.6$ eV and for lines
with greater excitation potentials we follow \citet{chen2000} and use
a value of 1.4.

As noted by \citet{carretta2000} the collisional damping parameters
are a concern for our Na\,{\sc i} lines. For the lines at 568.265 and
568.822 nm we use the cross sections as implemented in the code,
whilst for the lines at 615.422 and 616.075 nm we use a $\gamma_{\rm
  6}$ of 1.4. The mean difference between the two sets of lines (for
an LTE analysis) is 0.14 dex. This could indicate that the
$\gamma_{\rm 6}$ used for the two redder lines is too large, however,
NLTE is an additional concern for the determination of Na abundances
(see Sect.\,\ref{sect:naal}).

For the Si\,{\sc i} lines we use a $\gamma_{\rm 6}$ of 1.3.

If no other information is available for
the collisional broadening term we follow \citet{mackle1975} and
use a value of 2.5 (Mg, Al, Cr\,{\sc ii}, Ti\,{\sc ii}, and Zn).

\subsection{Stellar parameters}
\label{sect:par} 

\begin{table}
\caption[]{Reddening estimates for NGC\,6352 from the literature.}
\label{red.tab}
\begin{tabular}{lllllllll}
\hline\hline 
\noalign{\smallskip}
E(B-V) &  Ref. & Comment \\
\noalign{\smallskip}
\hline
\noalign{\smallskip}
 0.44           & \citet{alcaino1971} \\
 0.32$\pm$0.05  & \citet{hartwick1972} \\ 
                & \citet{hesser1976} \\
 0.29           & \citet{mould1984} \\
 0.21$\pm$0.03  & \citet{fullton1995}  \\
 0.33           & \citet{schlegelmap} &from NED$^{\rm a}$\\
\noalign{\smallskip}
\hline
\end{tabular}
\begin{list}{}{}
\item[$^{\mathrm{a}}$] The NASA/IPAC Extragalactic Database
at {\tt http://nedwww.ipac.caltech.edu/index.html}
\end{list}
\end{table}

\begin{table*}
  \caption[]{Stellar parameters. The first column identifies the stars
    (see Table\,\ref{phot.tab}), the second gives the colour corrected
    for the interstellar reddening, as described in
    Sect.\,\ref{sect:teff}. The third column gives the reddening
    corrected colour transfered to the standard system. It is this
    value that is used to derive the $T_{\rm eff}$ listed in the
    fourth column ($T_{\rm eff}^{\rm phot}$). Column five lists the
    $T_{\rm eff}$ derived from spectroscopy ($T_{\rm eff}^{\rm
      spec}$). Column six to eight list the finally adopted $\log g$,
    [Fe/H], and $\xi_t$ (as derived in Sect.\,\ref{sect:par}).  }
\label{par.tab}
\begin{tabular}{llllllll}
\hline\hline 
\noalign{\smallskip}
Star  & $(V-I)_{\rm 0, HST}$ & $(V-I)_{\rm 0}$ & $T_{\rm eff}^{\rm phot, a}$ &  $T_{\rm eff}^{\rm spec}$& $\log g_{\rm spec}$ & [Fe/H] & $\xi_{\rm t}$ \\
      &               &                & (K)                         & (K)                      &          &        & km\,s$^{-1}$\\
\noalign{\smallskip}
\hline
\noalign{\smallskip}
NGC\,6352-01 &  1.0257 & 1.0382 &  4706 & 4950 & 2.50 & --0.55 & 1.40 \\  
NGC\,6352-02 &  1.0709 & 1.0841 &  4609 & 4900 & 2.30 & --0.55 & 1.30 \\  
NGC\,6352-03 &  0.9238 & 0.9349 &  4947 & 5000 & 2.50 & --0.55 & 1.40 \\  
NGC\,6352-04 &  0.9470 & 0.9585 &  4890 & 4950 & 2.50 & --0.50 & 1.30 \\  
NGC\,6352-05 &  0.9164 & 0.9274 &  4966 & 4950 & 2.30 & --0.60 & 1.40 \\  
NGC\,6352-06 &  0.9056 & 0.9164 &  4994 & 4950 & 2.30 & --0.55 & 1.40 \\  
NGC\,6352-07 &  0.9379 & 0.9492 &  4912 & 5050 & 2.70 & --0.50 & 1.45 \\  
NGC\,6352-08 &  0.9060 & 0.9169 &  4992 & 5050 & 2.50 & --0.55 & 1.45 \\  
NGC\,6352-09 &  0.9566 & 0.9681 &  4866 & 4900 & 2.30 & --0.60 & 1.40 \\  
\noalign{\smallskip}
\hline
\end{tabular}

\begin{list}{}{}
\item[$^{\mathrm{a}}$] Based on Houdashelt et al. calibration
\end{list}
\end{table*}

\begin{figure}
\centering
\resizebox{\hsize}{!}{\includegraphics{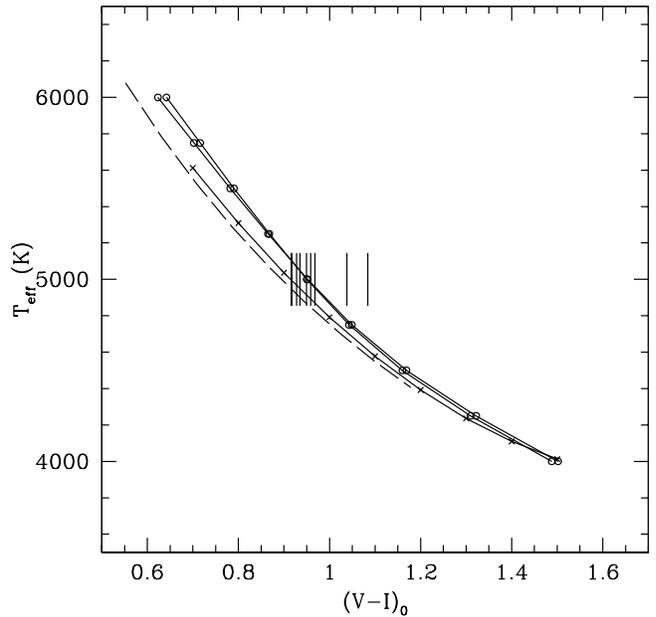}}
\caption{$T_{\rm eff}$ -- ($V-I$)$_{\rm 0}$ calibrations from \citet{alonso1999calib}
  (dashed line), \citet{houdashelt2000} (solid line with $\times$),
  and \citet{bessel1998} (solid line with $\circ$). The latter for models
  with and without overshooting. The colours of our target stars in
  NGC\,6352 are indicated by vertical lines, Table\,\ref{par.tab}.  }
\label{fig.teffvi}
\end{figure}

\subsubsection{Effective temperatures}
\label{sect:teff}
Initial estimates of the effective temperatures ($T_{\rm eff}$) for
the stars were derived using our HST/WFPC2 photometry, Table
\ref{phot.tab}. These magnitudes are in the in-flight HST/WFPC2 system
and must therefore be dereddened and then corrected to standard
Cousins colours before the temperature can be derived.

Estimates, from the literature, of the reddening towards NGC\,6352 are
collected in Table\,\ref{red.tab}. Reddening towards globular clusters
are often determined in relation to another cluster of similar
metallicity and with a well-known, and low, reddening value. For
NGC\,6352 47 Tuc has been considered a suitable match based on their
similar metallicities. In fact their metallicities might differ, such
that NGC\,6352 is somewhat more metal-rich. This would indicate that
the reddening relative to 47 Tuc is an upper limit.
\citet{fullton1995} provide the latest investigation of the reddening
estimate based on the cluster data themselves. Their determinations
are based on WFPC1 data.  They use two different techniques;
comparison with the RGB of 47 Tuc yielded $0.22\pm0.03$ mag and
solving for both metallicity and reddening, using the equations in
\citet{sarajedini1994}, yielded $0.21\pm0.03$ mag which is their
recommended value.  Another recent estimate, from the NED database,
based on the galactic extinction map of \citet{schlegelmap}, is 0.33
mag (see Table\,\ref{red.tab}). Given that this is a more general
evaluation than the study by \citet{fullton1995} we have opted for the
value in the latter study.

Differential reddening along the line-of-sight towards NGC\,6352 has
been estimated to be small. \citet{fullton1995} find it to be less
then 0.02 mag for WFPC1 CCD nos. 6--8 and less than 0.07 mag for CCD
No. 5. Given the various other error sources in the photometry: the
HST/WFPC2 reddening values (see below), the transformation to standard
values (Eq.\,\ref{prov}), and the temperature calibrations
(Fig.\,\ref{fig.teffvi}) we consider the reddening towards NGC\,6352 to
be constant for all our stars.

To deredden the colours in Table\,\ref{phot.tab} we used the relations
in \citet{holtzman1995b} Table 12. The reddening correction in
$V_{\rm 555} - I_{\rm 814}$ corresponding to $E(B-V)=0.21$ is thus
0.258, which was applied to all stars.

After correcting the magnitudes for extinction we can transform the
in-flight magnitudes to standard colours.  As we have used the
relations in \citet{dolphin2000} to calibrate our in-flight magnitudes
\footnote{We have used the most updated values that are available on
  A. Dolphin's web-site at
  http://purcell.as.arizona.edu/wfpc2\_calib/} we also use his
relations to transform our in-flight magnitudes and colours to
standard Cousins colours.

$$ V_{\rm 0} = V_{\rm 555, 0} - 0.052(V-I)_{\rm 0}+0.027(V-I)_{\rm 0}^{\rm 2}$$
$$ I_{\rm 0} = I_{\rm 814, 0} - 0.062(V-I)_{\rm 0}+0.025(V-I)_{\rm 0}^{\rm 2}$$

\noindent
Where, $ V_{\rm 0}$, $I_{\rm 0}$, and $(V-I)_{\rm 0}$ are the standard
magnitudes and colours, respectively, and $V_{\rm 555, 0}$ and $I_{\rm
  814, 0}$ are the dereddened in-flight magnitudes. Incidentally, for
these filters the coefficients by \citeauthor{dolphin2000} are
identical to those by \citet{holtzman1995b} (their Table\,7).

Solving for $(V-I)_{\rm 0}$ we obtain (the other solution
is un-physical)

\begin{equation}
(V-I)_{\rm 0} = \frac{0.99 - \sqrt{0.99^2 +  0.004(V_{\rm 555}-I_{\rm 814})_{\rm 0}}}{0.004}\,. \label{prov}
\end{equation}

Eq.\,\ref{prov} is then used to obtain the final $(V-I)_{\rm 0}$ to be
used to derive $T_{\rm eff}$.

In the literature several calibrations, both empirical and
theoretical, of colours in terms of $T_{\rm eff}$ are available. In
Fig.\,\ref{fig.teffvi} we compare one empirical and two theoretical
calibrations. In e.g. \citet{houdashelt2000} a more extensive
comparison is available. The calibration by \citet{alonso1999calib} is
originally calculated using colours in the Johnson system, while the
calibrations by \citet{bessel1998} and \citet{houdashelt2000} as
well as the HST/WFPC2 in-flight UBVRI system are in the Johnson-Cousin
system. The \citet{alonso1999calib} calibration was transformed to the
Johnson-Cousin system using the relations in \citet{fernie1983}.

It is noteworthy that all three calibrations, at the colours of our
stars, agree within less than 100 K. As we have no reason to believe
that either calibration is superior and, more importantly, our colours
most likely have large errors (since the various calibration steps
when going from in-flight HST/WFPC2 colours to standard colours are
not too well calibrated) we choose to use the \citet{houdashelt2000}
calibration for our starting values.  In Table\,\ref{par.tab} we list
the derived standard colours and the $T_{\rm eff}$ from the
\citet{houdashelt2000} calibration.

\subsubsection{The metallicity of NGC\,6352}
\label{sect:met}

\begin{table*}
\caption[]{Metallicity estimates for NGC\,6352 from the literature.}
\label{feh.tab}
\begin{tabular}{lllllllll}
\hline\hline 
\noalign{\smallskip}
[Fe/H] & Method & Ref. & Comment \\
\noalign{\smallskip}
\hline
\noalign{\smallskip}
$\geq 0.1 \pm 0.1$ & Two-colour diagram relative to Hyades & \citet{hartwick1972} &\\ 
$-1.3 \pm 0.1$     & Detailed abundance analysis & \citet{geisler1981} & 1 star \\
$-0.38 _{\rm 47 Tuc}$ & High-resolution spectra & \citet{cohen1983} & 8 stars, 47 Tuc at $-0.8$ dex\\
$-0.51\pm0.08$  & Based on $Q_{39}$ & \citet{zw84}  &  \\
$-0.50 \pm 0.2$    & TiO band strength & \citet{mould1984} & 8 stars\\
$-0.79\pm 0.06$  & Detailed abundance analysis & \citet{gratton1987}  & 3 stars \\ 
$-0.64\pm 0.06$& Re-analysis of $W_{\lambda}$ from \citet{gratton1987}  & \citet{cg97}  & $\sigma = 0.11$, 3 stars\\
$-0.80$ & High-resolution spectra  & \citet{francois1991} & 1 star\\
\hline
\noalign{\smallskip}
$-0.50\pm0.08$& Ca{\sc ii} triplet  & \citet{rutledge1997}&  23 stars, Based on ZW84 scale$^{\rm a}$ \\
$-0.70\pm0.02$&   & & 23 stars, Based on CG97 scale$^{\rm b}$ \\
\hline
\noalign{\smallskip}
$-0.78$  & {Re-calibration using Fe\,\sc{ii}}$^{\rm c}$ & \citet{kraft2003}& MARCS \\ 
$-0.70$  & & & Kurucz conv. overshoot\\
$-0.69$  & & & Kurucz no conv. overshoot \\
\noalign{\smallskip}
\hline
\end{tabular}
\begin{list}{}{}
\item[$^{\mathrm{a}}$] ZW84 $=$ \citet{zw84}
\item[$^{\mathrm{b}}$] CG97 $=$ \citet{cg97}.
\item[$^{\mathrm{c}}$] Three different types of model atmospheres were
  used in the re-calibration. These are indicated in the comment
  column. For a full discussion of these atmospheres as well as the
  results see \citet{kraft2003}.
\end{list}
\end{table*}

The metallicity of a globular cluster is often estimated from the
colour magnitude diagram. Several such estimates exist for
NGC\,6352. They are listed in Table\,\ref{feh.tab}.

Spectroscopy of stars as faint as those in NGC\,6352 is obviously
difficult with smaller telescopes, however, measurements of strong
lines like the IR Ca\,{\sc ii} triplet lines are useful tools and
\citet{rutledge1997} observed 23 stars in the field of NGC\,6352. They
derived a metallicity of $-0.5$ or $-0.7$ dex depending on which
calibration for the IR Ca\,{\sc ii} triplet they used.  Narrow-band
photometry of e.g. TiO can also provide metallicity estimates, see
e.g. \citet{mould1984} who found an iron abundance of $-0.50 \pm 0.2$
dex.

Detailed abundance analysis requires higher resolution and could thus
only be done for the brightest stars prior to the 8m-class telescopes.
This normally means that the stars under study will be rather cold
(e.g. around $4000-4300$ K). For such cool stars detailed abundance
analysis becomes harder as molecular lines become stronger when the
temperature decreases. In spite of these difficulties early studies
provide interesting results from detailed abundance
analysis. Analyzing the spectra of one star \citet{geisler1981}
derived an [Fe/H] of $-1.3 \pm 0.1$ dex while \citet{gratton1987}
analyzed three stars and found a value of $-0.79\pm 0.06$ dex (the
error being the internal error). Gratton's $W_{\lambda}$ were later
reanalyzed by \citet{cg97} using updated atomic data as
well as correcting the \citet{gratton1987} $W_{\lambda}$s. They
derived an [Fe/H] of $-0.64 \pm 0.06$ dex. \citet{cohen1983} analyzed
8 stars in NGC\,6352 using high-resolution spectra and found the
cluster to have a mean iron abundance of $+0.38$ relative to 47
Tuc. With 47 Tuc at $-0.8$ dex this puts NGC\,6352 at $-0.42$ dex.

Apart from the \citet{geisler1981} value all studies listed
in Table\,\ref{feh.tab} appear to point to an [Fe/H] for NGC\,6352
between $-0.5$ and $-0.8$ dex. To be perfectly sure we will explore a
somewhat larger range of [Fe/H] in our initial analysis
(Sect.\,\ref{sect:finalpar}).

\subsubsection{First estimate of $\log g$ }
\label{sect:logg}

Assuming that the metallicities in the literature are approximately
correct we can use stellar evolutionary models to get an estimate of
the range of surface gravities that our programme stars should have.
In particular we consulted the stellar isochrones by
\citet{girardi2002} for $Z=0.001, 0.004, 0.008$ which corresponds to
$-1.33, -0.70, -0.40$ dex, respectively, according to the calibration
given in \citet{bertelli1994}. In these models HB stars have $\log g$
between 2.2 and 2.4 dex and RGB stars at the same magnitude also have
$\log g$ in this range. So even if one or two of our stars are RGB
stars (which have less reddening than the HB stars) exploring the same
$\log g$ range will be enough. To be entirely safe we have explored a
range of $\log g$ from 1.7 to 2.5 dex.

\subsubsection{Derivation of final stellar parameters}
\label{sect:finalpar}

In this study we will assume that all the stellar parameters can be
derived from the spectra themselves, what is sometimes called a
detailed or fine abundance analysis. This means that we require:

\begin{itemize}

\item ionizational equilibrium, i.e. [Fe\,{\sc i}/H] = [Fe\,{\sc ii}/H],
this sets the surface gravity,

\item excitation equilibrium, 
that [Fe\,{\sc i}/H] as a function of  $\chi_{\rm l}$ should
show no trend, this sets  $T_{\rm eff}$,

\item that lines of different strengths should give the same abundance,
i.e. [Fe\,{\sc i}/H] as a function of $\log W/{\lambda}$ should show no 
trend, this sets the microturbulence, and  

\item last but not least important, the [Fe\,{\sc i}/H] derived
should closely match the metallicity used to create the model
atmosphere

\end{itemize}

Often in stellar abundance analysis the investigator has a set of
stellar parameters that are assumed to be rather close to the final
value and a model is created with those values and the trends
discussed above are inspected and the parameters changed in a
prescribed iterative fashion until no trends are found. Here, however,
we can not be very certain about our starting values, even if we have
done our best to find the most likely range (see
Sect.\,\ref{sect:teff}, \ref{sect:met}, and \ref{sect:logg}).
Especially the reddening poses a specific uncertainty.  The assumed
value for the reddening strongly influences $T_{\rm eff}$. We have
therefore opted for a slightly different approach by calculating
abundances for each star for a grid of model atmospheres spanning the
whole range of possible stellar parameters. After inspecting the first
grid we were then able to refine the grid around the most likely
values and produce a more finely spaced grid. This grid then allowed
the derivation of the stellar parameters for the final model.

First we constructed a grid of MARCS model atmospheres (Gustafsson et
al. 1975, Edvardsson et al. 1993, Asplund et al. 1997). The grid spans
the following range $T_{\rm eff}$= 4500, 4600, 4700, 4800, 4900, 5000,
5100 , [Fe/H]=$-0.25 -0.5, -0.75$, and $\log g$= 1.7, 2.0, 2.3, 2.5.
Using these models, the measured equivalent widths and the line
parameters discussed above we calculated Fe\,{\sc i} and Fe\,{\sc ii}
abundances for all models for each star and for three different values
of the microturbulence, $\xi_t$=$1.0, 1.5, 2.0$.  This grid of results
was inspected with regards to the criteria discussed above and it
turned out to be very straightforward to identify the range of
temperatures that were applicable. We then created a finer grid around
the appropriate temperatures and inspected the same criteria again and
from this inspection it was, again, straightforward to find the stellar
parameters that fulfilled all of the criteria listed above.

\begin{figure}
\centering
\resizebox{\hsize}{!}{\includegraphics[angle=-90]{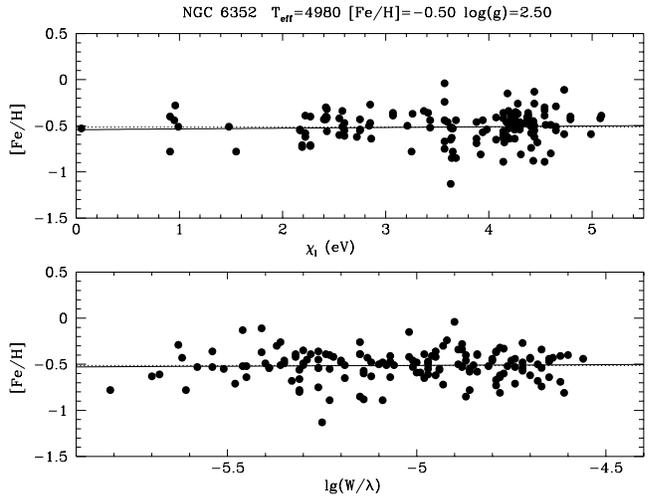}}
\caption{Diagnostic check that the final parameters for NGC\,6352-03
  give no trends for [Fe/H] as a function of $\chi_{\rm l}$ and $\log
  W/{\lambda}$. Parameters used to create the model atmosphere are
  indicated on the top. A $\xi_{\rm t}=1.40$ was used when deriving
  the stellar abundances.  The mean [Fe/H] is indicated with a dotted
  line in each panel and the trends of [Fe/H] vs $\chi_{\rm l}$ and
  $\log W/{\lambda}$ are indicated with full lines (the dotted and
  full lines almost completely overlap). The slopes are: vs $\chi_{\rm
    l}$ $-0.0037$ and vs $\log W/{\lambda}$ $+0.0077$ }
\label{fig.diag}
\end{figure}

An example of the final fit for NGC\,6352-03 of the slopes are given
in Fig.\,\ref{fig.diag}.  Here we see how well the excitation and
strength criteria are met by the set of final parameters. 

In the ideal situation the four criteria listed at the beginning of
this section should be ``perfectly'' met.  In practice we assumed that
ionizational equilibrium was met when $ |$[Fe\,{\sc i}/H]$ -
$[Fe\,{\sc ii}/H]$| < 0.025$, that the excitation equilibrium was
achieved when the absolute value of the slope in the [Fe\,{\sc i}/H]
vs $\chi_{\rm l}$ diagram was $\leq 0.005$. Similarly, that the
$\xi_{\rm t}$ was found when the slope in the [Fe\,{\sc i}/H] vs $\log
W/{\lambda}$ diagram was $\leq 0.005$. For some stars we relaxed the
criterion for the absolute value of the slope in the [Fe\,{\sc i}/H]
vs $\chi_{\rm l}$ diagram somewhat as it proved impossible to satisfy
that at the same time as satisfying the criterion for line strength
equilibria. The final slopes are listed in Table\,\ref{slopes.tab} and
the values for [Fe\,{\sc i}/H] and [Fe\,{\sc ii}/H] can be found in
Table\,\ref{ab.tab}. We note that with the method adopted here we did
only find one combination of stellar parameters that fulfilled all
four of our criteria, no degeneracies were found.

\subsubsection{A new reddening estimate and final $\log gs$ -- discussion}
\label{sect:justification}

\paragraph{New reddening estimate}

As discussed in Sect.\,\ref{sect:par} and summarized in Table
\ref{red.tab} the reddening estimates vary quite considerably between
different studies. We used the reddening to derive de-reddened colours
used to determine $T_{\rm eff}$ in Sect.\,\ref{sect:par} but these were
merely used as starting values and we subsequently found new $T_{\rm
  eff}$s. The difference between the first estimates and the final, adopted
$T_{\rm eff}$ is around +90 K. We may use this temperature offset to
derive a new estimate for the reddening.  The new reddening estimate
is found by changing the reddening such that we minimize the
difference between our spectroscopic $T_{\rm eff}$ and the photometric
$T_{\rm eff}$. We find a minimum difference of $0\pm20$ K if we add a
further 0.036 mag to the reddening as measured in the HST/WFPC2
in-flight system, which we found in Sect.\,\ref{sect:teff} to be
0.258. Thus $E(V_{\rm 555}-I_{\rm 814})=0.294$ which corresponds to
$E(B-V)=0.24$.

\paragraph{Surface gravity ($\log g$)}

We note that although we allow $\log g$ to vary freely 
we did indeed, by requiring ionizational equilibrium, derive
final $\log g$ values that are consistent with stellar evolutionary
tracks (e.g. Girardi et al. 2002). 

NGC\,6352-07 appears to have an unusually large $\log g$ for being
situated on the HB. From its location in the CMD the star appears as a
bona fide HB star (unless the reddening towards this particular star
is significantly less than towards the stars in general). The reason
for this is not clear to us.

As an additional test we have used infrared $K$ magnitudes from the
2MASS survey \citep{skrutskie2006} and the basic formula $\log g_{\rm
  star} = 4.44 + 4 \log {T_{\rm star}/T_\odot} + 0.4 (M_{\rm bol,
  star} - M_{\rm bol, \odot})$ to estimate the $\log g$s. For the Sun
we adopted a temperature of 5770 K and $M_{\rm bol, \odot} = 4.75 $.
For the stars we used a mass of 0.8 $M_{\odot}$ and the $T_{\rm eff}$
from Table\,\ref{par.tab}.  When infrared data are available they are
a better choice for deriving the bolometric magnitude than the visual
data as they suffer less from reddening and metallicity effects.  The
bolometric magnitudes were derived using ${M_{\rm bol}} = {M_{\rm K}}
+ {\rm BC_{\rm K}}$, where the bolometric correction was set to 1.83
(from Houdashelt et al. 2000).  Using this procedure we found a $\log
g$ around 2 for all our stars with $E(B-V)=0.21$ and $(m-M)=14.44$
\citep{harris1996}. However, as shown above our spectroscopically
derived $T_{\rm eff}$s appear to indicate a higher reddening, $\sim
0.24$. We also note that the error on the distance modulus is $\pm
0.15$ magnitudes \citep{fullton1995}.  Changing $(m-M)$ to $14.05$ and
adopting our new reddening estimate we derive $\log g$s of $\sim 2.2$
dex. However, as discussed in Sect.\,\ref{sect:errors} and summarized
in Tables\,\ref{errors.tab} and \ref{sloperrors.tab}, the effect on
the final elemental abundances from such a small change in $\log g$ is
negligible.

We may thus conclude that the $\log g$s derived by requiring
ionizational equilibrium for Fe is a valid method for abundance
analysis of the type of stars studied here.

\begin{table}
\caption[]{Slopes for [Fe/H] for individual lines
 as a function of $\chi_{\rm l}$ and $\log W/{\lambda}$, compare
Fig.\,\ref{fig.diag}}
\label{slopes.tab}
\begin{tabular}{lrrrrr}
\hline\hline 
\noalign{\smallskip}
Star & \multicolumn{1}{l}{Slope($\chi_{\rm l}$)} & \multicolumn{1}{l}{Slope($\log W/{\lambda}$)} \\
\noalign{\smallskip}
\hline
\noalign{\smallskip}
NGC\,6352-01& --0.0026 & --0.0265 \\
NGC\,6352-02&   0.0002 &   0.0062 \\
NGC\,6352-03& --0.0037 &   0.0077 \\
NGC\,6352-04& --0.0066 & --0.0225 \\
NGC\,6352-05& --0.0001 &   0.0120 \\
NGC\,6352-06& --0.0041 &   0.0140 \\
NGC\,6352-07& --0.0059 & --0.0400 \\
NGC\,6352-08& --0.0023 & --0.0040 \\
NGC\,6352-09&   0.0081 &   0.0068 \\
\noalign{\smallskip}
\hline
\end{tabular}
\end{table}

\subsection{Stellar abundances - error budget}
\label{sect:errors}

\begin{table*}
\caption[]{Error estimates for NGC\,6352--03. Investigation of the effect on the
 resulting abundances from changes of the stellar parameters. Here we 
  change $T_{\rm eff}$ with $-100$ K, $\log g$ with $+ 0.4$ dex,
  [Fe/H] with $+ 0.1$ dex and $\xi_{\rm t}$ with $\pm 0.20$ km s$^{\rm
    -1}$. The elemental abundances are given as [X/H], where X is the
  element indicated in the first column. For three elements we also
  include data for abundances derived from lines arising from singly
  ionized atoms (as indicated in the first column). The second column
  gives the final elemental abundances as reported in
  Table\,\ref{ab.tab}. Here also the one $\sigma$ (standard deviation)
  and the number of lines used are indicated.  The following columns
  report the changes in the abundances relative to the results
  reported in column two when the stellar parameters are varied as
  indicated in the table header. The differences are given in the
  sense ${\rm [X/H]}_{\rm Final}-{\rm [X/H]}_{\rm
    Modified}=\Delta{\rm [X/H]}$ and ${\rm [X/Fe]}_{\rm Final}-{\rm
    [X/Fe]}_{\rm Modified}=\Delta{\rm [X/Fe]}$, respectively, where X is any
  element. Hence the values for the modified models are equal to
  ${\rm[X/H]}_{\rm Final}+\Delta{\rm [X/H]}$ and ${\rm[X/Fe]}_{\rm
    Final}+\Delta{\rm [X/Fe]}$, respectively.}
\label{errors.tab}
\begin{tabular}{llrrrrrr|rrrrr}
\hline\hline 
\noalign{\smallskip}
 Element        &\multicolumn{2}{l}{Final abundances}     & \multicolumn{5}{c}{$\Delta$[X/H]}  & \multicolumn{5}{c}{$\Delta$[X/Fe]} \\ 
                &\multicolumn{2}{l}{}  &  $\Delta T_{\rm eff}$ & $\Delta \log g$ &$\Delta {\rm [Fe/H]}$ &\multicolumn{2}{c}{$\Delta \xi_{\rm t}$} &  $\Delta T_{\rm eff}$ & $\Delta \log g$ &$\Delta {\rm [Fe/H]}$ &\multicolumn{2}{c}{$\Delta \xi_{\rm t}$} \\
 &  &  & --100K &  +0.4 &   +0.1 &  +0.2 & --0.2  & --100K &  +0.4 &   +0.1 &  +0.2 & --0.2\\
\noalign{\smallskip}
\hline
\noalign{\smallskip}
 Na            & --0.16 $\pm$ 0.14 &  (4) &   +0.07 &  +0.05& --0.01&  +0.03& --0.05 & --0.02&  +0.04& --0.01&  --0.04&  +0.01\\
 Mg            & --0.08            &  (1) &   +0.08 &  +0.07& --0.01&  +0.06& --0.06 & --0.01&  +0.06& --0.01&  --0.01&   0.00\\
 Al            & --0.11 $\pm$ 0.01 &  (2) &   +0.06 &  +0.01&   0.00&  +0.01& --0.02 & --0.03&   0.00&   0.00&  --0.06&  +0.04\\ 
 Si            & --0.36 $\pm$ 0.16 &  (11)&   +0.00 & --0.05& --0.01&  +0.02& --0.02 & --0.09& --0.05& --0.01&  --0.05&  +0.04\\
 Ca            & --0.39 $\pm$ 0.08 &  (12)&   +0.10 &  +0.05&   0.00&  +0.09& --0.08 &  +0.01&  +0.04&   0.00&   +0.02&  --0.02\\ 
 Ti            & --0.39 $\pm$ 0.15 &  (40)&   +0.14 &  +0.02&   0.00&  +0.06& --0.06 &  +0.05&  +0.01&   0.00&  --0.01&   0.00\\          
 Ti{\sc ii}    & --0.24 $\pm$ 0.17 &  (14)&  --0.01 & --0.15& --0.03&  +0.08& --0.08 & --0.10& --0.16& --0.03&   +0.02&  --0.02\\         
 Cr            & --0.61 $\pm$ 0.11 &  (6) &   +0.14 &  +0.02&   0.00&  +0.07& --0.07 &  +0.05&  +0.01&   0.00&    0.00&  --0.01\\  
 Cr{\sc ii}    & --0.53 $\pm$ 0.15 &  (6) &  --0.05 & --0.16& --0.02&  +0.04& --0.04 & --0.14& --0.17& --0.02&  --0.03&  +0.02\\
 Fe            & --0.54 $\pm$ 0.16 & (193)&   +0.09 &  +0.01&   0.00&  +0.07& --0.06 &       &       &       &        & \\
 Fe{\sc ii}    & --0.55 $\pm$ 0.11 &  (18)&  --0.07 & --0.19& --0.05& --0.08&  +0.04 & --0.17& --0.20& --0.05&  --0.14&  +0.11\\
 Ni            & --0.60 $\pm$ 0.10 &  (26)&   +0.07 & --0.03& --0.01&  +0.05& --0.06 & --0.02& --0.04& --0.01&  --0.02&   0.00\\ 
 Zn            & --0.22 $\pm$ 0.05 &  (2) &  --0.03 & --0.08& --0.02&  +0.07& --0.07 & --0.12& --0.09& --0.02&    0.00&  --0.01\\
\noalign{\smallskip}
\hline
\noalign{\smallskip}
\end{tabular}
\end{table*}

\begin{table}
\caption[]{Slopes for NGC\,6352-03 for [Fe/H] for individual lines
 as a function of $\chi_{\rm l}$ and $\log W/{\lambda}$ for the same 
changes in stellar parameters as in Table\,\ref{errors.tab}}
\label{sloperrors.tab}
\begin{tabular}{llll}
\hline\hline 
\noalign{\smallskip}
Parameter & Change & \multicolumn{1}{l}{Slope\,($\chi_{\rm l}$)} & \multicolumn{1}{l}{Slope\,($\log W/{\lambda}$)} \\
\noalign{\smallskip}
\hline
\noalign{\smallskip}
Final slopes        &                    &   --0.0037 &   +0.0077\\
\noalign{\smallskip}
\hline
\noalign{\smallskip}
$\Delta T_{\rm eff}$  & --100 K            &  +0.0229 & --0.0356 \\
$\Delta \log g$      & +0.4 dex           & --0.0029 & --0.0625 \\
$\Delta {\rm [Fe/H]}$& +0.1 dex           & --0.0042 &  +0.0123 \\
$\Delta \xi_{\rm t}$  & +0.2 km s$^{-1}$   &  +0.0162 & --0.1466  \\
                     & --0.2 km s$^{-1}$   & --0.0237 &  +0.1620 \\
\noalign{\smallskip}
\hline
\end{tabular}
\end{table}

To investigate the effect of erroneous stellar parameters on the
derived elemental abundances we have for one star, NGC\,6352-03,
varied the stellar parameters and re-derived the elemental
abundances. The results are presented in Table\,\ref{errors.tab}.
Note that the Na abundances reported in this table have not
been corrected for NLTE effects (see Sect.\,\ref{sect:naal})

We see that, for lines from neutral elements, errors in the
temperature scale are in general the largest error source, whilst
changes in $\log g$ generally causes smaller changes. The opposite is
true for abundances derived from lines arising from singly ionized
species.

It is notable that an error in the temperature causes essentially the
same error in e.g. the Ca abundance as in the Fe abundance (from
neutral lines).  This means that the ratio of Ca to Fe remains
constant.  It is also interesting to note that the Si abundance
appears particularly robust against any erroneous parameter.  Changes
in metallicity in the model cause neglible changes in the final
abundances.

In Table\,\ref{sloperrors.tab} we list the slopes for the diagnostic
checks for excitation equilibrium and line strength equilibrium
(compare Fig.\ref{fig.diag} and Table\,\ref{slopes.tab}) for each of
the models used to calculate the error estimates in
Table\,\ref{errors.tab}. As can be seen changes in $T_{\rm eff}$ as
well as in $\xi_{\rm t}$ causes notable changes and these models would
hence easily be discarded as not fulfilling the prerequisite for a
good fit.  Changes in $\log g$ and [Fe/H] causes smaller changes in
the slopes. However, as can be seen in Table\,\ref{errors.tab} a
change in $\log g$ causes a real change in the ionizational
equilibrium and such a model would also thus be discarded. Finally,
even though a change in [Fe/H] in the model has very limited effect on
slopes as well as on (most) derived elemental abundances, we require
the model to have a [Fe/H] that is the same as that derived using the
final model. Hence, also models with offset [Fe/H] would be discarded.
 
In summary, these final considerations show that we have derived model
parameters that are self-consistent and that errors in [X/Fe], where
X is any element, are reasonably robust against errors in the 
adopted parameters (with the exception of singly ionized species and 
Zn, which all have at least one change in a parameter causing a change
in abundance larger than 0.1 dex. Table\,\ref{errors.tab}).

Additionally, we note that our internal line-to-line scatter
($\sigma$) is on par with what is found in other studies of HB and RGB
stars in metal-rich globular and open clusters (e.g. Sestito et
al. 2007, Carretta et al. 2001, Carretta et al. 2007, and clusters
listed in Table \,\ref{tab:compcl}).

\section{Results}
\label{sect:res}

\begin{table*}
\caption[]{Stellar abundances. For each star we give the mean
  abundance ([X/H], X being the element indicated in the first column),
  the $\sigma$ and the number of lines used in the final abundance
  derivation. The error in the mean is thus $\sigma$ divided by
  $\sqrt{N_{\rm lines}}$.  In the two last entries we give the mean
  and median values for the cluster. For the mean value we also give
  the $\sigma$. The mean and median values are based on all nine
  stars.  }
\label{ab.tab}
\begin{tabular}{lllllllllll}
\hline\hline 
\noalign{\smallskip}
El & NGC\,6352-01 & NGC\,6352-02 & NGC\,6352-03 & NGC\,6352-04 & NGC\,6352-05 &NGC\,6352-06 \\
\noalign{\smallskip}
\hline
\noalign{\smallskip}
Na I & --0.46 0.05   4& --0.38 0.15   4& --0.16 0.14   4& --0.49 0.12   4& --0.16 0.09   4 & --0.51 0.09   4&\\
Mg I & --0.18 0.00   1& --0.04 0.00   1& --0.08 0.00   1& --0.07 0.00   1& --0.09 0.00   1 & --0.07 0.00   1&\\
Al I & --0.22 0.21   2& --0.24 0.11   2& --0.11 0.01   2& --0.30 0.00   1& --0.30 0.23   2 & --0.22 0.08   2&\\
Si I & --0.31 0.13  13& --0.28 0.11  13& --0.36 0.16  11& --0.34 0.15  13& --0.40 0.14  12 & --0.41 0.15  11&\\
Ca I & --0.35 0.12  13& --0.19 0.11  12& --0.39 0.08  12& --0.36 0.10  12& --0.42 0.08  11 & --0.40 0.11  12&\\
Ti I & --0.40 0.14  35& --0.40 0.17  38& --0.39 0.15  40& --0.31 0.16  36& --0.45 0.13  36 & --0.44 0.15  38&\\
Ti II& --0.21 0.14  16& --0.27 0.16  15& --0.24 0.17  14& --0.09 0.39  16& --0.30 0.37  15 & --0.23 0.39  16&\\
Cr I & --0.59 0.11   8& --0.55 0.18   7& --0.61 0.11   6& --0.60 0.15   8& --0.70 0.07   7 & --0.69 0.09   7&\\
Cr II& --0.53 0.06   6& --0.56 0.12   4& --0.53 0.15   6& --0.54 0.12   3& --0.63 0.07   6 & --0.70 0.16   4&\\
Fe I & --0.53 0.17 201& --0.54 0.16 193& --0.54 0.16 193& --0.51 0.17 191& --0.60 0.15 194 & --0.57 0.16 196&\\
Fe II& --0.54 0.09  18& --0.55 0.11  16& --0.55 0.11  18& --0.51 0.14  17& --0.58 0.10  17 & --0.59 0.10  19&\\
Ni I & --0.54 0.24  28& --0.46 0.23  28& --0.60 0.10  27& --0.57 0.11  27& --0.65 0.09  28 & --0.65 0.09  26&\\
Zn I & --0.21 0.17   2& --0.07 0.45   2& --0.22 0.05   2& --0.29 0.40   2& --0.49 0.00   1 & --0.43 0.24   2&\\
\noalign{\smallskip}
\hline
\end{tabular}
\begin{tabular}{lllllllllll}
\hline\hline 
\noalign{\smallskip}
El   &  NGC\,6352-07    & NGC\,6352-08     & NGC\,6352-09       & NGC\,6352-mean & NGC\,6352-median\\
\noalign{\smallskip}
\hline
\noalign{\smallskip}
Na I & --0.48 0.06  4 & --0.27 0.05   4& --0.46 0.10   4  & --0.37 0.14& --0.46 \\
Mg I & --0.09 0.00  1 & --0.03 0.00   1& --0.07 0.00  1   & --0.08 0.05& --0.07 \\
Al I & --0.22 0.04  2 & --0.29 0.00   1& --0.16 0.00 1    & --0.23 0.06& --0.22  \\
Si I & --0.30 0.06 12 & --0.42 0.20  12& --0.31 0.11  12  & --0.35 0.05& --0.34 \\
Ca I & --0.38 0.05 11 & --0.38 0.09  12& --0.38 0.11 12   & --0.36 0.07& --0.38 \\
Ti I & --0.34 0.13 38 & --0.42 0.15  36& --0.42 0.14  38  & --0.40 0.05& --0.40 \\
Ti II& --0.08 0.37 16 & --0.16 0.38  14& --0.20 0.41  15  & --0.20 0.08& --0.21   &  \\
Cr I & --0.58 0.16  7 & --0.65 0.10   7& --0.64 0.12   7  & --0.62 0.05& --0.61 \\
Cr II& --0.45 0.07  5 & --0.64 0.08   4& --0.69 0.11   6  & --0.59 0.08 & --0.56\\
Fe I & --0.53 0.17 201& --0.55 0.17 196& --0.57 0.18 197  & --0.55 0.03& --0.54 \\		   		         
Fe II& --0.54 0.09 18 & --0.56 0.10  17& --0.59 0.11 18   & --0.55 0.03 & --0.55  \\		         
Ni I & --0.57 0.14  26& --0.65 0.11  26& --0.61 0.09  26  & --0.59 0.06& --0.60 \\
Zn I & --0.27 0.03  2 & --0.45 0.16   2& --0.28 0.04  2   & --0.30 0.13& --0.28\\
\noalign{\smallskip}
\hline
\end{tabular}
\end{table*}

\begin{figure*}
\centering
\resizebox{\hsize}{!}{\includegraphics{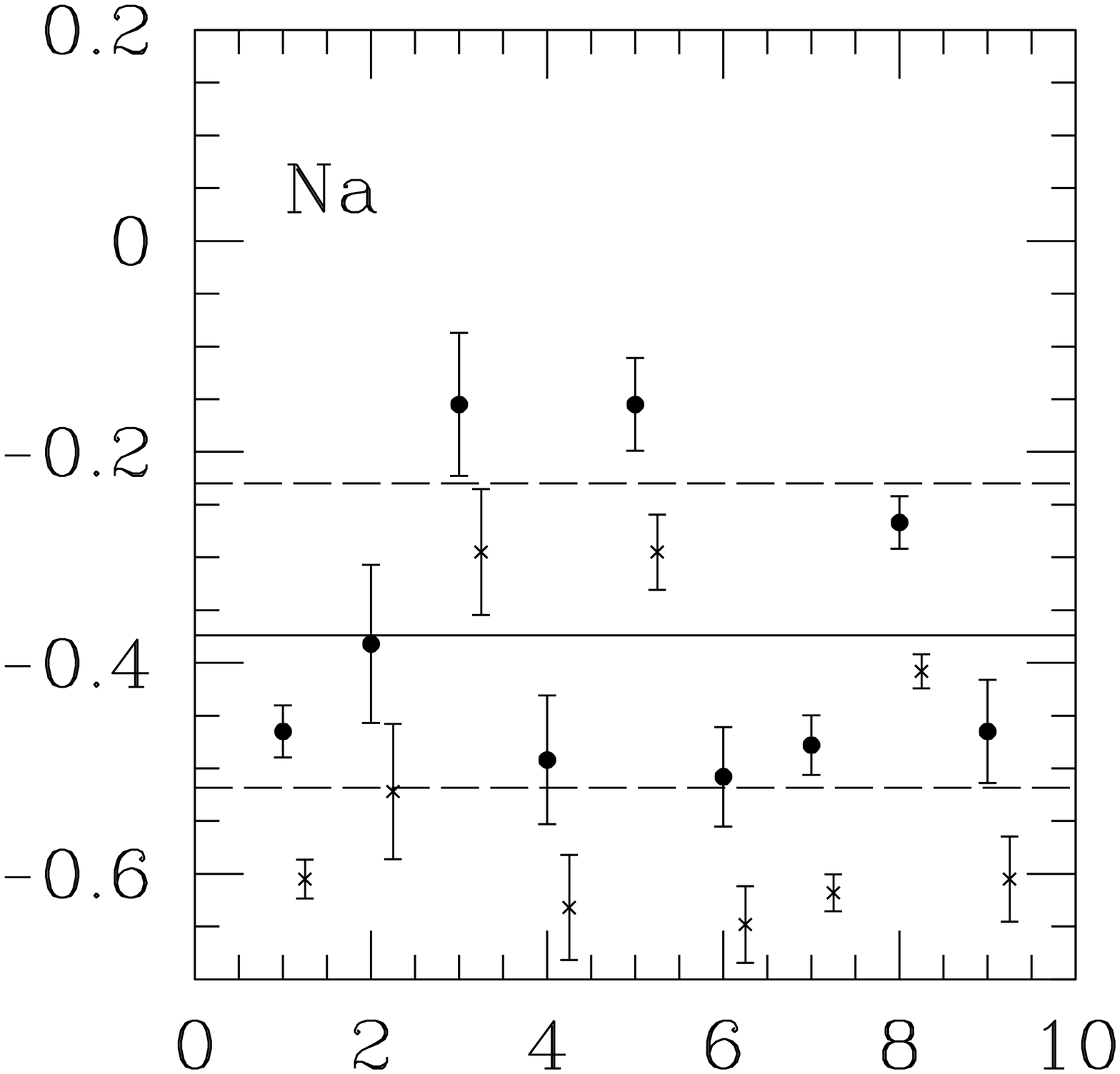}\includegraphics{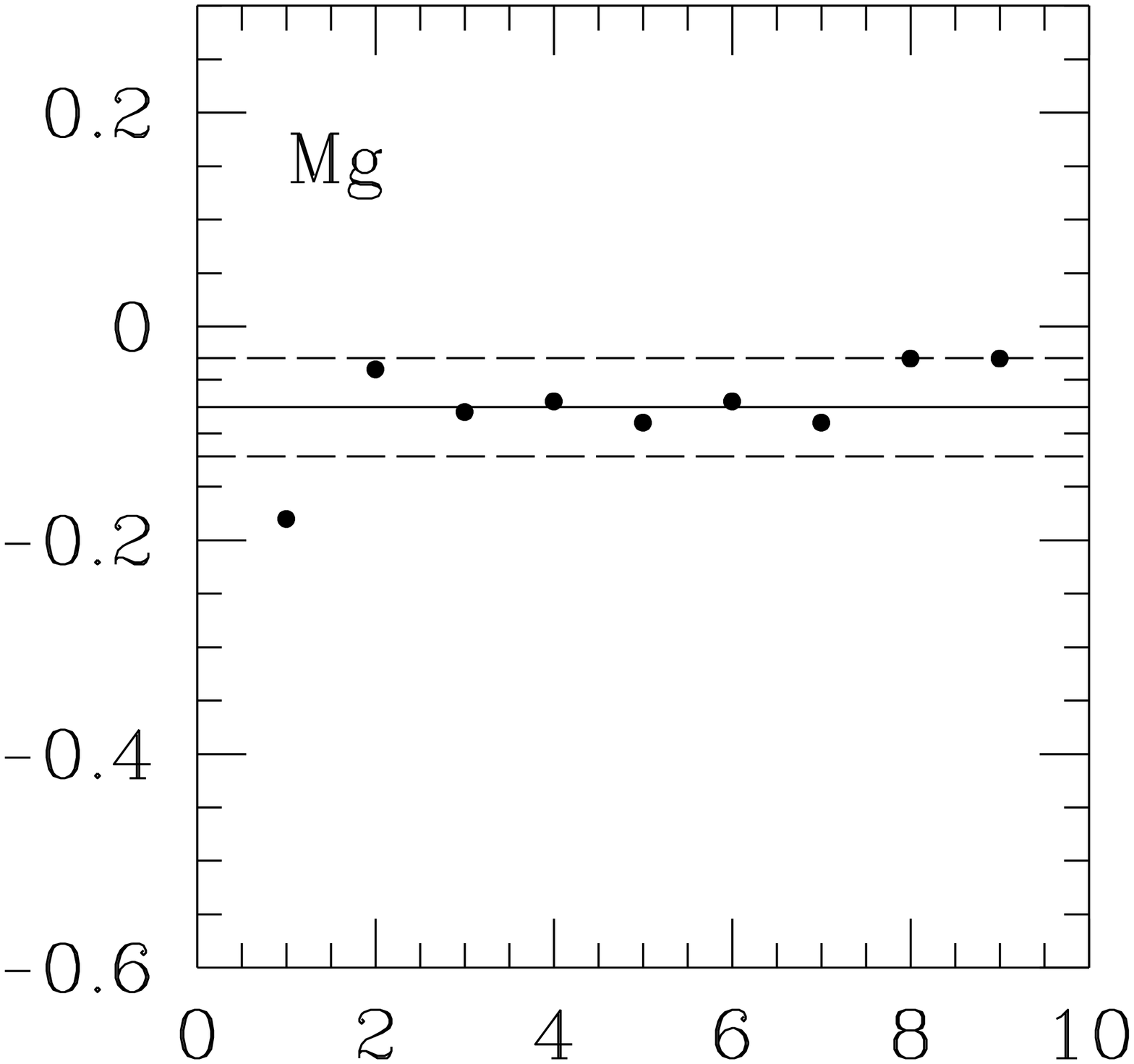}\includegraphics{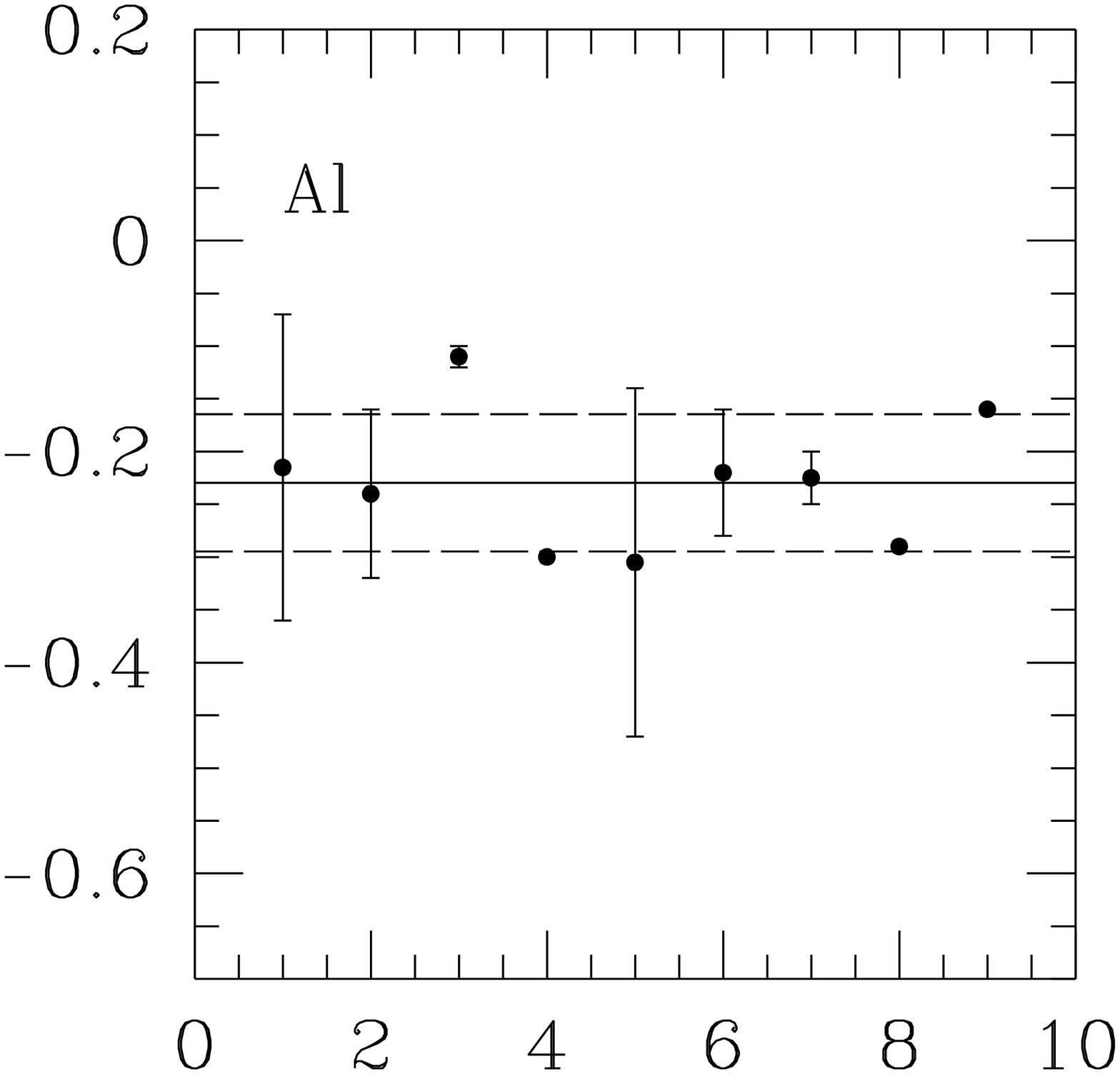}\includegraphics{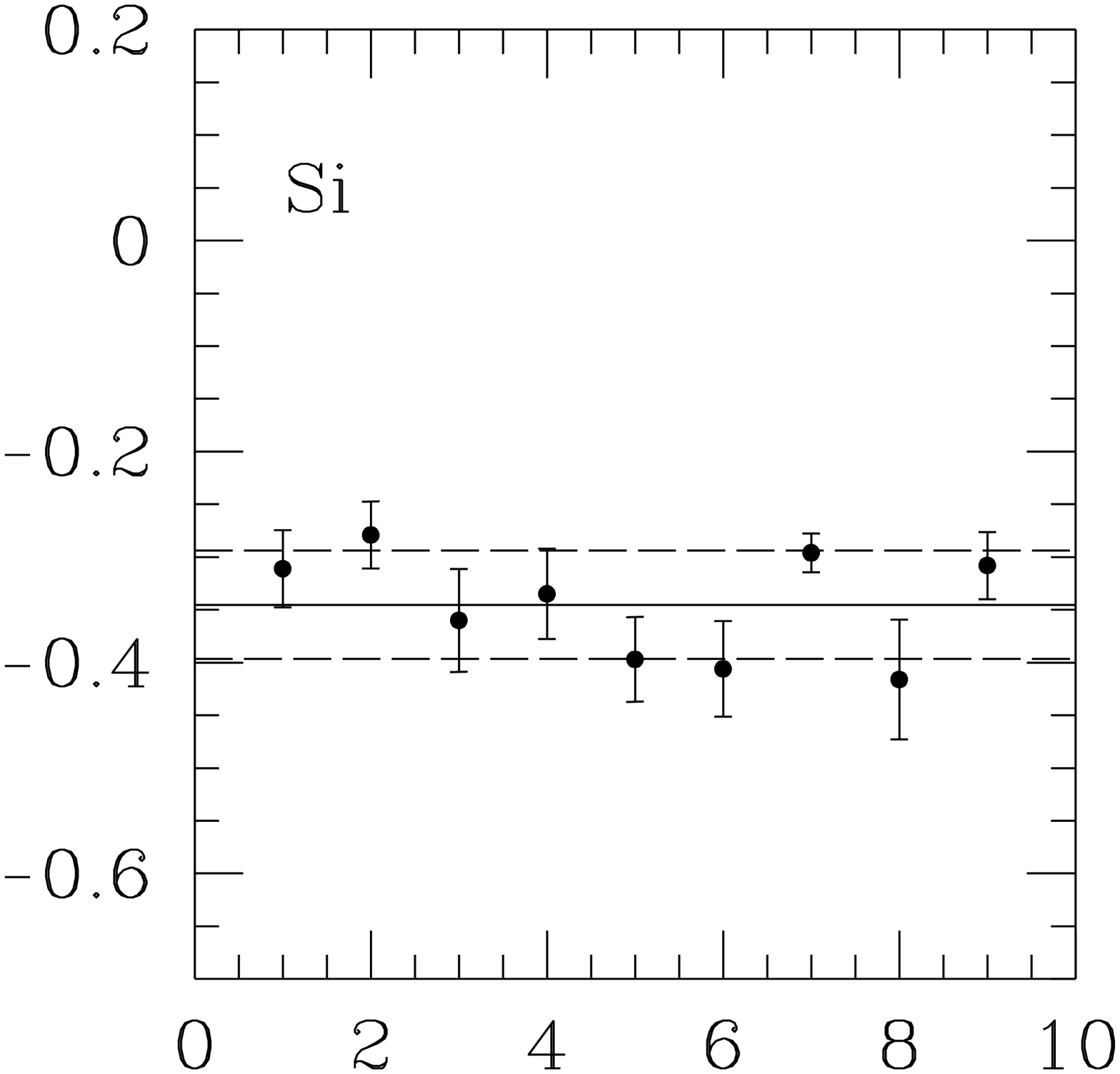}\includegraphics{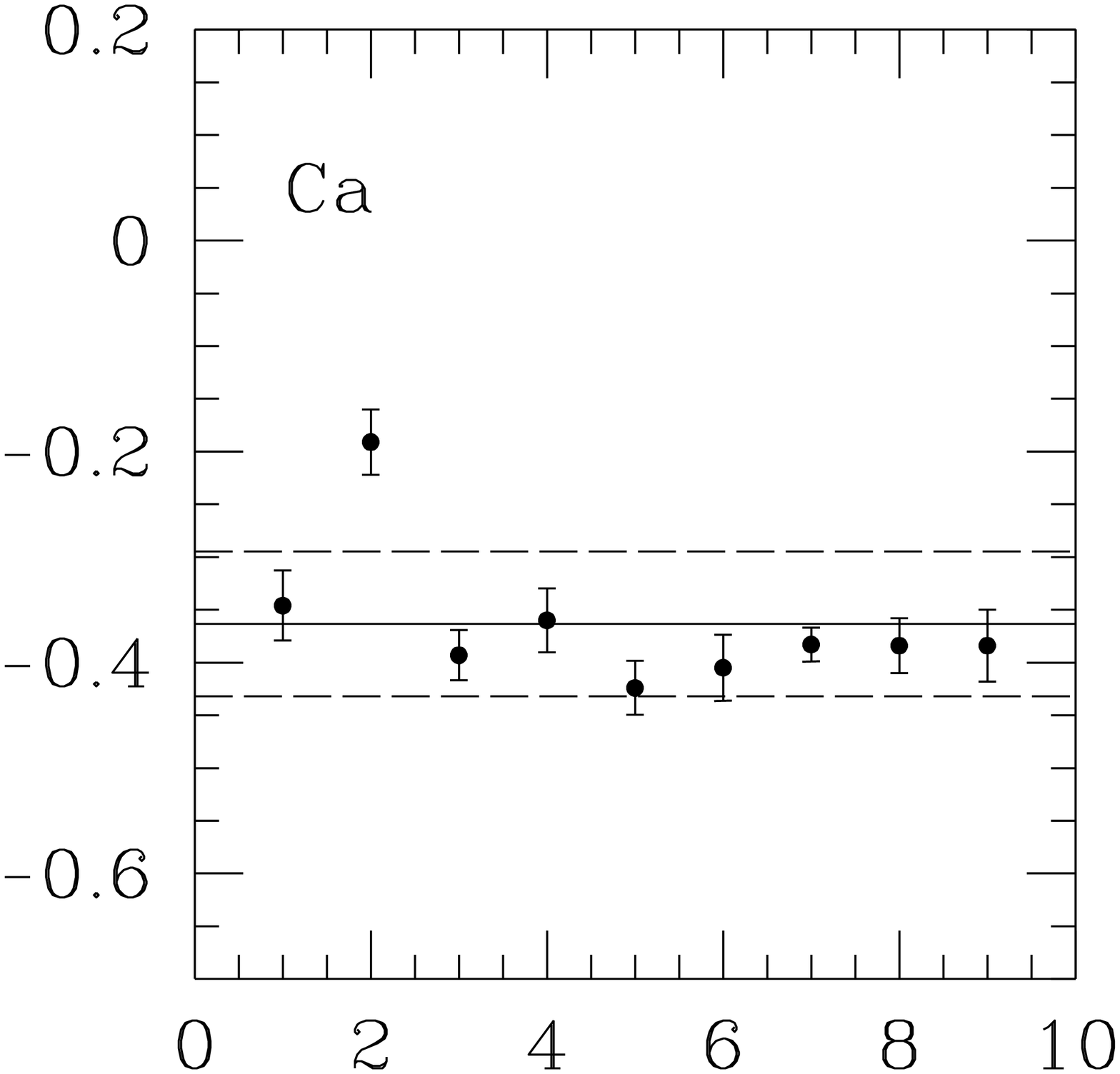}}
\resizebox{\hsize}{!}{\includegraphics{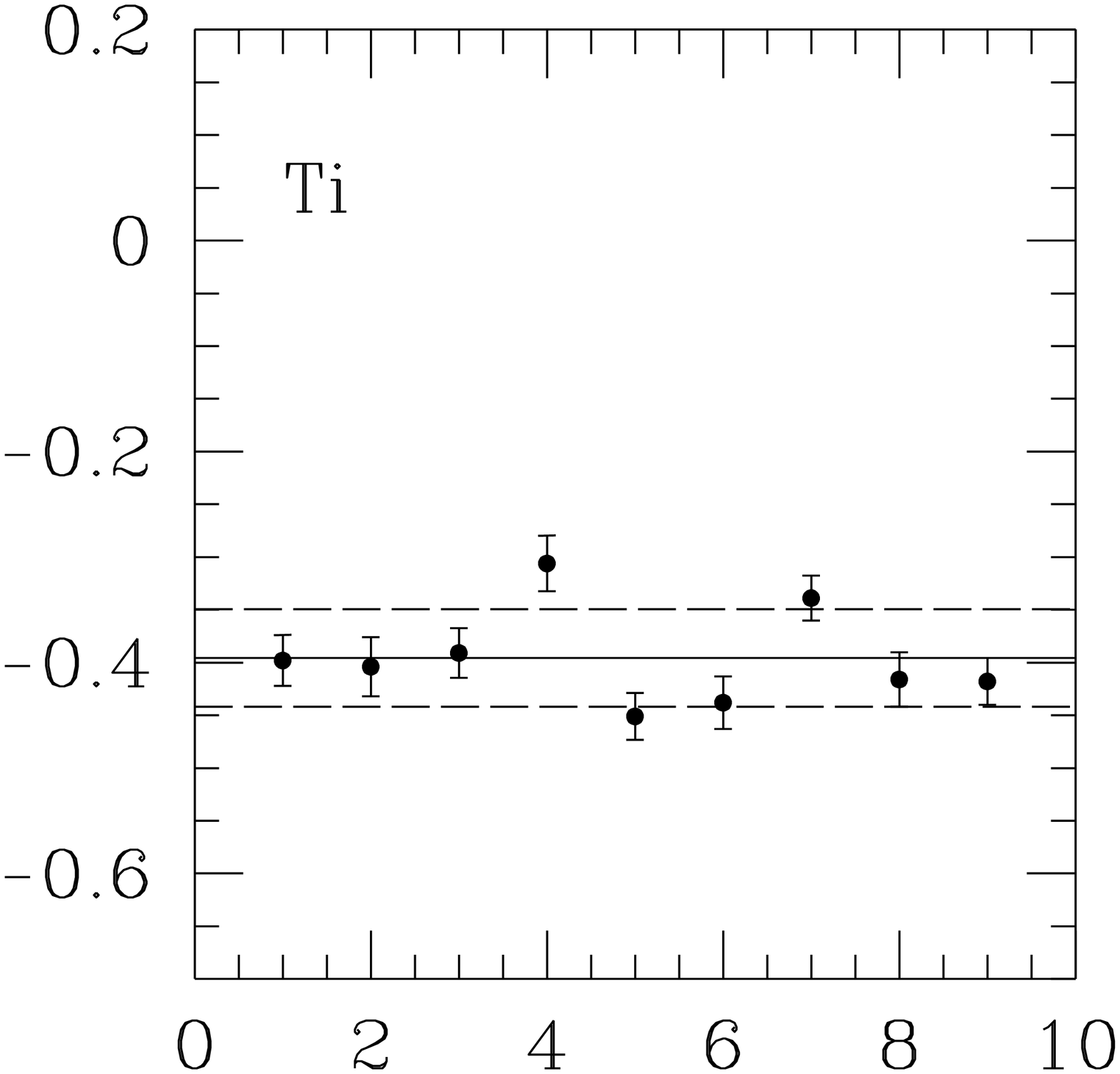}\includegraphics{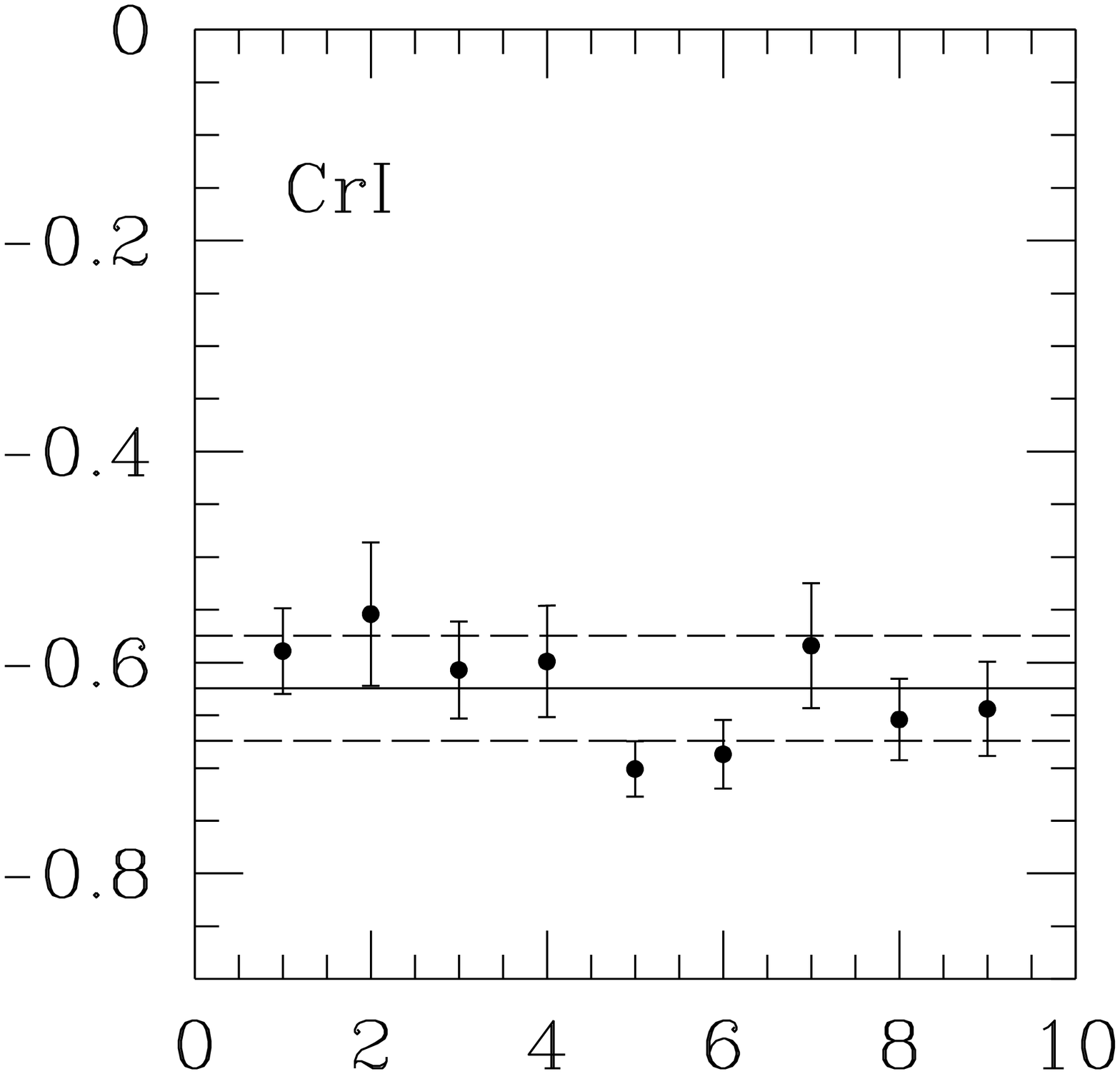}\includegraphics{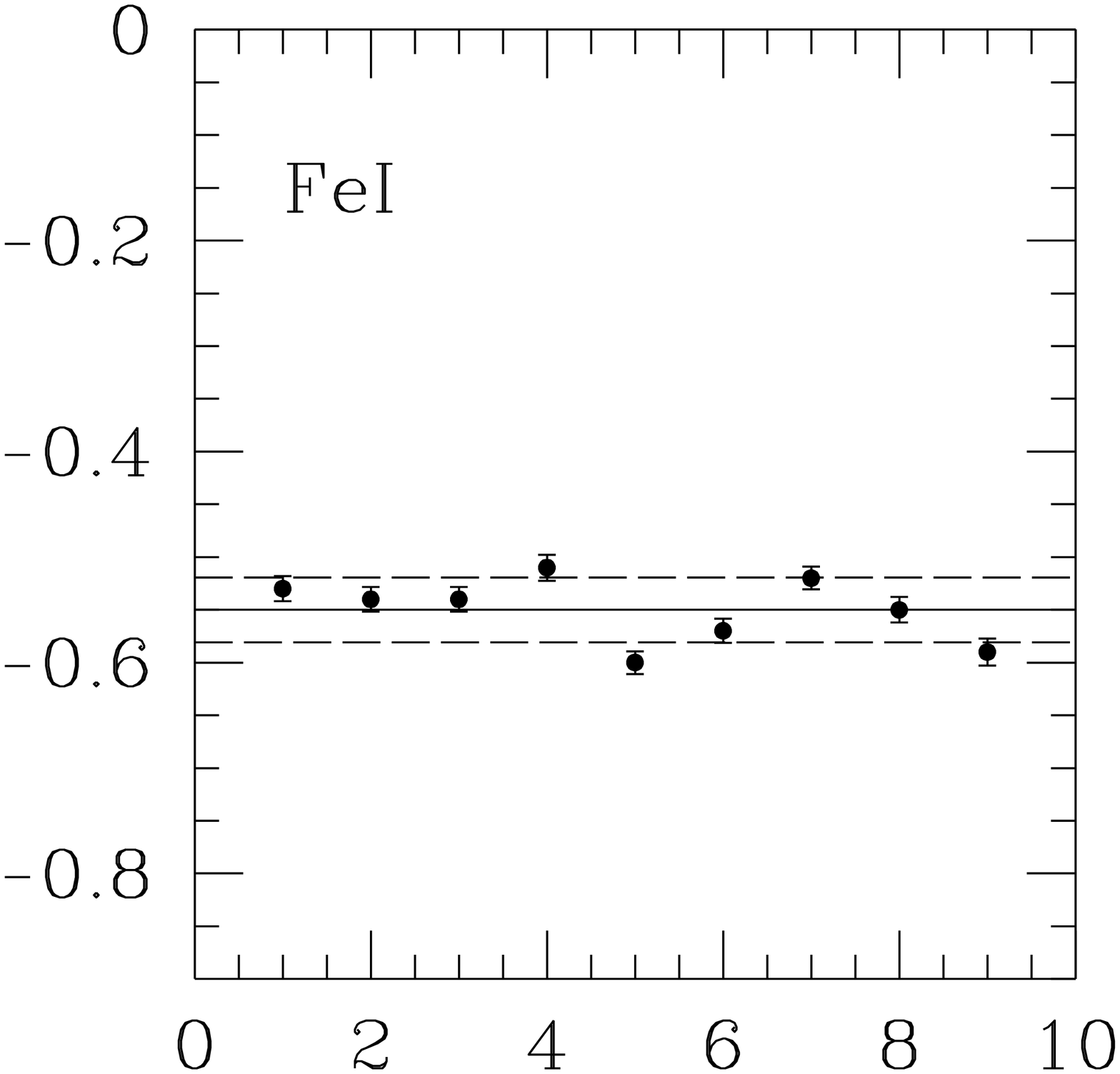}\includegraphics{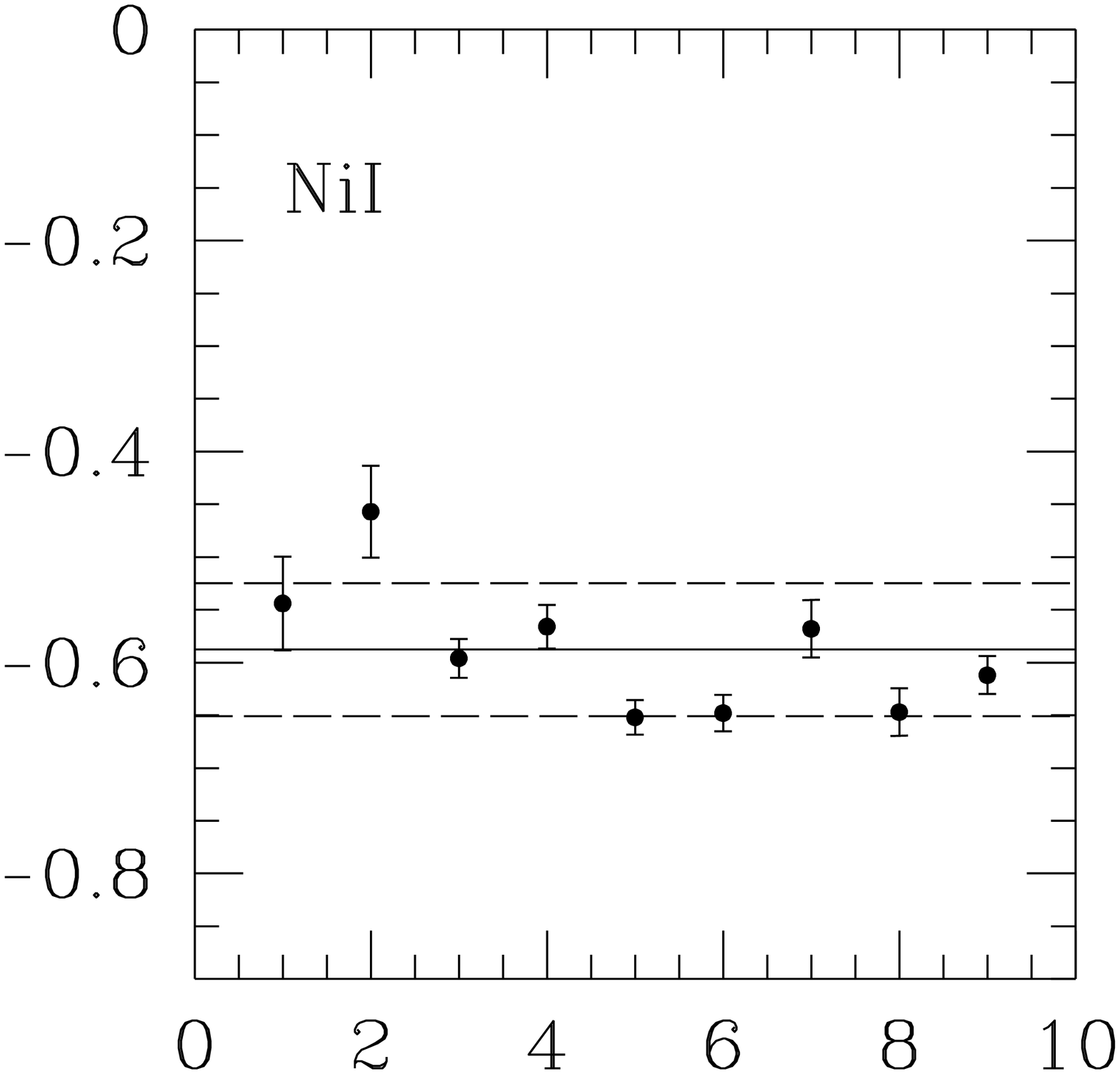}\includegraphics{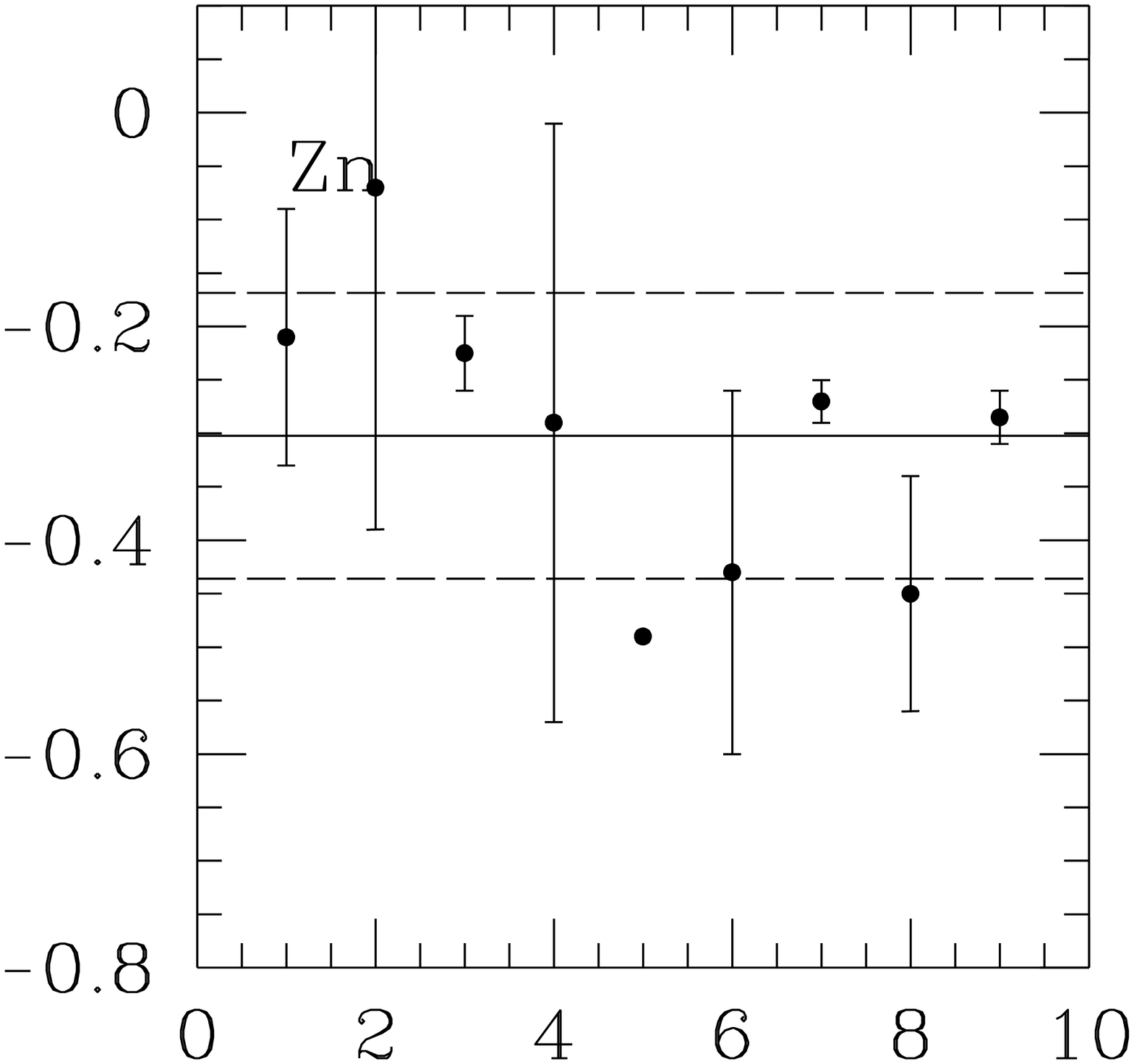}}
\caption{Elemental abundances for individual stars. On the $y$-axes we
  show [X/H], where X is the element indicated in the upper left hand
  corner in each plot. On the $x$-axes are the ID numbers of the stars
  (as defined in Table\,\ref{phot.tab}). For each star we also plot
  the error in the mean as an error-bar. For Na we also show the NLTE
  corrected data (Table\,\ref{phot.tab}) as $\times$.  For Mg we have
  only analyzed one line, hence no error-bar. The same is true for
  three stars as concerns Al. For each element we indicate the cluster
  mean with a solid line and the associated $\sigma$ (based on the
  values for all nine stars) with a dashed line above and below (see
  Table\,\ref{ab.tab}, penultimate column).  }
\label{fig.abind6352}
\end{figure*}

\begin{figure}
\centering
\resizebox{\hsize}{!}{\includegraphics{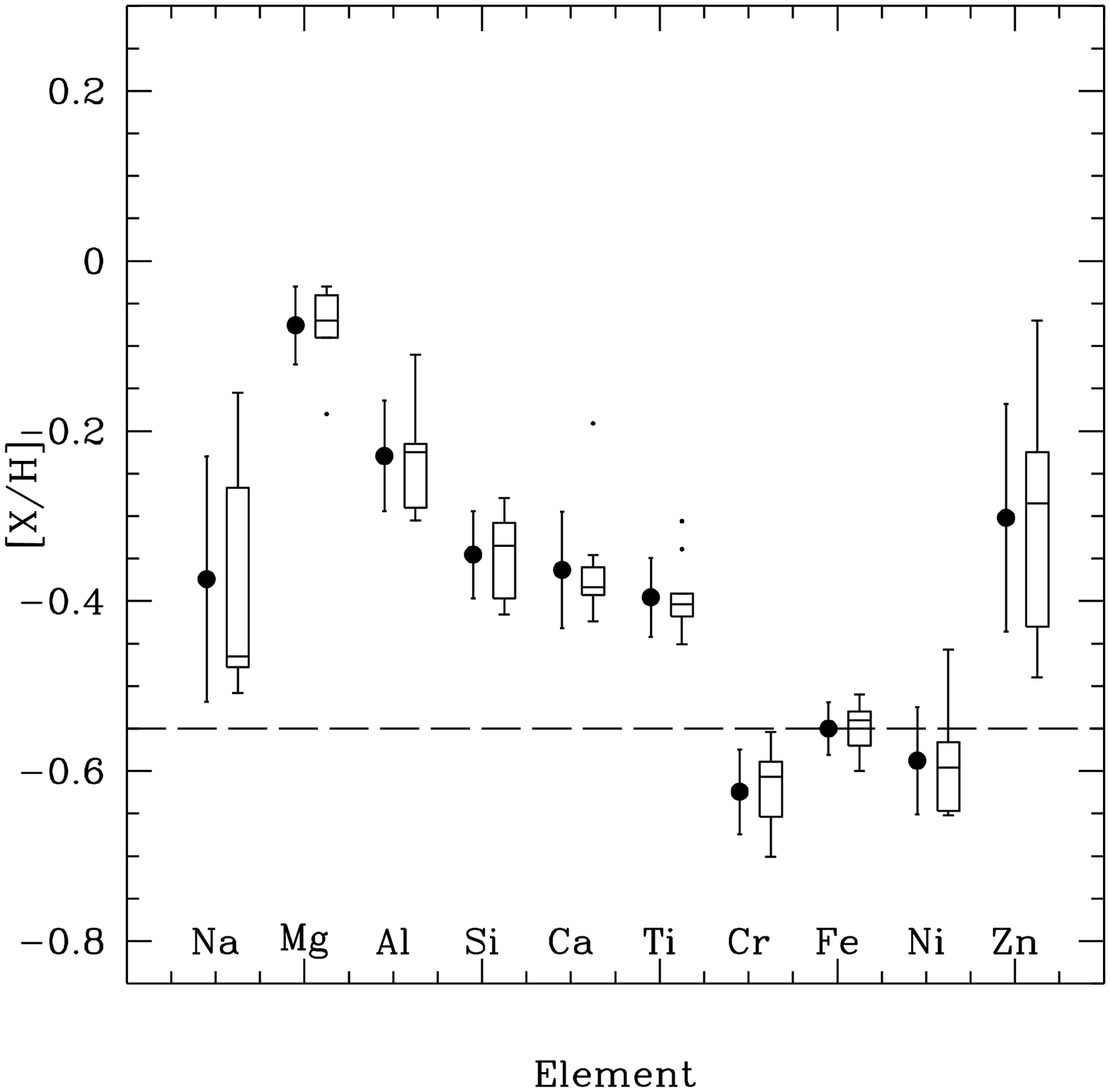}}
\caption{Elemental abundances for the globular cluster
NGC\,6352. $\bullet$ with error bars indicate the mean abundance for the
cluster with it's associated scatter (see Table\,\ref{ab.tab}).  The
dashed, horizontal line indicate the mean [Fe/H] value for the cluster
stars.  For each element we also plot the abundances using a so called
box-plot.  In the boxplots the central vertical line represents the
median value.  The lower and upper quartiles are represented by the
outer edges of the boxes, i.e.  the box encloses 50\% of the sample.
The whiskers extend to the farthest data point that lies within 1.5
times the inter-quartile distance.  Those stars that do not fall
within the reach of the whiskers are regarded as outliers and are
marked by dots.  }
\label{fig.ablightngc6352}
\end{figure}

We have derived elemental abundances for 9 horizontal branch stars in
NGC\,6352.  Our results are reported in Table\,\ref{ab.tab} and
Figs.\,\ref{fig.abind6352} and \ref{fig.ablightngc6352}. 

All our abundances have been determined based on a 1D LTE analysis,
though we did check the most up-to-date references on NLTE studies of
all the elements investigated here. When relevant, a note has been
added in the discussion below, but we note that most of the NLTE
investigations have been carried out for solar-type dwarf stars, hence
they rarely cover the parameter space spanned by our stars.

\subsection{Results -- Fe, Ni and other iron-peak elements} 

The mean iron abundance for NGC\,6352 (relative to the Sun) is
$-0.55\pm0.03$ dex. Although it is thought that \ion{Fe}{i} lines
suffer from NLTE effects \citep[e.g.][]{collet2005,thevenin1999},
 the magnitude of these effects are not yet fully
established. Opposing results (very small or very large effects) have
been found by different authors, even when studying the same
objects. NLTE effects are expected to be of the order of 0.05~dex in
stars like the Sun, and possibly increase at low metallicities and
gravities. Fe\,{\sc ii} lines remain the safest
solution, but since we have imposed the ionization balance in order to
derive $\log g$ for our stars, our metallicity scale has not been
corrected for any non-LTE effect.

Ni appears somewhat under-abundant compared to iron at [Ni/Fe]$=-0.04$,
and also Cr is slightly less abundant than iron at
[Cr/Fe]$=-0.07$. Both results are very compatible with what is seen
for local field dwarf stars at the same [Fe/H], one example is given
by \citet{bensby2005}, as well as with results for galactic bulge
stars (see Sect.\,\ref{sect:comp}).

NGC\,6352-02 has a higher [Ni/H] abundance than the rest of the stars
and NGC\,6352-01 and NGC\,6352-02 have a higher line-to-line scatter.
We have few direct explanations for these results, although it is
expected that the scatter in general should increase as we go to
cooler stars \citep[][their Fig.\,1]{luck2007} and NGC\,6352-02 is
our coolest star and NGC\,6352-01 one of the cooler ones. It is also
true that NGC\,6352-02 has the largest scatter in Cr\,{\sc ii}
abundances too.

Zn shows large error-bars. We note that \citet{bensby2003} found
that, for dwarf stars more metal-rich than the sun, one of the lines
started to give higher and higher Zn abundances while the other line
gave lower values. The reason for this is not clear but could have to
do with that either the line is blended or that the line experience
non-LTE effects as it gets stronger.  In the HB stars the line is
rather strong ($75-100$ m{\AA}).

As reported by \citet{asplund2005araa}, no NLTE analyses for iron-peak
elements (except iron) have been published so far.

\subsection{Results -- $\alpha$-elements}

The cluster is clearly enhanced in the $\alpha$-elements; for Si and
Ca the enhancement is around 0.2 dex relative to iron, (Fig.
\ref{fig.ablightngc6352}), while Ti is somewhat less enhanced and Mg
is more enhanced. The [Mg/Fe] should be taken with a pinch of salt as
we have only been able to measure one line and that line, although
clean and in a nice spectral region, is fairly strong in the HB stars
(112--118 {m\AA}). Nevertheless, these enhancements are typical for
dwarf stars in the solar neighbourhood that belong to the thick disk
and for galactic bulge stars (see Sect.\,\ref{sect:comp}).

We note that [Ca/H] for star NGC\,6352-02 deviates substantially from
those of the other stars. It appears that the difference is real as we
can not attribute it to e.g. continuum placement or significantly
different stellar parameters. We include this star in our mean
abundance for the cluster. If this star was excluded the resulting
abundance would be [Ca/H]$= -0.38 \pm 0.02$ as compared with $-0.36\pm
0.07$ if it is included.

Among these three $\alpha$-elements, only the abundances of magnesium
could be corrected for non-LTE effects, which for most lines are
positive (in the range 0.1--0.2~dex going from the Sun to metal-poor
stars). However, \citet{asplund2005} mentions a minor dependence of the
non-LTE effects on the effective temperature and gravity, which in
turn means that the abundances of our stars should have relatively
small corrections. Corrections for Si are expected to be negligible,
and the situation of Ca is highly uncertain.

\subsection{Results -- Na and Al}
\label{sect:naal}

\begin{table}
\caption[]{NLTE-corrected Na abundances. See Sect.\,\ref{sect:naal}
  for details of the correction. The first column identifies the stars
  according to Table\,\ref{phot.tab}. Column two and three gives the
  uncorrected Na abundances and those corrected for NLTE,
  respectively.  The last column gives the $\sigma$. }
\label{nanlte.tab}
\begin{tabular}{lllllllllll}
\hline\hline 
\noalign{\smallskip}
Star & [Na/H] & [Na/H] & $\sigma$\\
     & Uncorrected & NLTE corrected &\\
\noalign{\smallskip}
\hline
\noalign{\smallskip}
NGC\,6352-01 & --0.46 & --0.60 & 0.04 \\
NGC\,6352-02 & --0.38 & --0.52 & 0.13 \\
NGC\,6352-03 & --0.15 & --0.30 & 0.12 \\
NGC\,6352-04 & --0.49 & --0.63 & 0.10 \\
NGC\,6352-05 & --0.15 & --0.30 & 0.07 \\
NGC\,6352-06 & --0.50 & --0.65 & 0.07 \\
NGC\,6352-07 & --0.47 & --0.62 & 0.04 \\
NGC\,6352-08 & --0.26 & --0.41 & 0.03 \\
NGC\,6352-09 & --0.46 & --0.60 & 0.08 \\
\noalign{\smallskip}
\hline
\end{tabular}
\end{table}

Na is represented by four lines in each stellar spectrum, whilst Al
is represented by two lines in most of the stars (Table\,\ref{eqw.tab}
and \ref{ab.tab}).

Both Al and Na (as well as O) are known to vary from star to star in
globular clusters \citep[see e.g. review by][]{gratton2004}.  In fact,
for RGB stars several clusters show correlations between Al and Na
abundances \citep[see e.g. Fig.\, 14 in][for a compilation of several,
  mainly metal-poor, globular clusters]{ramirez2002} such that as
[Al/Fe] increases so does [Na/Fe]. The interpretation of this result
is complicated due to the fact that {\bf both} elements are subject to
NLTE effects, although the effect is largest at low metallicities.

For Na, different studies
\citep{baumuller1998,mashonkinaNa,takeda2003,shi2004} find very
similar results: non-LTE effects are stronger for warm, metal-poor
stars and for low gravity stars, and they depend on the lines employed
in the analysis. The smallest NLTE corrections apply to the
\ion{Na}{i} doublet at 615.4 and 616.0~nm (corrections are less than
0.1~dex for disk stars), and to the doublet at 568.2 and 568.8~nm (a
correction of $\approx$0.1~dex for dwarfs, though the correction seem
to increase for sub-giants).  \citet{mashonkinaNa} have studied the
statistical equilibrium of \ion{Na}{i} lines for a large range of
stellar parameters, including the ones characteristic of our
sample. Hence, for Na, we are in the position to be able to correct
our Na abundances with a certain confidence.  Based on Fig.~6 of
\citet{mashonkinaNa}, we have estimated non-LTE corrections of the
order of $-$0.12~dex for the 615.4/616.0~nm doublet and of $-$0.16~dex
for the abundances derived from the 568.4/568.8~nm doublet. We list
the revised Na abundances in Table\,\ref{nanlte.tab} and in
Fig.\,\ref{fig.abind6352} (Na panel) we show both sets of results.

For Al, instead, the situation is not as clear as for Na.  According
to \citet{baumuller1997}, non-LTE effects for the excited lines
at 669.6/669.8~nm (the ones we have used in this analysis) are smaller
than for the Al resonance lines, but they increase with decreasing
metallicity, and they are the highest at low gravities. Unfortunately,
no study of NLTE in Al has yet included mildly metal-poor giant stars,
hence it is very difficult to apply any correction to our
abundances. In Table~2 of \citet{baumuller1997}, the coolest
and lowest gravity object for which non-LTE effects for the excited
lines have been computed and found to be around $+$0.1~dex is a star
with effective temperature of 5630~K, $\log g$ = 3.08, and
[Fe/H]=$-0.18$. Because of these uncertainties, we have decided to
discuss both Na and Al as derived from our 1D LTE analysis, and only
show what would change in the [Na/Fe] vs. [Fe/H] diagram should 
we apply the corrections discussed above.

\begin{figure}
\centering
\resizebox{\hsize}{!}{\includegraphics{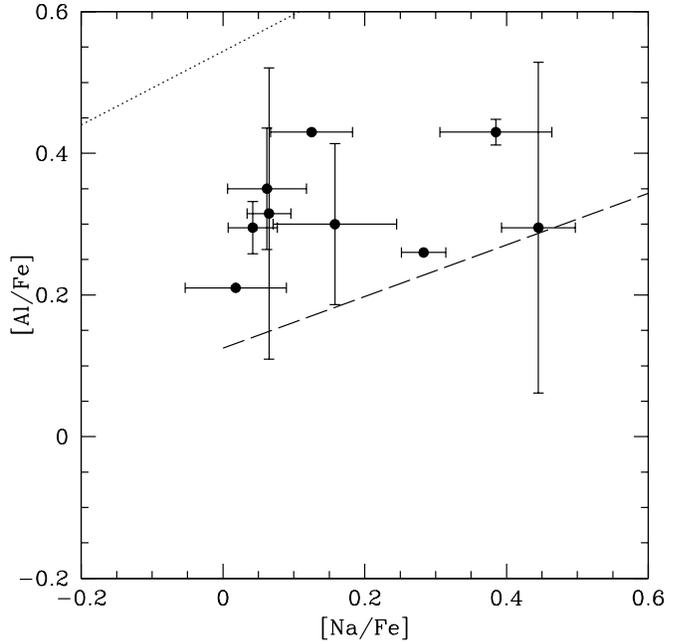}}
\caption{[Al/Fe] vs. [Na/Fe] for our stars. The dashed line 
indicates the relation found by \citet{ramirez2002}
for M\,71. The dotted line is also taken from that paper 
and represent the correlation for M\,4. Error-bars are
shown for all stars for [Na/Fe]. Three stars
only have one Al line measured. These do not have errorbars
 for [Al/Fe].
\label{fig.naal}}
\end{figure}

In Fig.\,\ref{fig.naal} we show our [Al/Fe] vs. [Na/Fe] and compare
them to those of \citet{ramirez2002} for the globular clusters M\,71
and M\,4 \citep[taken from][]{ramirez2002}. M\,71 is similar to
NGC\,6352 in that it has an intermediate metallicity ([Fe/H]=--0.71,
\citet{ramirez2002}. M\,4 is a metal-poor cluster \citep[at
  --1.18 dex,][]{harris1996}.  The data used to derive the relation
for M\,71 included one HB star, the rest are RGB stars, both above and
below the HB. Hence the comparison may be somewhat
unfair. Nevertheless we find that NGC\,6352 appears more enhanced in
Al than M\,71 and less enhanced in Na but the slope of the
correlation is similar. NGC\,6352 falls below the trend of
M\,4. Both M\,71 and M\,4 are less metal-rich than NGC\,6352. All data
in Fig.\,\ref{fig.naal} is without NLTE corrections.  Hence, there are
additional problems with this comparison in that different types of
stars (RGB vs HB) would have different corrections thanks to their
different $T_{\rm eff}$.

We can  conclude that it appears  that our data indicate  that also on
the  HB  there is  a  trend  in Al  and  Na  abundances,  and that,  in
metal-rich globular  clusters, these correlate in a  manner similar to
that found for stars on the RGB in other globular clusters.

\subsection{Comparison of elemental abundances with results from 
previous studies}
\label{sect:comp}

\begin{figure}
\centering
\resizebox{\hsize}{!}{\includegraphics{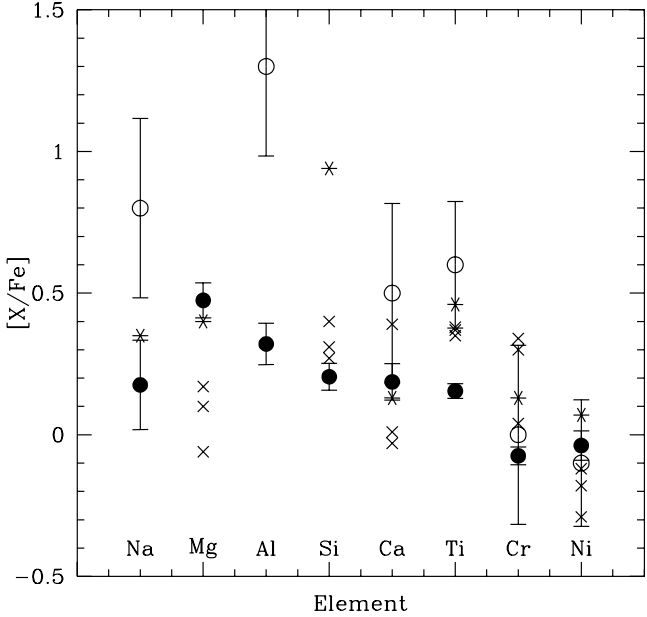}}
\caption{Comparison of our abundances ($\bullet$)
with those derived by \citet{geisler1981} ($\circ$)
and \citet{gratton1987} ($\times$), and \citet{francois1991} ($\ast$). 
The \citet{gratton1987} abundances have been 
corrected, see Sect.\,\ref{sect:comp}. Error bars on our data indicate
the star-to-star scatter while those on the \citet{geisler1981}
data indicate their quoted total errors. There are no error bars 
available for the two other data sets.
\label{fig.abcomp}}
\end{figure}

In Fig.\,\ref{fig.abcomp} we compare our abundances relative to Fe with
those derived by \citet{geisler1981}, \citet{gratton1987}, and
\citet{francois1991}.  The relative measure of [Element/Fe] should be
more robust against erroneous model parameters than [Element/H] (see
Table\,\ref{errors.tab}).  \citet{geisler1981} analyzed the spectrum
of one star (H37).  The \citeauthor{geisler1981} star was found to
have an effective temperature of 4200 K, a $\log g$ of 0.9 dex, and
[Fe/H]=$-1.3\pm0.1$.  Their abundances agree well with ours for Cr and
Ni and reasonably well for Ca and Ti while the lighter elements
differs significantly. This is probably mainly due to the small number
of lines available for those elements in the study by
\citet{geisler1981} which means that an error in $W_{\lambda}$ and/or
$\log gf$-value for a single line will have a larger impact than when
many lines are available. We note that their [Fe/H] differs
significantly from ours. The material available in the literature does
not allow a deeper investigation of this discrepancy.

\citet{gratton1987} analyzed spectra of five metal-rich globular
clusters.  He analyzed spectra of three NGC\,6352 stars and derived a
mean [Fe/H] of $-0.79$ dex. \citet{rutledge1997} later confirmed the
cluster membership for two of these stars (H111 and H142).
\citet{cg97} later reanalyzed the stars measured by
\citet{gratton1987}. Comparing $W_{\lambda}$s from their new and old
spectra (for a few of the clusters where such material was available)
they concluded that the $W_{\rm \lambda}$s in \citet{gratton1987} were
overestimated and derived a correction formula. Using the correct
$W_{\rm \lambda}$ they derived an [Fe/H] of $-0.64$ dex. They only
re-analyzed the Fe\,{\sc i} lines from \citet{gratton1987}. As
\citet{gratton1987} did not measure any Fe\,{\sc ii} lines we are not
in a position to re-analyse his data using our method as described in
Sect.\,\ref{sect:finalpar} as that requires ionizational
equilibrium. Instead we have derived a scaling of the abundances in
\citet{gratton1987} using the strength of the tabulated $W_{\rm
  \lambda}$s in his Table 6 using Eq. (1) in \citet{cg97}. Note that
this equation {\it is} valid for NGC\,6352 as it has essentially the
same metallicity as Arcturus (see their discussion). The applied
corrections are essentially $+0.1$ dex for all the elements apart from
Si\,{\sc i} which has a correction of $+0.2$ dex. This is due to that
Si\,{\sc i} is represented by weaker lines for which the correction
is larger.

In Fig.\,\ref{fig.abcomp} we compare our data with the data from
\citet{gratton1987} corrected as described above. For some elements,
e.g. Mg, Si, and Ti, the data for his three stars agree very well with
each other while for other elements, notably Ca and Cr, one of the
stars deviates significantly from the two other stars. Comparing with
our data the agreement is very good for Ca and Si but less good for
the lighter elements, i.e. Mg. We also note that there is a large
discrepancy between the Cr and Ti abundances from the two studies. As
before most of this is likely attributable to the few lines available
for the light elements and for Cr (in Gratton only one line, we use
six lines). We are more concerned about the discrepancy between the Ti
abundances. One possible explanation could be the different treatment
we use for the collisional broadening.

\citet{francois1991} derived elemental abundances for six giant stars
in three globular clusters (four stars in NGC\,1904 and one star in
NGC\,5927 and NGC\,6352, respectively). The comparison with our data
(Fig.\,\ref{fig.abcomp}) shows an overall agreement in that
$\alpha$-elements are enhanced while iron group elements are
solar. There is one notable difference: Si. There is not enough
information available to further investigate this discrepancy.

Overall we find that the agreement between our results and results
from earlier investigations is remarkably good considering the
difficulties facing the study of faint, metal-rich stars in globular
clusters. This compassion further strengthens our confidence in our
abundance analysis and the conclusions that NGC\,6352 is clearly
enhanced in [$\alpha$/Fe] and have roughly solar [Cr/Fe] and [Ni/Fe].

\section{Discussion -- putting NGC\,6352 into context}
\label{sect:disc}

\begin{figure*}
\centering
\resizebox{\hsize}{!}{\includegraphics[angle=-90]{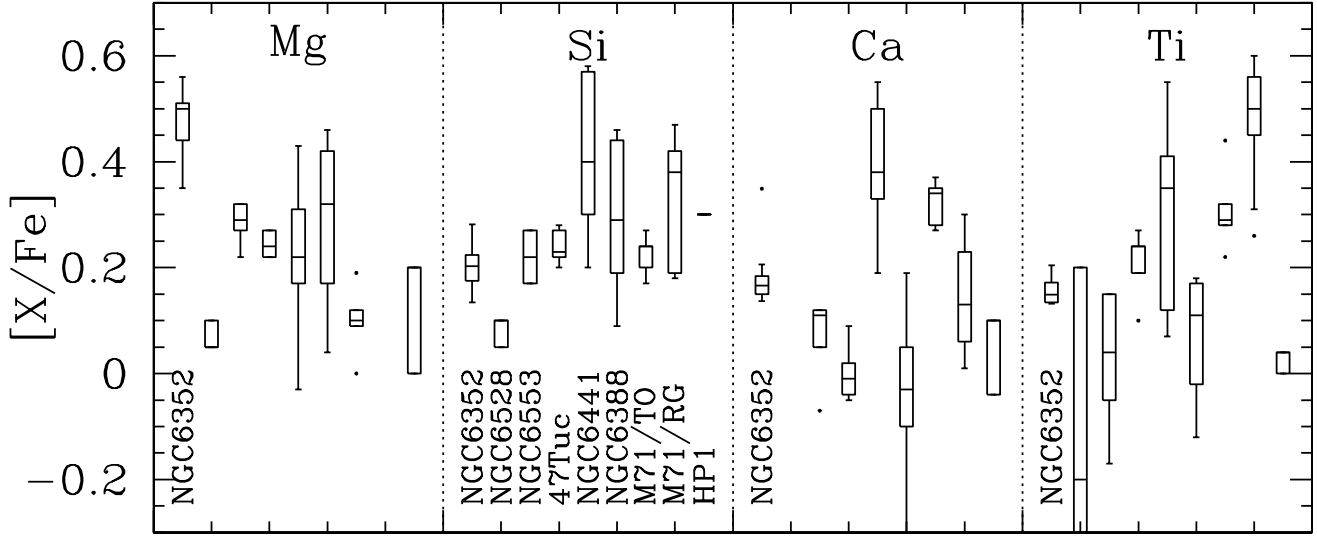}}
\caption{Comparison of [Element/Fe], where Element is Mg, Si, Ca, and
  Ti, for NGC\,6352 with the clusters listed in
  Table\,\ref{tab:compcl}. Our data for NGC\,6352 is indicated in all
  four panels, whilst the other clusters are identified in the second
  panel ([Si/Fe]). Table\,\ref{tab:compcl} gives the references and the
  number of stars included from each
  study. Section\,\ref{sect:selection2} discusses our selection of
  comparison data. The data for each study are shown in the form of
  box-plots. The lower and upper quartiles are represented by the
  outer edges of the boxes, i.e. the box encloses 50\% of the sample.
  The whiskers extend to the farthest data point that lies within 1.5
  times the inter-quartile distance. Those stars that do not fall
  within the reach of the whiskers are regarded as outliers and are
  marked by solid circles. There is no Mg data for the RG sample for
  M\,71 and the two stars in HP-1 have the same [Si/Fe] abundance. }
\label{fig.compgc.alpha}
\end{figure*}

We now attempt a first comparison of the elemental abundances 
we find in NGC\,6352 with those in other globular clusters as well
as for stars in the field (solar neighbourhood and the Galactic Bulge).
Our selection of comparison clusters is outlined below and then
follows a brief discussion putting NGC\,6352 into context.

\subsection{Selection of studies of other metal-rich globular clusters to compare NGC\,6352 with}
\label{sect:selection2}

When compiling stellar abundances from different studies there are a
number of considerations to take into account. For giant stars there
are two main issues that stands out: {\bf a)} increasing importance of
molecular lines in the stellar spectra as the stars get cooler
\citep{fulbright2006}, and {\bf b)} the need to include the sphericity
of the stars in the calculation of model atmospheres and elemental
abundances \citep{heitereriksson}.

In our study of NGC\,6352 we have only included HB stars to avoid the
issue of molecular lines (as they are warmer than the RGB stars). HB
stars are also in the region where plane parallel stellar models can
be used \citep{heitereriksson}. A first consideration would therefore
be to only compare our elemental abundances with those of other
studies of HB stars in globular clusters. This, it turns out, is
however, rather limiting as few studies have focused on HB stars.

An additional concern when selecting studies to compare with is the
different methods used by different studies to derive the stellar
parameters. In our study we have used ionizational equilibrium to
derive $\log g$ (i.e. requiring that iron abundances derived from
Fe\,{\sc i} and Fe\,{\sc ii} lines yield the same iron abundance). As discussed
in Sect.\,\ref{sect:justification} this method is valid for our stars.
We have therefore chosen to use only data from studies that employ the
same methods as we do when deriving the stellar parameters or studies
that even though the route is different their analysis yields
ionizational equilibrium. For the latter type of studies we have only
included stars for which ionizational equilibrium is achieved.
Obviously, through this process a number of studies were excluded. We
would like to note that this decision and hence exclusion of some
studies should not be taken as judgment regarding these studies.  We
believe that it is more interesting to make a comparison between
studies that use methods that are closely related and hence that
systematic differences between the studies will be minimized and we
will thus be in a position to make an (almost) differential
comparison.

We used Harris' catalogue \citep[][]{harris1996} to source a list of all
globular clusters with [Fe/H]$>-1$ and searched the literature (with
the help of ADS and ArXiv/astro-ph) for spectroscopic studies of the
stars in these clusters.  The clusters, and number of stars selected
from each study, are listed in Table\,\ref{tab:compcl}.

Additionally, there is an emerging literature were NIR
spectra are deployed.  This is, of course, especially beneficial for
the study of heavily obscured clusters and clusters with differential
reddening. However, for our comparison we decided not to include these
studies, as it would be difficult to make comparisons with the data
obtained from visual spectra.

We have not attempted to normalize the elemental abundances that we have taken
from the different studies. Although the studies all give ionizational
equilibrium they have not all used the same type of model atmosphere nor 
the same set of atomic line data. As there are no stars overlapping between
the different studies a normalization becomes difficult and it might in 
the end only add noise to the data. We have chosen to look at the data
``as is'' as we are especially concerned with general trends rather than
detailed comparisons or very small differences we believe that this approach
is the more advisable at this stage.

\begin{table*}
\caption[]{References for the clusters used in
  Figs.\,\ref{fig.compgc.alpha} and \ref{fig.compgc.nicr}. The first
  column gives the cluster name, the second to fourth list the number
  of various types of stars: turn-off (TO), horizontal branch (HB), and
  red giants/asymptotic giant branch stars (RGB/AGB) taken from the
  study and used in our comparison, the fifth column lists the mean
  [Fe/H] quoted in the study (i.e. this includes all stars in their
  study, we may be using a subset of those stars, compare
  Sect.\,\ref{sect:selection2}), and the reference is given in the
  penultimate column with additional comments in the last column.}
\label{tab:compcl}
\begin{tabular}{lcccllllllllll}
\hline\hline 
\noalign{\smallskip}
Cluster & \multicolumn{3}{l}{$\#$ of stars}  &$<$[Fe/H]$>$ & Reference & Comment\\
        & TO & HB & RGB/AGB  &\\
\noalign{\smallskip}
\hline
\noalign{\smallskip}
47\,Tucanae  & & 1 & 4 &--0.66$\pm$0.12 & \citet{alvesbrito2005}\\
NGC\,6528    & & 1 & 2 & --0.10$\pm$0.20 & \citet{2004AandA...423..507Z} &\\
NGC\,6388    & & &8 &--0.80 & \citet{wallerstein2007} & Used the data for which ionizational equilibrium \\
             & &  & & &  &   was used  to derive $\log g$\\
NGC\,6441    & &  &9 & --0.34$\pm$0.02 & \citet{2007AandA...464..953G} & Only stars where ionizational equilibrium occurred \\
   & &  & & &  &are included (see Sect.\,\ref{sect:selection2}) \\
NGC\,6553    & & 3 & 1 &--0.20 & \citet{alvesbrito2006} & NMARCS \citep{plez1992}\\
M\,71 & & & 10 & --0.79$\pm$0.01& \citet{1994AJ....107.1773S} & \\
M\,71 & 5 & & & --0.80$\pm$0.02& \citet{2005ApJ...629..832B}  & \\
HP--1 & & &2  & --1.00$\pm$0.20 &  \citet{barbuy2006} & NMARCS \citep{plez1992}\\
\noalign{\smallskip}
\hline
\end{tabular}
\end{table*}

\begin{figure}
\centering
\resizebox{\hsize}{!}{\includegraphics[angle=-0]{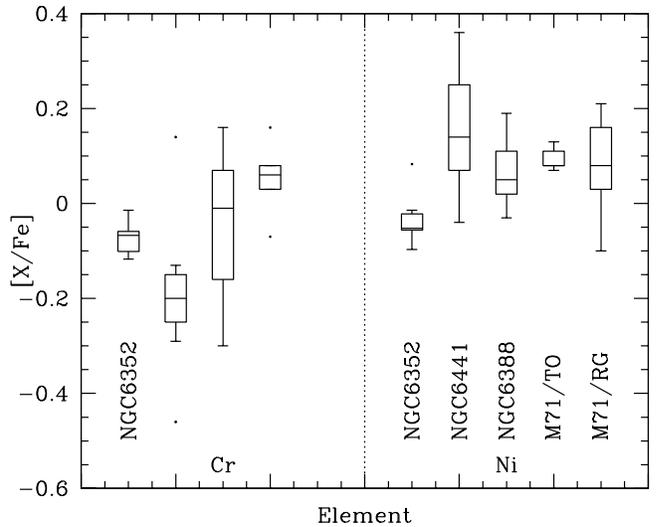}}
\caption{Comparison of [Cr/Fe] and [Ni/Fe] for NGC\,6352 with the
  clusters listed in Table\,\ref{tab:compcl} and that have Cr and Ni
  abundances measured. Our data for NGC\,6352 is identified in both
  panels and the other clusters are identified in the panel that shows
  the Ni abundances. The number of stars from each study are given in
  Table\,\ref{tab:compcl}.  The data for each study are shown in the
  form of box-plots. The lower and upper quartiles are represented by
  the outer edges of the boxes, i.e. the box encloses 50\% of the
  sample.  The whiskers extend to the farthest data point that lies
  within 1.5 times the inter-quartile distance. Those stars that do
  not fall within the reach of the whiskers are regarded as outliers
  and are marked by solid circles. There are no Cr data for the RG
  stars in M\,71. }
\label{fig.compgc.nicr}
\end{figure}

In Figs.\,\ref{fig.compgc.alpha} and \ref{fig.compgc.nicr} we compare
our results for NGC\,6352 with elemental abundances relative to Fe for
the clusters in Table\,\ref{tab:compcl}.  [X/Fe] is preferred to [X/H]
(where X is any element) as that ratio is relatively more robust
against errors in the stellar parameters (compare
Sect.\,\ref{sect:errors}).

\subsection{Discussion}

\begin{figure}
\centering
\resizebox{\hsize}{!}{\includegraphics[angle=-0]{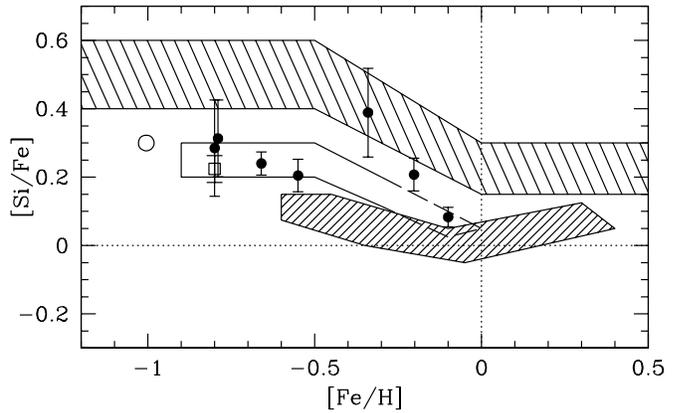}}
\caption{Comparison of abundances in metal-rich globular cluster with
  elemental abundance trends in the field (solar neighbourhood) and
  the Galactic bulge. The globular clusters are the same as in
  Table\,\ref{tab:compcl} and Figs.\,\ref{fig.compgc.alpha} and
  \ref{fig.compgc.nicr}. Here we show the mean value of [Si/Fe] for
  the selected stars as a function of [Fe/H] (as listed in the
  table). The TO stars in M71 are shown with a $\Box$ and the two
  stars in HP-1 with an $\circ$. The unshaded area marked with a solid
  and dashed line shows the trend for the thick disk and the more
  densely dashed area that of the thin disk
  \citep[e.g.][]{bensby2005}. The trend in the galactic Bulge is the
  other dashed area and this is based on \citet{fulbright2007}. No
  attempt to normalize the studies has been done. }
\label{fig.sinew}
\end{figure}

The major features of the elemental abundances in metal-rich globular
clusters is that they are enhanced in the $\alpha$-elements
(Fig.\ref{fig.compgc.alpha}) and that Ni and Cr closely follow Fe
(Fig.\ref{fig.compgc.nicr}). This appears to be the case regardless of
the [Fe/H] for the clusters (see Table\,\ref{tab:compcl}). Thus the
abundance patterns in the metal-rich globular clusters over-all
resembles that found in the halo, the thick disk, and the Bulge
\citep[e.g.][respectively, for the halo, thick disk and
  bulge]{arnone2005, bensby2005, fulbright2007} with the exception of
NGC\,6528 which shows consistent solar values for all
$\alpha$-elements. The observation that the metal-rich globular
clusters are enhanced in the $\alpha$-elements indicates that the
stars formed in these clusters were formed out of gas that had been
rapidly enriched in heavy elements produced in SN\,II but to lesser
extent, if at all, from SN\,Ia and hence more resemble the halo and thick disk
than the thin disk (compare Fig.\,\ref{fig.sinew}).

A few old, metal-rich open clusters have been studied
\citep[e.g.][]{carretta2007,sestito2007,yong2005}.  For NGC\,6253 and
NGC\,6791 \citet{carretta2007} find both $\alpha$-elements as well as
iron group elements to follow Fe. Thus they more resemble the
metal-rich thin disk \citep[compare plots
  in][]{carretta2007,bensby2005}.  It is interesting to note that the
most metal-rich stars in the thin disk in the solar neighbourhood not
necessarily are the youngest ones \citep[compare Fig. 4
  in][]{bensby2007} and hence the mean age of those field stars are
rather compatible with what is found for the old open clusters
discussed here.

This would indicate that, unless self-enrichment is a key element for
globular clusters, globular clusters in the Milky Way (in general)
trace the older stellar populations (as their ages also would
indicate) and, apparently, to no extent that of the thin disk. Whilst
the open clusters (at least the most metal-rich ones) follow the same
abundance pattern as that of the metal-rich thin disk.

In the Milky Way $\sim$150 globular clusters have been detected. They
present a bimodal metallicity distribution \citep[e.g.][]{zinn1985},
which may point to a period of enhanced cluster formation perhaps
triggered by a merger (compare e.g.  models and discussion in
\citet{casuso2006}. All the globular clusters in the Milky Way appear
to be old \citep[see e.g.][]{deangeli2005,rosenberg1999}.

\citet{zinn1985} divided the globular clusters in the Milky Way into
two groups according to their metallicity and showed that the majority
have metallicities peaking at $-1.6$ dex and are spatially and
kinematically distributed in a fashion similar to the halo stars. On
the other hand the clusters with [Fe/H]$\geq -0.8 $ dex peak at $-0.5$
dex and are strongly concentrated around the galactic nucleus, see
\citet{vandenbergh1993} for an excellent figure. This system is
thought to be physically and kinematically distinct from the more
metal-poor clusters \citep{zinn1985,armandroff1989}.  Further
divisions of the metal-rich clusters into disk and bulge clusters have
been discussed but this remains an open question
\citep[e.g.][]{minniti1995,zinn1996,harris1998}. Recently,
\citet{bica2006} found that the metal-rich globular clusters in the
Milky Way have a spatial distribution that is spherical which thus
appear to point more to a bulge than a thick disk connection. This is
somewhat in contradiction with the results by \citet{dinescu2003}
who, using the full spatial velocity for a set of globular clusters
find that at least one of them (NGC\,6528) is associated with the bar.

For NGC\,6352 we do not have the full set of space velocities as no
proper motion study of this cluster has ever been attempted. NGC\,6352
is situated outside the bar but in the Galactic plane, it has a
measured radial velocity along the line of sight ($V_{\rm LSR}=-120$
km\,s$^{-1}$). Thus its position and velocity (as far as we know) are
consistent with disk membership. At 5.4 kpc away from the bulge it is 
sufficiently far away that a Bulge membership can not be confirmed,
at least not until proper motions have been obtained. 

\section{Summary}
\label{sect:sum}

We present a study of elemental abundances for $\alpha$- and iron-peak
elements for nine HB stars in the metal-rich globular cluster
NGC\,6352. The elemental abundances are based on high-resolution, high
signal-to-noise spectra obtained with the UVES spectrograph on
VLT. The elemental abundances have been derived using standard LTE
calculations and stellar parameters have been derived from the
spectra themselves by requiring ionizational as well as excitational
equilibrium.

Our major findings are that the cluster:

\begin{itemize}
\item has {[Fe/H]$= -0.55$}
\item is enhanced in  the $\alpha$-elements
\item shows solar values for the iron peak elements
\end{itemize}

NGC\,6352 is a bulge/disk cluster. The final classification of this
cluster (based on its kinematic properties) must await the measurement
of proper motions and hence the derivation of the full space velocity
vector. However, the elemental abundances may appear to indicate a
disk rather than a bulge membership (if we believe that the clusters
accurately trace the underlying stellar populations).

Based on the stellar parameters derived from spectroscopy we suggest
that the reddening towards NGC\,6352 is $\sim 0.24$ and that the
distance modulus is $\sim 14.05$, which is somewhat smaller than the
nominal value of 14.44 quoted in the literature. However, our new
suggested distance modulus and reddening estimate are well within the
error-bars of previous estimates.

This is a first paper in a series of papers dealing with the elemental
abundances and ages of metal-rich globular clusters. We therefore
spent time on creating a homogeneous line-list that could be used for
all clusters. During this work we noted that there is a lack of
homogeneous data sets of line data for several of the iron group
elements. In particular do we lack laboratory data for Ni\,{\sc i} as
well as Cr\,{\sc i} and Cr\,{\sc ii} for lines that are useful in the
studies of HB stars.

When evaluating the available $\log gf$-values for Fe\,{\sc i} lines
we found that the correction factor to the \citet{may1974}
oscillator strengths suggested by \citet{fuhr1988} is not needed
for the lines we are employing in our abundance analysis and we hence
recommend the usage of the \citet{may1974} data as is.


\begin{acknowledgements}
We would like to thank Bengt Gustafsson, Bengt Edvardsson, and Kjell
Eriksson for usage of the MARCS model atmosphere program and their
suite of stellar abundance (EQWIDTH) and synthetic spectrum generating
programs.  We would also like to acknowledge the staff at Paranal and
at the ESO headquarters who, as part of their job, took and reduced
our stellar spectra.  SF is a Royal Swedish Academy of Sciences
Research Fellow supported by a grant from the Knut and Alice
Wallenberg Foundation.  This work has made use of the NED, SIMBAD, and
VALD databases. This publication makes use of data products from the
Two Micron All Sky Survey, which is a joint project of the University
of Massachusetts and the Infrared Processing and Analysis
Center/California Institute of Technology, funded by the National
Aeronautics and Space Administration and the National Science
Foundation.
\end{acknowledgements}

\bibliographystyle{aa}
\bibliography{referenser}

\clearpage
\Online
\clearpage

\end{document}